# Wide-Field InfraRed Survey Telescope
# WFIRST
# Interim Report

## Science Definition Team

J. Green[1], P. Schechter[2]
C. Baltay[3], R. Bean[4], D. Bennett[5], R. Brown[6], C. Conselice[7], M. Donahue[8], S. Gaudi[9], T. Lauer[10], S. Perlmutter[11], B. Rauscher[12], J. Rhodes[13], T. Roellig[14], D. Stern[13], T. Sumi[15], A. Tanner[16], Y. Wang[17], E. Wright[18], N. Gehrels[12], R. Sambruna[19], W. Traub[13]

Consultants
J. Anderson[6], K. Cook[20], P. Garnavich[21], L. Hillenbrand[22], C. Hirata[22], Z. Ivezic[23], E. Kerins[24], J. Lunine[4], M. Phillips[25], G. Rieke[26], A. Riess[21], R. van der Marel[6], D. Weinberg[9]

Project Office
R.K. Barry[12], E. Cheng[27], D. Content[12], K. Grady[12], C. Jackson[12], J. Kruk[12], M. Melton[12], N. Rioux[12]

James Green, SDT Co-Chair — Date: 7-11-11

Paul Schechter
Digitally signed by Paul Schechter
DN: cn=Paul Schechter, c=US, o=Massachusetts Institute of Technology, ou=Kavli Institute for Astrophysics and Space Research, email=schech@mit.edu
Location: MIT Cambridge MA 02139
Date: 2011.07.11 12:48:52 -04'00'

Paul Schechter, SDT Co-Chair — Date

1 University of Colorado/Center for Astrophysics and Space Astronomy
2 Massachusetts Institute of Technology
3 Yale University
4 Cornell University
5 University of Notre Dame
6 Space Telescope Science Institute
7 University of Nottingham
8 Michigan State University
9 Ohio State University
10 National Optical Astronomy Observatory
11 University of California Berkeley/Lawrence Berkeley National Laboratory
12 NASA/Goddard Space Flight Center
13 Jet Propulsion Laboratory/California Institute of Technology
14 NASA/Ames
15 Osaka University
16 Georgia State University
17 University of Oklahoma
18 University of California Los Angeles
19 NASA Headquarters
20 Lawrence Livermore National Laboratory
21 Johns Hopkins University
22 California Institute of Technology
23 University of Washington
24 University of Manchester
25 Las Campanas Observatory
26 University of Arizona
27 Conceptual Analytics

# EXECUTIVE SUMMARY

The Wide Field Infrared Survey Telescope (WFIRST) is the highest ranked recommendation for a large space mission in the recent New Worlds, New Horizons (NWNH) in Astronomy and Astrophysics 2010 Decadal Survey. The most pressing scientific questions in astrophysics today require a very wide-field survey in order to be answered, and existing telescopes such as the Hubble Space Telescope (HST), James Webb Space Telescope (JWST) or the Keck telescope cannot make these kinds of observations due to their optical designs and narrow fields of view. The first generation of digital wide-field surveys from the ground (e.g. Sloan) have resulted in significant advances in astronomy and astrophysics, and their value is recognized by the additional decadal recommendation for continuing development of new ground-based telescopes designed for survey work (e.g. LSST). Bringing these wide-field design principles into space will revolutionize astronomical surveys in much the same way that the Hubble Space Telescope revolutionized imaging of individual astronomical objects and galaxies. The absence of atmospheric distortion and absorption, and the darkness of space enable space-based surveys to cover the near infrared band and go deeper, with more precision and accuracy, than can ever be possible from the ground. WFIRST will be the most sensitive near infrared space telescope designed for wide-field survey work, and will produce the most powerful and informative astronomical data set for probing the nature of dark energy, cataloguing the variety of exoplanet systems, and mapping the distribution of matter across cosmic time. The mission addresses the most fundamental issues in astrophysics utilizing existing, proven technologies, and could move into design and development immediately if a new start were to be approved.

The WFIRST Science Definition Team (SDT) was formed to refine the science case for the mission, optimize the design and implementation scheme, and, with the WFIRST Project, develop a Design Reference Mission (DRM) to serve as the basis for further programmatic and technical review. This document is the interim report of the SDT and Project; accordingly, the interim DRM (IDRM) presented herein does not represent the final or only mission concept that could meet the requirements of the decadal recommendations. However, it does serve as an "existence proof" that an in-scope mission can be developed which successfully addresses all of the scientific objectives implicit in the decadal recommendation.

The WFIRST IDRM uses an unobscured, non-cryogenic, 1.3m three mirror anastigmat to feed a single instrument. The instrument contains three channels, an imager and two identical, slitless spectrometers. All three channels use Mercury-Cadmium-Telluride (HgCdTe) near-infrared sensor devices. The imager covers 0.60-2.0 μm with a pixel scale of 0.18 arcsec and the spectrometers cover 1.1-2.0 μm with a scale of 0.45 arcsec/pixel. An L2 orbit minimizes concerns with stray light from the Earth or Moon, provides an unobstructed view of the sky and a thermally stable environment. No new technologies are required to build WFIRST, which can be ready for launch in 2020. The IDRM design example has a simpler, more robust overall design with increased field of view than the JDEM-Omega design.

The SDT has refined the scientific objectives of WFIRST to three cornerstone goals, of equal importance:

1. Complete the statistical census of planetary systems in the Galaxy, from habitable Earth-mass planets to free floating planets, including analogs of all of the planets in our Solar System except Mercury.

2. Determine the expansion history of the Universe and the growth history of its largest structures in order to test explanations of its apparent accelerating expansion including Dark Energy and modifications to Einstein's gravity.

3. Produce a deep map of the sky at near-infrared (NIR) wavelengths, enabling new and fundamental discoveries ranging from mapping the Galactic plane to probing the reionization epoch by finding bright quasars at $z>10$.

In the following sections, the science case for each of these questions is detailed, and the scientific gain relative to our current understanding is quantified through the use of metrics, "Figures of Merit", or FOM. These broadly accepted metrics, coupled with WFIRST's ability to test dark energy with all of the three main probes using multiple consistency checks for systematic control, demonstrate that WFIRST is unparalleled in its capacity to address the fundamental science issues listed above, and is the next logical step in the advancement of space-based astronomical research.

The SDT also studied the broader aspects of the mission implementation, including the question of whether the mission should be reduced in scope or redesigned in response to a decision by the European Space Agency (ESA) to fly the Euclid mission. Consideration was given to the appropriate response to any potential NASA/ESA collaboration on a joint mission. The SDT has reached the following conclusions on these matters:

1. Any implementation of WFIRST should include all of the science objectives and utilize all of the techniques outlined in the NWNH decadal recommendation:

- Baryon Acoustic Oscillation Survey with Redshift Space Distortion information
- Exoplanet Microlensing Survey
- Supernova Ia Survey
- Weak Lensing Survey
- Near Infrared Sky Survey – including a survey of the Galactic Plane
- Guest Investigator Program

2. Due to the importance of the scientific questions, and the need for verification of the results, WFIRST should proceed with all of its observational capabilities intact regardless of the ESA decision on Euclid. The actual observation program and time allocations may be re-optimized in light of Euclid's selection or in response to any Euclid or ground based results prior to WFIRST's launch.

3. Should NASA and ESA decide to pursue a joint mission or program, all of the capabilities currently included in WFIRST must be included in the joint effort.

WFIRST represents an opportunity to re-write the text books by performing the first space-quality near infrared astronomical surveys over very wide fields. The most pressing questions in astrophysics today require observation of tens of millions of objects over a significant fraction the total available sky. Telescopes such as HST and JWST were not designed for these types of observations, and could never address the scientific questions addressed by WFIRST. No other planned or existing instrumentation, including non-US planned missions, can address the breadth of science, nor approach the quality of the data, that WFIRST provides. WFIRST will be the pre-eminent space astronomical science machine of the 2020's if approved for development. The cost of WFIRST is anticipated to be similar to the NWNH cost estimate, a fraction of the cost of its sister telescopes, HST and JWST. It is the unanimous recommendation of the WFIRST SDT that the preliminary studies we have begun be allowed to proceed to development and flight as rapidly as programmatic realities allow. The SDT affirms the NWNH decision that development of WFIRST is the highest priority for space astronomy in the coming decade.

# 1 INTRODUCTION

The New Worlds, New Horizons (NWNH) in Astronomy and Astrophysics 2010 Decadal Survey prioritized the community consensus for ground-based and space-based observatories. Recognizing that many of the community's key questions could be answered with a wide-field infrared survey telescope in space, and that the decade would be one of budget austerity, WFIRST was top ranked in the large space mission category. In addition to the powerful new science that could be accomplished with a wide-field infrared telescope, the WFIRST mission was determined to be both technologically ready and only a small fraction of the cost of previous flagship missions, such as HST or JWST. In response to the top ranking by the community, NASA formed the WFIRST Science Definition Team (SDT) and Project Office. The SDT was charged with fleshing out the NWNH scientific requirements to a greater level of detail. NWNH evaluated the risk and cost of the JDEM-Omega mission design, as submitted by NASA, and stated that it should serve as the basis for the WFIRST mission. The SDT and Project Office were charged with developing a mission optimized for achieving the science goals laid out by the NWNH report. The SDT and Project Office opted to use the JDEM-Omega hardware configuration as an initial starting point for the hardware implementation. JDEM-Omega and WFIRST both have an infrared imager with a filter wheel, as well as counter-dispersed moderate resolution spectrometers.

The primary advantage of space observations is being above the Earth's atmosphere, which absorbs, scatters, warps and emits light. Observing from above the atmosphere enables WFIRST to obtain precision infrared measurements of the shapes of galaxies for weak lensing, infrared light-curves of supernovae and exoplanet microlensing events with low systematic errors, and infrared measurements of the H$\alpha$ hydrogen line to be cleanly detected in the $1<z<2$ redshift range important for baryon acoustic oscillation (BAO) dark energy measurements. The Infrared Astronomical Satellite (IRAS), the Cosmic Background Explorer (COBE), Herschel, Spitzer, and Wide-field Infrared Survey Explorer (WISE) are all space missions that have produced stunning new scientific advances by going to space to observe in the infrared.

This interim report describes progress as of June 2011 on developing a requirements flowdown and an evaluation of scientific performance. An Interim Design Reference Mission (IDRM) configuration is presented that is based on the specifications of NWNH with some refinements to optimize the design in accordance with the new scientific requirements. Analysis of this WFIRST IDRM concept is in progress to ensure the capability of the observatory is compatible with the science requirements. The SDT and Project will continue to refine the mission concept over the coming year as design, analysis and simulation work are completed, resulting in the SDT's WFIRST Design Reference Mission (DRM) by the end of 2012.

Box 1

"WFIRST is a wide-field-of-view near-infrared imaging and low-resolution spectroscopy observatory that will tackle two of the most fundamental questions in astrophysics: Why is the expansion rate of the universe accelerating? And are there other solar systems like ours, with worlds like Earth? In addition, WFIRST's surveys will address issues central to understanding how galaxies, stars, and black holes evolve. ….. WFIRST will settle fundamental questions about the nature of dark energy, the discovery of which was one of the greatest achievements of U.S. telescopes in recent years. It will employ three distinct techniques—measurements of weak gravitational lensing, supernova distances, and baryon acoustic oscillations—to determine the effect of dark energy on the evolution of the universe. An equally important outcome will be to open up a new frontier of exoplanet studies by monitoring a large sample of stars in the central bulge of the Milky Way for changes in brightness due to microlensing by intervening solar systems. This census, combined with that made by the Kepler mission, will determine how common Earth-like planets are over a wide range of orbital parameters. It will also, in guest investigator mode, survey our galaxy and other nearby galaxies to answer key questions about their formation and structure, and the data it obtains will provide fundamental constraints on how galaxies grow."

From the New Worlds, New Horizons Decadal Survey in Astronomy and Astrophysics

## 1.1 SCIENCE REQUIREMENTS

The SDT and Project have developed a requirements flowdown matrix for the mission. The three top-level scientific objectives for WFIRST are:

- Complete the statistical census of planetary systems in the Galaxy, from habitable Earth-mass planets to free floating planets, including analogs of all of the planets in our Solar System except Mercury.

- Determine the expansion history of the Universe and the growth history of its largest structures in order to test explanations of its apparent accelerating expansion including Dark Energy and modifications to Einstein's gravity.

- Produce a deep map of the sky at NIR wavelengths, enabling new and fundamental discoveries ranging from mapping the Galactic plane to probing the reionization epoch by finding bright quasars at z>10.

These objectives then drive the requirements for the observatory capabilities and design. A top-level flow-down of the WFIRST requirements is given in Figure 1. The Science Objectives above are the highest level science requirements and appear at the top of the page. The derived scientific survey capability requirements of the observatory are listed in the left-hand boxes and data set requirements in the middles boxes. The top-level Observatory design/operations parameters are listed in the right-hand boxes. A more detailed discussion of the basis for the requirements is given in Appendix A and Appendix B.

WFIRST Science Objectives:

1) Complete the statistical census of planetary systems in the Galaxy, from habitable Earth-mass planets to free floating planets, including analogs of all of the planets in our Solar System except Mercury.
2) Determine the expansion history of the Universe and its growth of structure so as to test explanations of its apparent accelerating expansion including Dark Energy and modifications to Einstein's gravity.
3) Produce a deep map of the sky at NIR wavelengths, enabling new and fundamental discoveries ranging from mapping the Galactic plane to probing the reionization epoch by finding bright quasars at z>10.

## WFIRST Survey Capability Rqts

### Exoplanet (ExP) Microlensing Survey

- Planet detection capability to ~0.1 Earth mass ($M_\oplus$)
- Detects ≥ 125 planets of 1 $M_\oplus$ in 2 year orbits in a 500 day survey, with the masses of ≥ 90 of these planets being determined to better than 20% *
- Detects ≥ 25 habitable zone† planets (0.5 to 10 $M_\oplus$) in a 500 day survey *
- Detects ≥ 30 free floating planets of 1 $M_\oplus$ in a 500 day survey *

* Assuming one such planet per star

† 0.72-2.0 AU, scaling with the square root of host star luminosity

### Dark Energy Surveys
### BAO/RSD Galaxy Redshift Survey

- ≥11,000 deg$^2$ sky coverage per dedicated year ("WIDE" Survey mode)
- Goal of ≥2,700 deg$^2$/yr "DEEP" Survey acquired during the WL Survey
- A comoving density of galaxy redshifts at z=2 of 4.9x10$^{-5}$ Mpc$^{-3}$ (WIDE) or 2.1x10$^{-4}$ Mpc$^{-3}$ (DEEP). [The source density is higher at lower redshifts, peaking at z=1 at 2.2x10$^{-4}$ Mpc$^{-3}$ (WIDE) or 5.9x10$^{-4}$ Mpc$^{-3}$ (DEEP)]
- Redshift range 0.7 ≤ $z$ ≤ 2
- Redshift errors $\sigma_z$≤0.001(1+$z$), equivalent to 300 km/s rms
- Misidentified lines ≤TBD% per source type, ≤10% overall; contamination fractions known to 0.2% (TBR)

### Supernova SNe-Ia Survey

- >100 SNe-Ia per Δz=0.1 bin for most bins for 0.4 < z < 1.2, per dedicated 6 months
- Redshift error σ ≤ 0.005 per supernova
- Relative instrumental bias ≤0.005 on photometric calibration across the wavelength range
- Distance modulus error (from lightcurve) $\sigma_\mu$ ≤0.02 per Δz=0.1 bin

### WL Galaxy Shape Survey *

- ≥ 2,700 deg$^2$ sky coverage per dedicated year (in a "DEEP" Survey mode)
- Effective galaxy density ≥30/amin$^2$, shapes resolved plus photo-zs
- Additive shear error ≤3x10$^{-4}$
- Multiplicative shear error ≤1x10$^{-3}$
- Photo-$z$ error distribution ≤0.04(1+$z$), error rate <2%

* Goal: The WL Galaxy Shape Survey shall be taken in a manner such that concurrent spectroscopy also meets the BAO survey requirements on source density, redshift errors and fraction of misidentified lines.

### Near Infrared Survey

- Identify ≥100 quasars at redshift $z$ ≥7
- Extend studies of galaxy formation and evolution to $z$ > 1 by making sensitive, wide-field images of the extragalactic sky at near-infrared wavelengths, thereby obtaining broad-band spectral energy distributions of ≥ 1e9 galaxies at z>1
- Map the structure of the Galaxy using red giant clump stars as tracers
- Enable a robust Guest Investigator (GI) program, with at least 10% of the mission lifetime available to the community through peer-reviewed, open competition

## WFIRST Data Set Rqts

### Exoplanet Data Set Rqts

- Observe ≥ 2 square degrees in the Galactic Bulge at ≤ 15 minute sampling cadence
- S/N ≥100 for J-band magnitude ≤20.5 sources
- ≤0.3" imaging angular resolution
- Sample light curves with filter W149
- Monitor color with filter F087, 1 exposure every 12 hours
- Minimum continuous monitoring time span: ~60 days
- Separation of ≥4 years between first and last observing seasons

### Dark Energy Data Sets
### BAO/RSD Data Set Rqts

- Spectrometer
  - Slitless prism
  - Dispersion $R_\theta$ = 195 (TBR) - 240 arcsec
  - S/N ≥7 for $r_{eff}$ = 300 mas for Hα emission line flux at 2.0 μm ≥1.5x10$^{-16}$ erg/cm$^2$-s (DEEP) or 3.1x10$^{-16}$ erg/cm$^2$-s (WIDE)
  - Bandpass 1.116μm ≤λ≤ 2.0 μm
  - Pixel scale ≤ 450 mas
  - System PSF EE50% radius 400 mas at 2 μm
  - ≥3 dispersion directions required, two nearly opposed
- Imager (for redshift zero reference)
  - S/N≥10 for $H_{AB}$≤23.5
  - Approximately equal time in filters F141 and F178

### Supernova Data Set Rqts

- Minimum monitoring time-span for an individual field: ~2 years with a sampling cadence ≤5 days
- Cross filter color calibration ≤0.005
- Four filters: F087, F111, F141, F178
- Slitless prism spec (P130) 0.6-2 μm λ/Δλ ~75 (S/N ≥ 2 per pixel bin) for redshift/typing
- Peak lightcurve S/N ≥15 at each redshift
- Dither with 15 (TBR) mas accuracy
- Low Galactic dust, E(B-V) ≤0.02

### WL Data Set Rqts

- From Space: 2 shape/color filter bands, F141 and F178, and 1 color filter band, F111
- S/N ≥18 per shape/color filter for galaxy $r_{eff}$ = 250 mas and mag AB = 23.7
- PSF second moment ($I_{xx}$ + $I_{yy}$) known to a relative error of ≤ 9.3x10$^{-4}$ rms (shape/color filters only)
- PSF ellipticity ($I_{xx}$-$I_{yy}$, 2*$I_{xy}$)/ ($I_{xx}$ + $I_{yy}$) known to ≤ 4.7x10$^{-4}$ rms (shape/color filters only)
- System PSF EE50 radius ≤170 mas for filter F141, and ≤193 mas for filter F178
- 5 random dithers required for shape/color bands, and 4 for F111 at same dither exposure time
- From Ground: 4 color filter bands ~0.4 ≤ λ ≤ ~0.97μm
- Provide an unbiased spectroscopic Photo-z Calibration Survey (PZCS) training data set containing ≥ 100,000 galaxies ≤ mag AB = 23.7 for F141 and F178 and covering at least 4 uncorrelated fields; redshift accuracy required is $\sigma_z$<0.01(1+$z$)

### Near Infrared Data Set Rqts

- Image ≥ 2500 deg$^2$ of high latitude sky in three near-infrared filters to minimum depths of mag AB = 25 at S/N=5. Fields must also have deep (ground-based) optical imaging
- Image ≥ 1500 deg$^2$ of the Galactic plane in three near-infrared filters

WFIRST IDRM Design/Operations Overview

Key WFIRST IDRM Observatory Design Parameters

- Off-axis focal telescope; 1.3m diameter telescope aperture
- ≤240 K telescope optical surfaces
- Bandpass 0.6 – 2.0 μm
- Pointing jitter ≤40 mas rms/axis
- Coarse Pointing Accuracy <~3 arcsec rms/axis
- Fine (Relative/Revisit) Pointing Accuracy <~25 mas rms/axis [TBR]
- ACS telemetry downlinked for pointing history reconstruction

Imager Channel (ImC): No re-imager; ~180K Pupil Mask

- ImC Pupil Mask stop diameter: 1.275 m
- ImC Effective Area: 0.811 $m^2$ (avg for all filters including QE and roll off)
- 5 band parfocal filter set on wheel, driven by ExP, SNe, WL
- R=75 (2-pix) parfocal, zero deviation prism + "dark" (TBD) position in same wheel
- ImC R75 Slitless Prism Effective Area: 0.782 $m^2$
- ImC FPA: 4x7 HgCdTe 2k x 2k SCAs, 2.1μm, ≤120K, 180 mas/pix
- FOV (active area) = ~0.291 $deg^2$; Bandpass 0.6 – 2.0 μm
- 4 Outrigger FGS SCAs mounted to ImC Focal Plane Assy (FPA)
- WFE is diffraction limited at 1μm
- TBD requirement on Intrinsic PSF ellipticity ... relate to knowledge rqt

Slitless Spec Channels (SpCs): ~180K Pupil Mask

- 2 oppositely dispersed SpCs provided, otherwise ~identical
- Optical Path: pupil mask stop, focal prism, refractive focal length reducer
- Bandpass 1.1-2.0μm; $R_\Theta$ = 160 (TBR) - 210 arcsec
- SpC Pupil Mask stop diameter: 1.27 m
- SpC Effective Area: 0.731 $m^2$ (average including QE)
- SpC FPA (1 of 2): 2x2 HgCdTe 2kx2k SCAs, 2.1μm, ≤120K, 450 mas/pix
- FOV (active area) = ~0.260 $deg^2$ (per SpC)

Aux FGS: 2 SCAs: controls Pitch/Yaw when ImC prism is in use

| ImC Filter | Bandpasses (μm) |
|---|---|
| F087 | 0.76 – 0.97 |
| F111 | 0.97 – 1.24 |
| F141 | 1.24 – 1.57 |
| F178 | 1.57 – 2.00 |
| W149 | 0.97 – 2.00 |
| ImC Prism | |
| P130 | 0.60-2.00 |

Key WFIRST IDRM Operations Concept Parameters

- 5-year mission life, but consumables required for 10 yrs
- Science Field of Regard (FOR): 54˚ to 126˚ pitch off the Sun, 360˚ yaw
- Roll ±10˚; SNe observations inertially fixed for ~90 days for viewing near the ecliptic pole(s)
- Gimbaled antenna allows observing during downlinks
- Slew/settle times: ~16 s for dithers, ~38 s for ~0.7˚ slews

SNe-Ia Survey (~11.5 $deg^2$-yrs to $z$ = 0.8, ~2.9 $deg^2$-yrs to $z$ = 1.2)

- A sample 2-tiered survey capability (given 6 months dedicated time) is shown, each tier optimized for a different $z$ range
- All tiers monitored on a 5-day cadence in selected filters + prism
- ~Square SNe fields are inertially fixed for ~90 days
- Tier 1 (to $z$=0.8): 5.76 $deg^2$; F111, F141, F178 (300 s @), P130 (1300 s)
- Tier 2 (to $z$=1.2): 1.44 $deg^2$, F111, F141, F178 (1100 s @), P130 (5300 s)
- SNe dedicated time is distributed in a 5-day cadence over ~2 years to provide suitable light curve tracking and accurate host galaxy references (e.g. if 6 months are dedicated SNe, 30 hrs of SNe field monitoring would be done every 5 days for ~2 years)
- SNe fields are monitored from end of one ExP Galactic Bulge season until the start of the 4th following ExP season (~1.8 yrs)
- Fields located in low dust regions ≤20˚ off an ecliptic pole (N and S fields not required)
- 4-9 sub-pixel dithers, accurate to ~15 mas, performed at each pointing

ExP Survey (~2.04 $deg^2$ monitored every 15 min, 144 days/yr)

- The Galactic Bulge is observable for two 72-day seasons each year
- The short revisit cadence impacts other observing modes while exoplanet data sets are being acquired. This, in combination with the field monitoring time span required for SNe and the 60 days required for ExP, limits the max number of Galactic Bulge seasons useable for exoplanet observations to seven (over 5 yrs)
- In each season 7 fields are revisited on a 15 min cadence, acquiring 88s exposures (filter W149) for light curve tracking except for one visit every 12 hours that uses filter F087 for color
- Fields are revisited to an accuracy of 1 pixel rms; no precise dithers

"DEEP" Galaxy Shape + Galaxy Redshift Survey (~2,700 $deg^2$/yr)

- 3 smooth-filled ImC maps are acquired, each rolled (Goal) by ~5˚ relative to each other to meet BAO/RSD survey roll requirements
- 5 ImC random dithers are acquired over the full field for the shape/color bands, and 4 random dithers for the color-only band
- 800 s (160 s per dither) integration time in each of 2 shape/color bands
- 640 s (160 s per dither) integration time in one color-only band
- SpC integrations are acquired simultaneously during ImC exposures, and more than meet BAO/RSD requirements
- ~7 $deg^2$ are completed each day (ground observations also required)
- ~5 ImC or SpC dispersed fields, observed at multiple roll angles over 1-2 weeks meet the PZCS training data set requirements. Ground observations of these fields are required, and locations accessible to both N and S hemisphere telescopes will be chosen

"WIDE" Galaxy Redshift Survey (~11,000 $deg^2$/yr, ~30 $deg^2$/day)

- 8 SpC maps are acquired during 4 separate passes (2 oppositely-dispersed maps are acquired each pass, one from each of two SpCs)
- The first two passes are offset from each other to rough fill both SpC and ImC SCA gaps, and the same is done for the next two passes at an ~5˚ different roll angle and with a different ImC filter
- Each pass consists of a series of eight 150s exposures, yielding a total SpC exposure time of 1200s. But the rough-filled nature of the exposures results in only 6 exposures being accumulated over ~96% of the observed sky, so only 6x150s = 900s of SpC time is creditable
- ~30 $deg^2$ are completed each day, with good N/S hemisphere flexibility

Figure 1: WFIRST Requirements Flowdown Overview

# 2 SCIENCE

## 2.1 Dark Energy Science

*Primary Dark Energy Science Objective: Determine the expansion history of the Universe and the growth history of its largest structures in order to test possible explanations of its apparent accelerating expansion including Dark Energy and modifications to Einstein's gravity.*

The dramatic progress in astronomy of the past two decades has included several unexpected results. Nothing has been more surprising than measurements that indicate that the expansion of the universe (discovered by Edwin Hubble nearly a century ago) is accelerating. The apparent accelerating expansion could be due to: (1) a constant energy density (the "cosmological constant") that may arise from the pressure of the vacuum, (2) an evolving universal scalar field, or (3) a flawed or incomplete understanding of gravity as described by Einstein's General Theory of Relativity. Any of these possibilities has profound consequences for our basic understanding of physics and cosmology. The future of the universe will be determined largely by whatever force or property of space is causing the acceleration, making the nature of the accelerating expansion one of the most profound and pressing questions in all of science. The imperative is to distinguish between these possibilities by carrying out a careful set of measurements designed to characterize the underlying source of the so-called "dark energy" that drives the accelerated expansion of the universe. Recent measurements reveal that about 75% of the total mass-energy of the universe is dark energy. In other words, dark energy is most of our universe today, yet we do not know what it is. One of WFIRST's primary mission goals is to understand the nature of the dark energy: "WFIRST will settle fundamental questions about the nature of dark energy, the discovery of which was one of the greatest achievements of U.S. telescopes in recent years." [NWNH]

Dark energy affects the universe in two significant ways. First, the *expansion history* (or geometry) of the universe is determined by the energy density of dark energy over cosmic time. The *growth of cosmic structures* from the density perturbations we see in maps of the Cosmic Microwave Background (CMB) to the galaxies and galaxy clusters we see today is governed by the attractive force of gravity and the repulsive dark energy, which inhibits the structure growth. Within the confines of General Relativity, measurements of the expansion history and growth of structure will give consistent results. Discrepancies between these two types of measurements might indicate a breakdown of General Relativity. Only by measuring both the expansion history and the growth of structure can we distinguish between the three broad classes of explanations for the acceleration outlined above. WFIRST has been designed to measure each of these with multiple independent techniques. As described in NWNH, WFIRST "will employ three distinct techniques—measurements of weak gravitational lensing, supernova distances, and baryon acoustic oscillations—to determine the effect of dark energy on the evolution of the universe". In addition to these three methods, WFIRST adds a fourth method, redshift space distortion (RSD), which provides an alternative measure of the growth of structure. The independence of these four techniques is crucial to verifying the accuracy of the measurements.

Box 2

"Why should WFIRST employ all three methods? Supernovae (in particular, type SNe Ia) give the best measurements of cosmic acceleration parameters at low redshift due to their greater precision per sample or per object. BAO excels over large volumes at higher redshift. Together SNe Ia and BAO provide the most precise measurements of the expansion history for $0 < z < 2$ and place significant constraints on the equation of state. Weak lensing provides a complementary measurement through the growth of structure. Comparing weak-lensing results with those from supernovae and BAO could indicate that "cosmic acceleration" is actually a manifestation of a scale-dependent failure of general relativity. Combining all three tests provides the greatest leverage on cosmic-acceleration questions. WFIRST can do all three."

From the Panel Reports--New Worlds, New Horizons in Astronomy and Astrophysics

### 2.1.1 *Supernovae*

Dark energy was first discovered by using distant supernova as beacons to measure how the Universe has grown. WFIRST will build on this methodology by taking it to a significantly higher level of precision than can be done with ground-based observations. These measurements are done using a particular form of supernova, those of Type Ia, which are the explosions of white dwarf stars of a common known mass. These supernovae can be seen at vast distances across the Universe, with their light travelling to us from a time when the Universe was a fraction of its present age. From studies of nearby Type Ia supernovae, we know that they all have about the same peak brightness after rescaling to the width of the lightcurve (and are therefore called “standard candles”), thus measuring how faint a distant supernova is tells us how far away it is. When we combine this information with the redshift, which tells us how much the Universe has expanded since the time the light was emitted, we probe the history of how the Universe expanded with time, see Figure 2.

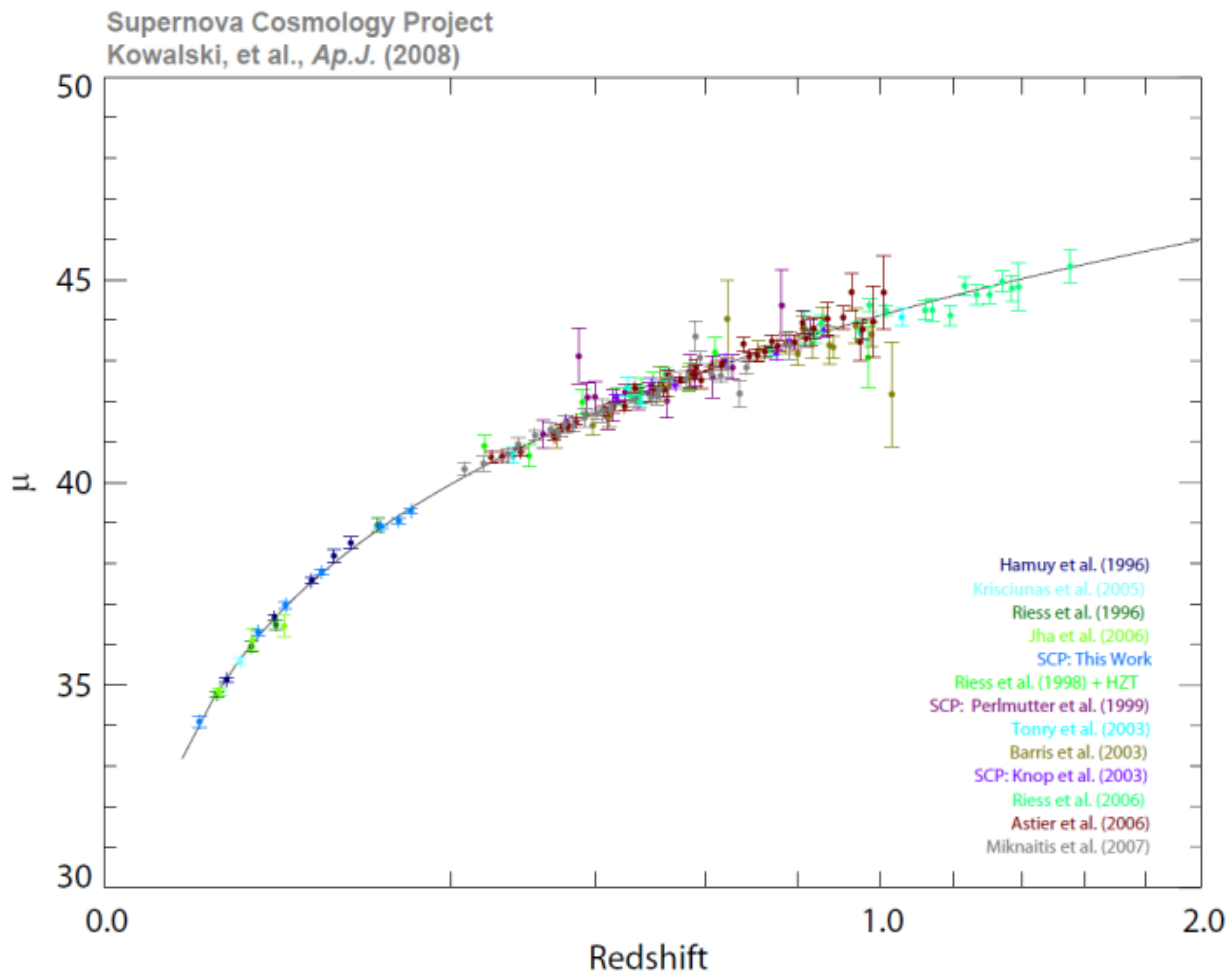

Figure 2: The Hubble diagram for SN Ia observations from the ground and space (HST). Plotted is the distance modulus, $\mu$ = apparent magnitude – absolute magnitude as a function of redshift.

In practice, using Type Ia supernovae as dark energy probes requires several kinds of observations, all of which will be conducted by WFIRST. The first step is to discover the supernova. This will be done by repeatedly observing a large patch of sky, every five days, looking for new objects that were not seen before.

Once the supernovae are discovered, they have to be classified to sift out the Type Ia’s from other types of supernovae, or even other variable objects. Classification is done both by following how the candidate supernovae change in brightness over time (their ‘light curves’), but also by verifying the spectral types of the candidates. The spectra obtained to confirm the classification will also provide the redshifts of the supernovae. All type Ia supernova that have been observed in the infrared exhibit a double peak in their brightness with time. No other type of supernova exhibits this behavior.

WFIRST’s dark energy goals require high and uniform accuracy measurements across the entire timespan of the Universe’s history being probed by the supernovae. A crucial part of this high accuracy comes from repeated observations of each supernova over its “life span”. While the Type Ia supernovae have roughly the same intrinsic explosion luminosity, there are important variations within this type, with fainter Type Ia’s fading away faster than the brighter ones. WFIRST uses the light curve to see how fast they fade, thus providing corrections to the peak luminosities. The supernovae occur in distant galaxies, which may have dust, that dims and reddens the light reaching us; observing supernovae over a wide range of colors allows the dust to be detected and the light curves corrected for it. By observing in the infrared, WFIRST supernova results will be less affected by dust– a key advantage of WFIRST.

During the dedicated Supernova Survey, images are used both to discover new supernovae and to measure the light curves from ones already discovered as they rise and fall in brightness. These images are interspersed with spectroscopic observations to obtain redshift information. The series of images must be done with a certain cadence to most efficiently use WFIRST’s capability to measure the expansion history of the Universe using supernovae.

### 2.1.2 *Weak Gravitational Lensing*

When light from galaxies propagates across the Universe, its path is slightly deflected by the gravity of other galaxies, an effect called “weak lensing” (WL), see Figure 3. Weak gravitational lensing has emerged as a unique probe of the growth of structure. Furthermore, weak lensing observables are sensitive to the expansion history as well. Thus, weak lensing plays a particularly powerful role in quantifying possible deviations from General Relativity, and is an excellent complement to the other dark energy techniques planned for WFIRST. Weak lensing has been quantified by surveys of increasing statistical power over the last decade using both ground and space-based telescopes. However, well away from massive clusters of galaxies, the subtle variations of the measured galaxy shapes (of order 1%) are difficult to measure using ground–based observatories with their large and time-variable instrumental point spread functions (PSFs). WFIRST has been designed to mitigate these effects with a thermally stable platform in space, enabling surveys spanning thousands of square degrees that will be statistics–limited, as opposed to upcoming ground-based surveys that may be limited by systematic effects over similar areas. The WFIRST weak lensing survey requires measurements of galaxy shapes in two filters. This second filter is useful for addressing wavelength-dependent PSF issues in galaxy shape measurements (Voigt et al. 2011; Cypriano et al. 2010) in order to ensure we achieve the low systematic errors needed for the WL experiment to succeed. This is an important cross check on systematic errors that other proposed experiments have not included. The WL signal is achromatic but possible systematic errors are not necessarily achromatic, so a second filter will significantly increase the robustness of a WL result. A third filter is required to get an additional photometric band (color information) that will then be combined with ground-based data to calculate a photometric redshift for each galaxy.

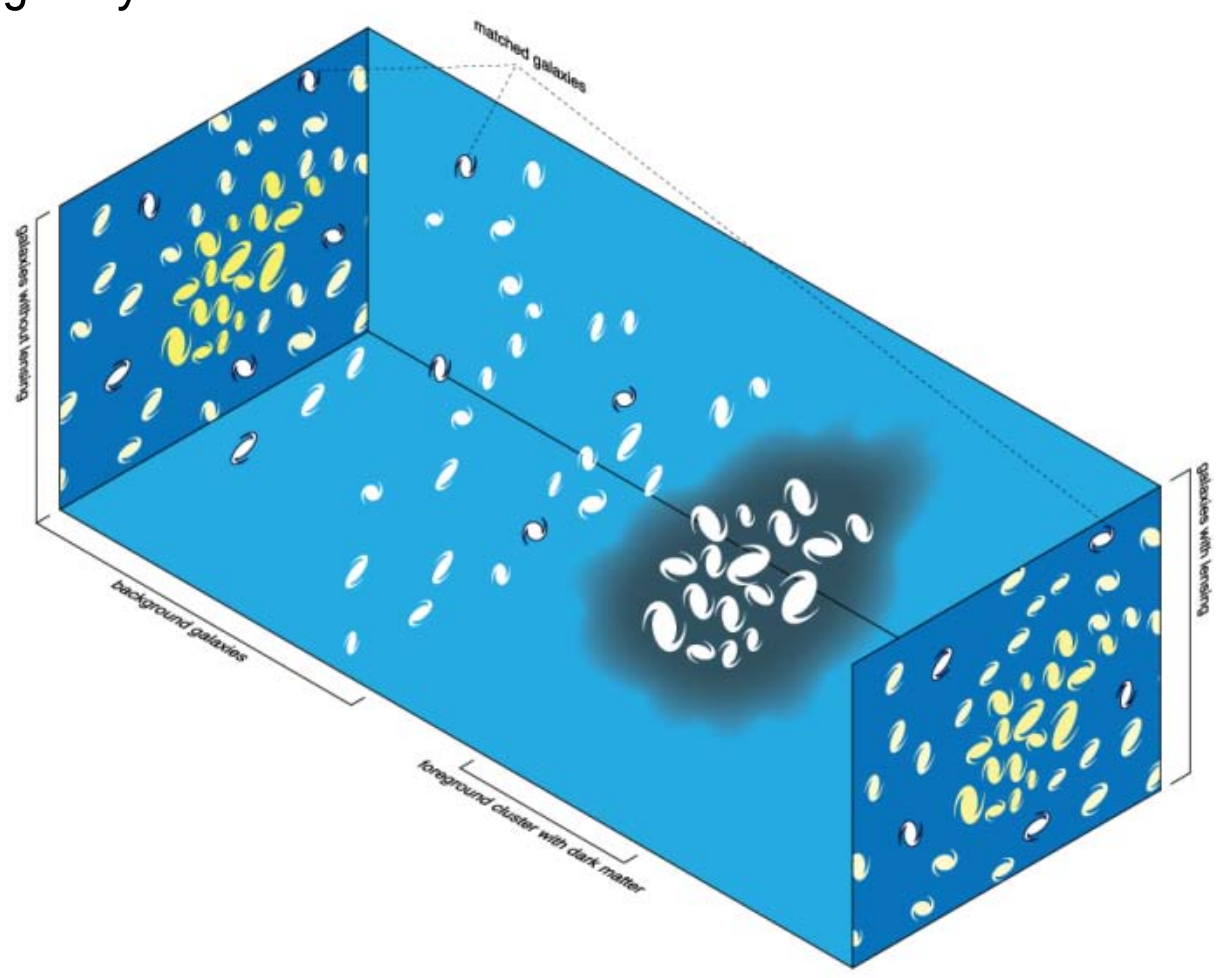

Figure 3: An illustration of weak lensing showing the effect of foreground galaxies on the light from distant background galaxies.

### 2.1.3 *Baryon Acoustic Oscillation (BAO)*

Until about 380,000 years after the Big Bang, photons and matter interacted frequently. Hot plasma - protons, electrons, and photons - sloshed about as sound waves in ever-deepening dark matter potential wells. When the plasma cooled and formed neutral atoms, the photons streamed freely through the universe, and we see them today as cosmic microwave background (CMB). After this final interaction of photons with electrons, the sound waves (acoustic oscillations) stopped in place, and left their signature on both the cosmic microwave background (CMB) (verified by the observed acoustic peaks in the CMB power spectrum) and on the matter distribution (verified by the BAO signature in the galaxy distribution power spectrum). BAO in the observed galaxy power spectrum (the three dimensional galaxy distribution) have a characteristic scale determined by the distance sound has traveled since the Big Bang by the time the photons last interacted with matter, a time which is precisely measured by the CMB anisotropy data. Comparing the observed BAO scales with the expected values gives the cosmic expansion rate in the radial direction, and the angular diameter distance $d_A(z)$ through an integral in the transverse direction, see Figure 4. Put another way, the scale of the BAO peak in the power spectrum acts as a "standard ruler", expanding along with the Universe from the time of the CMB to the present day. By measuring this scale, we determine the expansion history of the Universe.

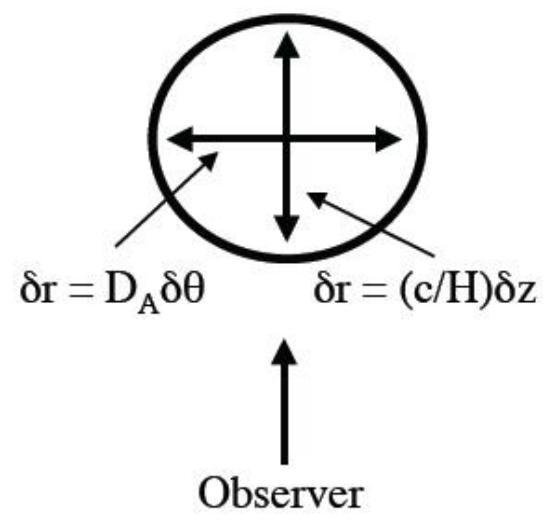

Figure 4: A representation of the BAO spherical shell with H(z) and $D_A(z)$ measurements.

Baryon acoustic oscillations are just one of several physical phenomena that affect the galaxy power spectrum. The same galaxy redshift survey that will be used to measure BAO will also be used to study a phenomenon called redshift space distortion (RSD) and to locate a feature in the galaxy power spectrum due to the transition from radiation domination to matter domination (TRMD) 100,000 years after the big bang. As effects of both of these on the galaxy power spectrum are sensitive to the assumed cosmology, they provide additional leverage, making WFIRST a yet more powerful mission for studying cosmic acceleration. In the discussion that follows, we distinguish among these different measurements, reserving the term baryon acoustic oscillations for the phenomenon described in the previous section. While RSD and TRMD are expected to sharpen WFIRST's dark energy results, they do not drive the requirements beyond those for BAO measurements.

Astronomers often use Hubble's law to calculate distances from redshifts, but this estimate does not take into account peculiar velocities, which are local gravita-tional deviations from the overall expansion of the universe. These local velocities will cause galaxies at equal distances to have slightly different redshifts. These slight differences elongate the redshift distribution for galaxies clustered along the line of sight. Gravity causes galaxies to fall in towards overdense regions. This pattern makes galaxies between us and the overdensity appear to be farther away, based on distance estimates using their redshifts and Hubble's Law. Galaxies on the far side of the overdensity appear closer. The net effect is to enhance the overdensity. These effects are collectively called "redshift space distortion" (RSD). This RSD measurement can be obtained from the same data required for BAO and used to constrain the growth of cosmic structure in a complementary way to WL. BAO along with RSD measurements allow us to simultaneously measure the cosmic expansion history and the growth rate of large scale structure, thus enabling precise measurement of dark energy as well as testing general relativity.

#### 2.1.4 *Complementary Dark Energy Constraints*

In the local universe (redshift $z<0.5$), supernovae are unparalleled in their ability to constrain dark energy for two reasons. First, there is insufficient cosmic volume for more statistical techniques (WL & BAO) to work. Second, cosmic structure gets increasingly nonlinear (and hard to model) as the redshift gets small. Dark energy was originally discovered using supernovae at relatively modest redshifts. At the highest redshifts ($z\sim1100$), we have the CMB, which establishes the BAO 'standard ruler' scale. The BAO provide a link between the expansion history of the Universe at $z=1100$ and $z=0.5$. The BAO systematic errors decrease with increasing redshift (due to decreased nonlinear effects), while the SN systematic errors increase with redshift. Using the SN and BAO connects the whole expansion history of the universe. Weak lensing fundamentally measures the three-dimensional distribution of matter, which can be compared with theoretical expectations. It is sensitive to both expansion and growth of structure, and has the potential to be the most powerful single dark energy technique. However, weak lensing measurements are difficult to make because the very small shape distortions induced in faint galaxies must be deconvolved from the much larger effects of the telescope and optics (the so-called Point Spread Function, or PSF). Successful measurements require high spatial resolution with an excellent quality and stability, and a known small-scale PSF combined with very large sky areas. In contrast, the BAO data are relatively easy to acquire since we only need to measure the positions and velocities of galaxies. In addition, we get the RSD information on the growth of structure for free.

The three methods employed by WFIRST to explore dark energy, weak lensing, BAO (with RSD), and supernovae, together give us three complementary probes of the cosmic expansion history, and two complementary probes of the growth history of cosmic large scale structure. WFIRST is an extremely capable dark energy probe able to make very robust measurements with significant opportunity for cross-checks and verification of results.

### 2.2 Exoplanet Science

*Primary Exoplanet Science Objective: Complete the statistical census of planetary systems in the Galaxy, from habitable Earth-mass planets to free floating planets, including analogs of all of the planets in our Solar System except Mercury.*

#### 2.2.1 *Introduction/Retrospective*

The first discovery of a planetary companion to a sun-like star by Mayor & Queloz in 1995 was, along with the discovery of dark energy, one of the greatest breakthroughs in modern astronomy. In the intervening 15 years, hundreds of exoplanets have been discovered, mostly by the radial velocity technique, and over 1200 candidates have been detected in transit by the Kepler space telescope (Borucki et al. 2011). From the very first discovery, it has been clear that nature hosts an enormous and unexpected diversity of exoplanetary systems, containing planets with physical properties and orbital architectures that are radically different from our own solar system. Theories of planet formation and evolution, originally developed to explain our solar system, have struggled to keep up with the flood of new planets. This ferment indicates a vibrant research domain, which is drawing talented young researchers to careers in astronomy and planetary science.

A second facet of the explosion of interest in extrasolar planets is the search for planets that could host life, those similar in temperature and mass to Earth. This goal, which is profound in both scientific and cultural terms, has captured the imagination of researchers, political decision makers, and people around the world. Microlensing, combined with transit surveys (see below), will tell us whether Earth-mass planets in habitable orbits are a common or uncommon outcome of planet formation. The answer will change humankind's view of the cosmos. It will also help determine the feasibility and scope of future direct detection missions to see such habitable planets around nearby stars and search for indications of life with spectroscopy.

Microlensing in space fulfills the Primary Exoplanet Science Objective of WFIRST for the following reasons. (a) WFIRST's exoplanet (ExP) microlensing survey "completes the statistical census of planetary systems" as a perfect complement to Kepler's survey. For example, we know that Kepler's ability to detect planets is ten times smaller at separations of 1 AU than 0.1 AU. It is also proportionally smaller for Earth-size planets than Jupiter-sized planets. As a result, even though Kepler is designed explicitly to detect terrestrial-size planets in the habitable zones of solar-type stars, this is close to the limit of its capabilities. Fortunately, the converse is true for ExP on WFIRST, where the region of high sensitivity runs from roughly the habitable zone outwards. Also, the microlensing signal is far less sensitive to the mass of the planet than is Kepler's transit technique to planet size. In sum, detecting terrestrial planets is well within microlensing's capability. Thus the sum of Kepler plus WFIRST's ExP survey will yield a composite census of planets on both sides of the habitable zone, overlapping at almost precisely that zone. (b) WFIRST's ExP program can detect "free-floating" planets below an Earth-mass in numbers sufficient to test planet-formation theory, a task not possible from the ground. It is only in space that the statistics of these detections will be adequate. (c) Exoplanet microlensing from the ground has had well-documented successes, but just as is the case for transits, space is a vastly better place to carry out statistically significant observations. Space-based microlensing is superior because of the steady viewing without interruptions for weeks at a time, with high angular resolution, stable images over wide-area fields.

For these reasons, WFIRST's ExP microlensing survey is both the most efficient and affordable technique for quantifying the statistical distribution of planets over a wide range of masses and semi-major axes. Its capabilities have been amply demonstrated from the ground through the first discovery of a "super-Earth", only six times the mass of our own planet (Beaulieu et al. 2006), and the detection of a new population of Jupiter-mass planets loosely bound or unbound to any host star (Sumi et al. 2011). This latter study also concluded that "free-floating" planets may outnumber the stars in our galaxy by two to one. The sheer number of stars in the galactic bulge make it possible to tease out those few planetary microlensing events from the innumerable, random passages of intervening stars across the background of stars in the Galactic Bulge. But ground-based systems have intrinsic limitations that limit their sensitivity and efficiency.

Box 3

### Projected WFIRST Exoplanet Discoveries

If each of the Exoplanet Microlensing Survey Requirements in Figure 1 are met, our current best estimates of exoplanet prevalence imply that WFIRST will detect:

3250 total bound exoplanets in the range of 0.1-1000 Earth masses and separations in the range 0.1-40 AU, including 320 sub-Earth-mass planets and 1500 sub-ten-Earth-mass planets.

2080 free-floating exoplanets, including 190 sub-Earth-mass free-floating planets and 480 sub-ten-Earth-mass free-floating planets.

If each star hosts planets with the masses and same semi-major axes as those in our Solar System, WIFRST will detect 280 terrestrial planets (Venus/Earth/Mars analogs), 3200 gas giants (like Jupiter/Saturn), and 84 ice giants (like Neptune/Uranus).

Therefore, microlensing in space—from WFIRST—is the best prospect we have for gaining critical knowledge on planet formation and evolution, and for learning the occurrence and parameter distribution of Earth-mass planets. The Exoplanet Task Force (Lunine et al. 2008) recognized this, embedding into its strategy a medium-cost space-based microlensing mission to complement the census of close-in planets currently underway with spaceborne transit techniques. The Decadal Survey for Astronomy and Astrophysics elevated microlensing to an even higher level of importance (NWNH, 2011), placing the science of exoplanets in the top two or three science areas of importance to astrophysics in the next decade. Space-based microlensing plays a pivotal role in the Decadal Strategy. As evidenced by the wide range of parameter space it will explore (see Figure 5), and the size and diversity of its projected exoplanet discoveries (see Box 3), WFIRST is the means to understanding the range of planetary system architectures (masses, orbital parameters) and to determining how many Earth-mass worlds inhabit the crucially important region from the inner portion of the classic habitable zone outward. It will also extend the search for free-floating planets down to the mass of Earth and below, addressing the question of whether ejection of planets from young systems is a phenomenon associated only with giant planet formation (disk instability?) or also involves terrestrial planets. It is also the only technique available to detect planets as small as the mass of Mars. Since Mars-mass bodies are thought to be the upper limit to the rapid growth of planetary "embryos", determining the planetary mass function down to a tenth the mass of the Earth uniquely addresses a pressing problem in understanding the formation of terrestrial-type planets.

### 2.2.2 *WFIRST Exoplanet Science Motivation*

*Exoplanet Survey Question #1: How do planetary systems form and evolve?*

In the most general terms, planet formation theories should describe all of the relevant physical processes by which micron-sized grains grow through 13-14 orders of magnitude in size and 38-41 orders of magnitude in mass to become the terrestrial and gas-giant planets we see today. These physical processes are ultimately imprinted on the final distributions of planet frequencies, compositions, and orbits. Thus by measuring these distributions, i.e., by determining the demographics of exoplanets, it is possible to gain insight into the physical processes that drive planet formation.

The discovery of gas giant planets orbiting at periods of only a few days, as well as evidence for the migration of the giant planets in our own solar system, have highlighted the fact that planet formation theories must also account for the possibility of large-scale rearrangement of planet positions through gravitational and gas dynamical effects during and after the epoch of planet formation, and thus must also track the planets through billions of years of evolution. Many of these theories also predict a substantial population of "free-floating" planets that have been ejected from their planetary systems through interactions with other planets. Indeed, evidence for such a population was recently found using microlensing (Sumi et al. 2011). The interpretation of the final exoplanetary system architectures that we observe today must account for these dynamical processes.

The exoplanet microlensing survey of the WFIRST mission will provide an integral and essential component of a coordinated plan by the exoplanet community to answer Exoplanet Survey Question #1. In particular, WFIRST will provide the only way to complete the statistical census of planets begun by Kepler, by measuring the demographics of planets with masses larger than that of Mars and separations of greater than 0.5 AU. This includes analogs to all the Solar System's planets except for Mercury, as well as most types of planets predicted by planet formation theories thus far. The number of such discoveries will be large with > 3000 bound and 2000 free-floating planet discoveries expected. Whereas Kepler is sensitive to close-in planets but is unable to detect the more distant ones, WFIRST is less sensitive to close-in planets, but surveys beyond 0.5 AU better than Kepler. WFIRST is sensitive to unbound planets with masses as low as the Earth, offering the only possibility to constrain the frequency of these planets, which may have been ejected during the planet formation process. *Thus, WFIRST and Kepler complement each other, and together they cover the entire planet discovery space in mass and orbital separation (See*

Figure 5*) and provide the comprehensive understanding of exoplanet demographics necessary to fully understand the formation and evolution of planetary systems.*

*Exoplanet Survey Question #2: How common are potentially habitable worlds?*

The age-old question of whether or not there is life on other worlds has gained even more relevance and immediacy with the discovery of a substantial population of exoplanets. The first step in determining how common life is in the universe is to determine the frequency of potentially habitable worlds, commonly denoted $\eta_\oplus$. While of course interesting in its own right, an accurate measurement of $\eta_\oplus$ also provides a crucial piece of information that informs the design of direct imaging missions intended to characterize potential habitable planets around nearby stars and search for biomarkers. Indeed, the primary goal of Kepler is to provide a robust measurement of $\eta_\oplus$. Early results from Kepler indicate that terrestrial planets are likely to be very common, at least for short periods of <50 d (Borucki et al. 2011, Howard et al. 2011), thus bolstering the case that $\eta_\oplus$ might be high. However, these data also indicate that a robust measurement of $\eta_\oplus$ with Kepler will require an extended mission, primarily because of larger-than-expected intrinsic stellar variability (Gilliland et al. 2011). Furthermore, it will be difficult or impossible to measure the masses of the potentially habitable planets detected by Kepler because of the faintness of the host stars.

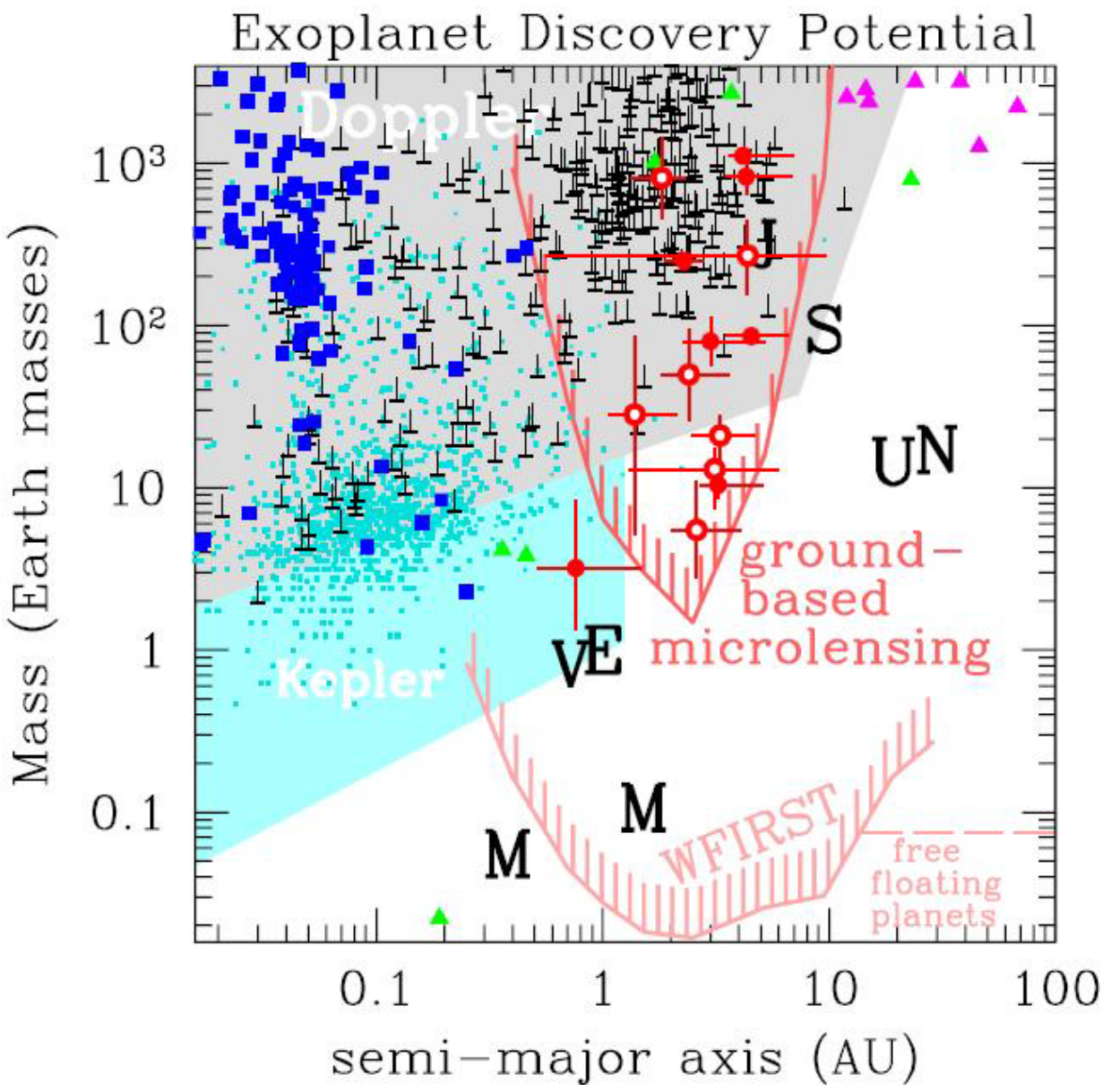

Figure 5: The distribution of known exoplanet masses and semi-major axes is compared to the expected final sensitivity of the Kepler mission and the sensitivity of a 500-day WFIRST Exoplanet survey. Ground based microlensing planets are show as red circles, radial velocity planets are brown inverted T's, planets discovered via their transits are blue squares, and planets found by timing and direct imaging are green and magenta triangles, respectively. The cyan region shows the expected sensitivity of Kepler, and the small darker cyan squares are Kepler's candidate planets. The purple curve shows the sensitivity of a 500-day WFIRST exoplanet program assuming 20% of stars have a planets of indicated mass and semi-major axis, and the dashed "free-floating planets" line indicates the sensitivity limit for free floating planets. Some planets beyond 10 AU will be discovered without a microlensing signal of their host stars.

The exoplanet survey of the WFIRST mission will directly address Exoplanet Survey Question #2 by providing an independent and complementary determination of $\eta_\oplus$. In particular, WFIRST will detect 27 $\eta_\oplus$ habitable planets orbiting solar-type stars, similar to the number from Kepler. In contrast to Kepler, however, WFIRST is sensitive to planetary mass, rather than planetary radius, thus enabling the statistical determination of the densities and surface gravities of habitable terrestrial planets. Furthermore, while Kepler is most sensitive to planets in the inner part of the Habitable Zone, and begins to lose sensitivity to planets in the outer habitable zone, WFIRST is most sensitive to planets in outer part of the habitable zone. Thus, the combination of WFIRST and Kepler data will make it possible to robustly interpolate into the habitable zone from regions just outside of it, even if the frequency of habitable planets turns out to be small.

In summary, WFIRST will provide crucial empirical constraints on planetary systems that will allow us to address the two fundamental questions: "*How do planetary systems form and evolve?*" and *"How common are potentially habitable worlds?".* Of course, these two questions are not independent. In particular, it is likely that the habitability of a given planet cannot be considered in isolation. The suitability of a planet for life depends on the average surface temperature, which determines if the planet resides in the habitable zone. However, there are many other factors that also may be important, such as the presence of sufficient water and other volatile compounds necessary for life (Raymond et al. 2004; Lissauer 2007). Therefore, the habitability of a planet likely depends on the formation and evolutionary history of its planetary system, and a reasonable understanding of planet formation is an important foundation for the search for nearby habitable planets and life.

### 2.3 Surveys and Guest Investigator Science

*Primary NIR Survey Science Objective: Produce a deep map of the sky at NIR wavelengths, enabling new and fundamental discoveries ranging from mapping the Galactic plane to probing the reionization epoch by finding bright quasars at z>10.*

In addition to dark energy and exoplanet science, WFIRST will provide a unique and powerful platform for a broad range of compelling infrared survey science. Some of these studies will use the data acquired for the dark energy and exoplanet surveys. Such legacy science is best exemplified by the low redshift Sloan Digital Sky Survey (SDSS) where the number of research papers now published using the archival SDSS data - originally collected to study the large-scale structure of the Universe - far exceeds the number of journal articles written by the original SDSS collaboration. To date, thousands of papers have been written using SDSS public data making it one of the most successful surveys of the Universe ever undertaken.

Other science will come from separate dedicated observations. This latter point is repeatedly emphasized in the *NWNH* report – e.g., "… the committee considers the guest investigator program to be an essential element of the mission …" (*NWNH*, pg. 207) and "As a straw-man example for the first 5 years ... the panel imagines ... a galactic-plane survey of one-half year, together with about 1 year allocated by open competition …" (EOS Panel Report, pg. 274). Indeed, the Decadal Survey constructed WFIRST out of three separate missions with similar hardware but very different scientific goals. One of these missions, the Near-Infrared Sky Surveyor (Stern et al. 2010), emphasized the vast scientific potential of a space-based, wide-field infrared survey observatory for a broad suite of Galactic and extragalactic science. We now briefly discuss some of this science, beginning with ancillary uses of core WFIRST observations and then discussing additional (guest) observational programs.

### 2.3.1 *Ancillary Science from the High Latitude Survey (HLS)*

The WFIRST dark energy program is likely to be the single largest component of the 5-year mission plan with a set of nested surveys. At the widest level, the BAO program will obtain shallow observations of ≥10,000 $deg^2$ with relatively uniform spectroscopic coverage but non-uniform imaging in multiple filters. The weak lensing (WL) program will obtain deep, uniform three-filter images of ≥ 2500 $deg^2$, and the supernova (SN) program will obtain extremely deep, cadenced observations of a few square degrees. For this report, we focus on the BAO and WL surveys, which together comprise the High Latitude Survey (HLS), though we note that significant ancillary science is expected from both the SN survey and the microlensing survey.

The HLS will identify unprecedented numbers of quasars at very high redshift. Quasars are among the most luminous objects in the universe, observable to the earliest cosmic epochs. They are thought to be supermassive black holes at the centers of galaxies, converting mass into energy 20 times more efficiently than stars. They provide fundamental information on the earliest phases of structure formation in the universe and are unique probes of the intergalactic medium. The discovery of a Gunn-Peterson (1965) trough in the spectra of several quasars at redshift $z > 6$ implies that the universe completed reionization near that redshift (e.g., Fan et al. 2001), though poor sample statistics, the lack of higher redshift quasars, and the coarseness of the Gunn-Peterson test make that inference somewhat uncertain. Currently, the most distant known quasar is at $z = 7.1$ (Mortlock et al., 2011), and only a dozen quasars have been confirmed at $z > 6$. Ambitious surveys will push this number to around 100 in the next few years, likely identifying one or two quasars at $z \sim 8$. WFIRST will fundamentally change the landscape of early universe investigations. Based on the Willott et al. (2010) quasar luminosity functions, WFIRST will identify thousands of quasars at $z > 6$, hundreds of quasars at $z > 7$, and push out to $z \sim 10$ should quasars exist at those redshifts (see Table 1). Such discoveries will directly measure the first epoch of supermassive black hole formation in the universe, probe the earliest phases of structure formation, and provide unique probes of the intergalactic medium along our line of sight to these distant, luminous sources. The large numbers of quasars identified in the first Gyr after the Big Bang will enable clustering analyses of their spatial distribution (e.g., Coil et al. 2007, Myers et al. 2007). Optical surveys are not capable of identifying quasars above $z \sim 6.5$ since their optical light is completely suppressed by the redshifted Lyman break and Lyman-alpha forests. Though JWST will be extremely sensitive at near-infrared wavelengths, it will not survey nearly enough sky to find the rarest, most distant, luminous quasars, which have a surface density of just a few per 10,000 $deg^2$. WFIRST will be the definitive probe of the first phase of supermassive black hole growth in the universe.

The HLS will also be a very powerful probe of galaxy evolution, reaching depths comparable to the Cosmic Evolution Survey (COSMOS; Scoville et al. 2007), but covering a thousand times more sky, enabling a breadth of statistical analyses similar to those undertaken on the low redshift SDSS. In particular, the size and scale of the dataset will allow for accurate measurements of the clustering of different types of galaxies (e.g., as a function of stellar mass, size, morphology, etc.) critical for testing hierarchical models of galaxy evolution and probing the details of how galaxies trace the underlying dark matter distribution at high redshifts. Though the spatial resolution will not be as exquisite as that of the Hubble Space Telescope, WFIRST is still expected to resolve ~80% of galaxies to $H$~25 (AB). These data will provide fundamental information on the formation and evolution of galaxies. In particular, near-infrared data is crucial for studying galaxies at $1 < z < 3$, which corresponds to the peak epoch of star formation in the universe. Near-infrared data is essential for deriving photometric redshifts for galaxies at these epochs, redshifts necessary for both WL and galaxy evolution science. WFIRST will significantly enhance ground-based optical surveys such as LSST, enabling galaxy evolution studies to reach to the critical $z$~2 cosmic epoch. The wide area of the HLS will also allow unique studies of distant galaxy clusters, which are powerful probes of both cosmology and galaxy evolution.

| Survey | Area ($deg^2$) | Depth (5-sigma, AB) | z>7 QSO's | z>10 QSO's |
|---|---|---|---|---|
| UKIDSS-LAS | 4000 | Ks=20.3 | 8 | - |
| VISTA-VHS | 20,000 | H=20.6 | 40 | - |
| VISTA-VIKING | 1500 | H=21.5 | 11 | - |
| VISTA-VIDEO | 12 | H=24.0 | 1 | - |
| Euclid, wide (5 yr.) | 15,000 | H=24.0 | 1406 | 23 |
| WFIRST, deep (1 yr.) | 2700 | F3=25.9 | 904 | 17 |
| WFIRST, wide (1 yr.) | (4730) | F3 = 25.3-25.5 | 1148 | 21 |

Table 1: Number of high-redshift quasars predicted for various ground- and space-based near-infrared surveys, based on the quasar luminosity function of Willott et al. (2010). Note: For the WFIRST wide survey, we only consider the 4730 $deg^2$ (out of 11,000 $deg^2$ total) that are imaged with at least two exposures in both filters.

Finally, the HLS will also be a powerful probe of cool stars and even cooler failed stars (brown dwarfs) within the Milky Way. For the coolest brown dwarfs, methane absorption in the *L*-band makes observatories such as the Wide-field Infrared Survey Explorer (WISE) (Wright et al. 2010) that are sensitive at 3 to 5 microns ideal for identifying objects at the star-planet boundary. However, the wide, sensitive, and high spatial resolution HLS survey will be a powerful tool for studying cool dwarfs in the Galaxy. The HLS will also be sensitive to cool white dwarfs (WDs) in the Milky Way halo, particularly to helium-atmosphere WDs, which become redder as they cool. [The formation of molecular hydrogen in hydrogen-atmosphere WDs causes such sources to become bluer when their temperatures drop below 5000 K (Hansen 1999).] WFIRST will derive the helium WD luminosity function, which provides unique, independent and direct measurements of the age of the Galactic halo and disk (e.g., Ferrario et al. 2005).

Many of the ancillary science objectives will benefit from having 3 (or more) filters in the imaging survey. With the vast number of objects detected, color information will aid in identifying sources. False positives will be reduced allowing the interesting rare new objects to be found with higher confidence.

### 2.3.2 *WFIRST Guest Investigator Science*

Competed guest investigator studies with WFIRST will significantly enhance the scientific return of this unique platform, providing the flexibility to respond to new scientific developments. The Decadal Survey suggested that ~6 months of the baseline 5 year mission be spent on a near-infrared survey of the Milky Way Galaxy. Such a survey would reach ~3 mag deeper than 2MASS (Skrutskie et al. 2006) and ~1.5 mag deeper than the UKIDSS Galactic Plane Survey (Lucas et al. 2008) before reaching the confusion limit. Mapping the entire 360 deg x 2 deg Spitzer GLIMPSE survey (Churchwell et. al, 2009) region would only require ~1 month of observations. Among the various scientific goals enabled by such a survey, with the appropriate choice of filters, a WFIRST Galactic plane survey could map the structure of the Milky Way by locating essentially all of the luminous red giant stars in the Galaxy, thereby identifying the edge of the Galactic disk and measuring the shape and the extent of the Galactic warp. Two-epoch observations of Galactic clusters, separated by ~5 years, would identify common proper motion members and detect sources well below the brown dwarf / free floating planet mass boundary of 10 – 15 Jupiter masses.

WFIRST will also be a powerful tool for obtaining deep near-infrared images (and spectroscopy) of nearby galaxies, complementing many large Treasury/Legacy programs on Hubble and Spitzer such as the Spitzer Infrared Nearby Galaxies Survey (SINGS) (Kennicutt et al. 2003), the ACS Nearby Galaxies Survey Treasury (ANGST) (Dalcanton et. al, 2009) and the Panchromatic Hubble Andromeda Treasury (PHAT) survey. Such data are useful for studying stellar astrophysics, mapping galaxy interactions and accretion, studying galaxy formation, and probing how the stellar mass function depends on environment. In a ~1 month program, WFIRST could resolve the stellar populations of ~100 nearby galaxies, allowing detailed studies delving into their star formation and merger histories.

The SDT has considered many additional guest investigator programs which are beyond the space limitations of the current report, ranging from exoplanet transit studies that would proceed alongside the exoplanet monitoring to measurements of the lensing mass profiles and stellar mass content of the most massive galaxy clusters at high redshift beyond the virial radius. In summary, operating beyond the reaches of the Earth's atmosphere, free of its limiting absorption and thermal background, WFIRST will deeply map the sky at near-infrared wavelengths, thereby enabling new and fundamental discoveries that address nearly the full breadth of astrophysics, reaching from the local neighborhood of the Milky Way Galaxy to the epoch of reionization.

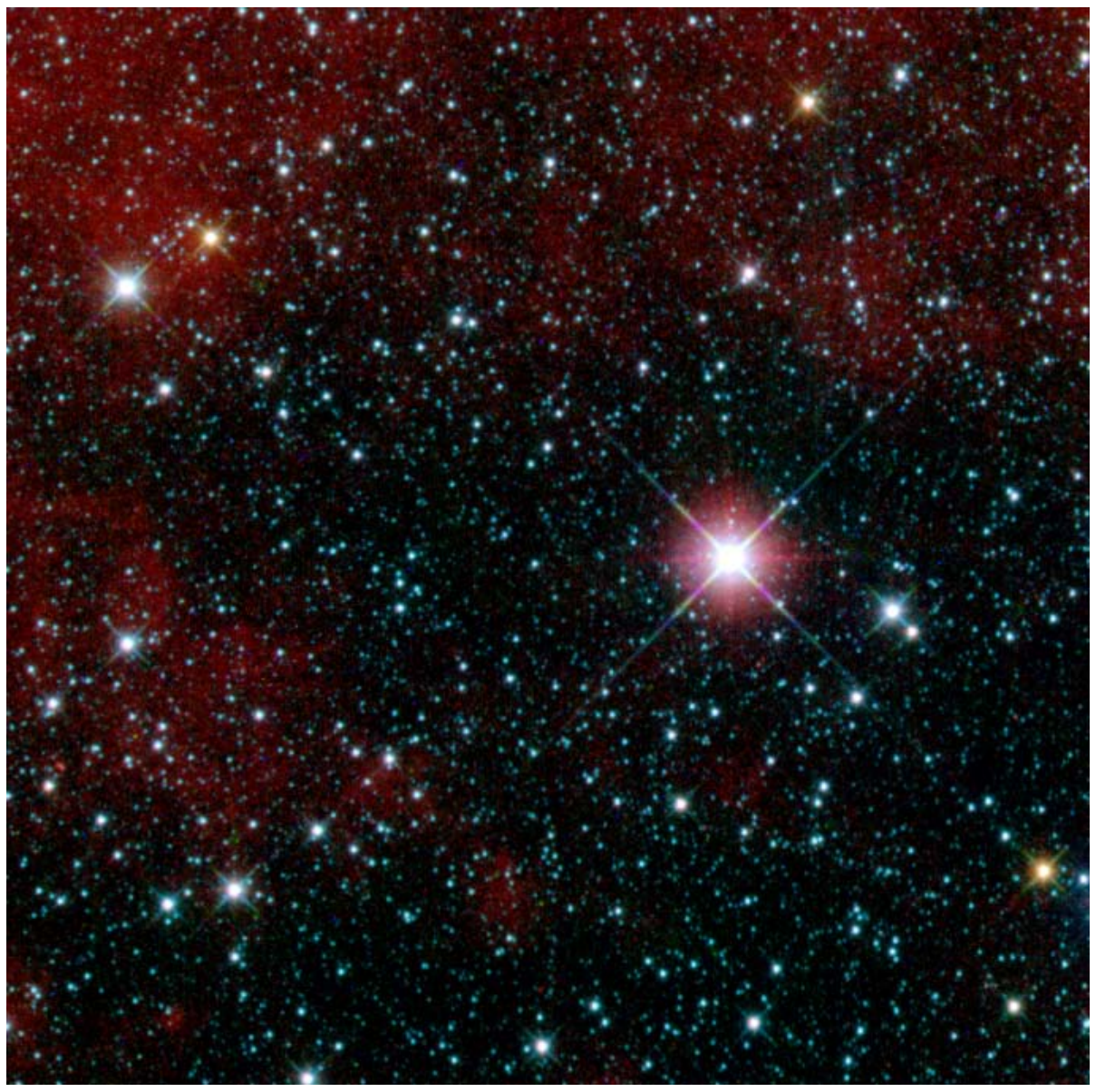

Figure 6: Sky image from the WISE space telescope of the Carina constellation region of the sky.

# 3 FIGURES OF MERIT

## 3.1 Figure of Merit (FoM) Introduction

In a mission with multiple science objectives there is competition for finite resources: telescope time, pixel size and number, field of view, and filter definitions are among the most obvious. As a tool to inform the allocation of these resources and to evaluate tradeoffs we have adopted a set of quantitative figures of merit (FoM) and have used these to allocate resources in a balance that closely follows the notional mission described in NWNH. Relative risk and cost will of necessity enter into the ultimate choice of the WFIRST DRM.

The decadal survey report describes the JDEM-Omega design as being nearly ideal for the mandated science, but identified a far broader science case than that of dark energy alone. The 2010 report suggested a five year mission during which two years were allocated to a high latitude survey of imaging and spectroscopy of galaxies for weak lensing, BAO and RSD measurements. Six months were allocated to Type Ia supernovae identification and follow-up. One and a half years were allocated to an exoplanet microlensing study, leaving one year for a survey of the Galactic plane and GI programs.

While the panel suggested a balance between programs, it also clearly anticipated the need to adapt to evolution in the scientific priorities of the community as new discoveries emerge. For example, the RSD method has gained substantially more credibility in the community in the two years since the community first responded to the Astro2010 request for information. Future revisions to the proposed strawman time allocations are certain.

Care should be taken when using a FoM as any FoM is only as good as the assumptions upon which it is based. Additionally, head to head comparisons of FoMs are valid only when comparably realistic assumptions are made.

## 3.2 Exoplanet FoM

### 3.2.1 *Exoplanet FoM Description*

The exoplanet FoM provides a quantitative measure of the ability of the WFIRST mission to achieve the primary exoplanet science goals outlined above. A given value of the FoM realized by a given mission design is related in a well understood, quantitative way to the science deliverables of the mission. The FoM thus provides a way to quantitatively assess the impact of changes in the mission design to the science return. The primary FOM depends on the product of four separate metrics that are tied directly to the four primary measurement requirements, and is defined as:

$$FoM_{ExP} = (N_{\oplus} N_{ff} N_{HZ} N_{20\%})^{3/8},$$

where the terms are:

- $N_{\oplus}$: The number of detected planets with M = $M_{\oplus}$ and period P = 2 years, assuming every main sequence star has one such planet. This choice is designed to be a diagnostic of the overall planet sensitivity and yield for the experiment. If the mission can detect these planets, it is guaranteed to detect more distant planets at the Einstein ring radius and beyond. Thus $N_{\oplus}$ quantifies the ability to address ExP survey requirement #2a (see Appendix A). This is also a region of parameter space difficult to access from the ground. Planets at fixed period are considered rather than at fixed semi-major axis, a, because $P/R_E$ is a weaker function of primary mass than $a/R_E$.

- $N_{20\%}$: The number of planets detected with M = $M_{\oplus}$ and period P = 2 yr for which the primary mass can be determined to 20%. Addresses survey requirement #2b.

- $N_{ff}$: The number of free-floating $1M_{\oplus}$ planets detected, assuming one free floating planet per star. Addresses exoplanet survey requirement #4.

- $N_{HZ}$: The number of habitable planets detected assuming every F, G and K star has one, where habitable means 0.1-10$M_{\oplus}$, and [0.7-2 AU]$(L/L_{\odot})^{1/2}$. Addresses exoplanet survey requirement #3.

These quantities are then multiplied and taken to the 3/8 power, so that the FoM is proportional to the three halves power of the observing time, all else being equal. This was chosen to match the scaling of the Dark Energy FoM.

WFIRST probes regions of parameter space (the habitable zone and beyond) that are in large part inaccessible to other methods, and our FoM is designed to reflect these unique capabilities. The current and future priors on our FoM are likely to be quite weak. We consider two sets of priors: Stage 1, corresponding to the number of planets currently known in each of the four categories above, and Stage 2, corresponding to our projections for the number of planets known in each category at the time of launch, for definiteness taken to be 2021, as shown in Table 2.

With the numbers shown in Table 2, we find that $FOM_{ExP}$ = 410 without any priors and $FOM_{ExP}$ = 539 with the assumed Stage 2 priors. Without WFIRST, we expect $FOM_{ExP}$ = 6 (with a significant chance of $FOM_{ExP}$ = 0).

| | $N_{\oplus}$ | $N_{20\%}$ | $N_{HZ}$ | $N_{ff}$ |
|---|---|---|---|---|
| Stage 1 (2011) | 0 | 0 | ~3 (ref 1) | 0 |
| Stage 2 (2021) | ~5 (ref 2) | ~2 (ref 2) | 25 (ref 3) | ~0.5 |
| WFIRST | 127 | 90 | 27 | 30 |

Table 2: Stage 1 and Stage 2 priors and WFIRST results for the exoplanet FoM [1. Borucki et al. 2010; 2. Bennett et al. 2010: MPF Astro2010 RFI, Figure 4; 3. http://kepler.nasa.gov/Science/about/ScienceGoals/expectedResults/, and Exoplanet task force final report (Lunine et al. 2008)]

#### 3.2.2 *FoM Evaluation of IDRM Microlensing Science Return*

The FoM has been calculated with an updated version of the space-based microlensing simulations developed by Bennett and Rhie (2002). Although WFIRST takes advantage of the wide-field, high resolution images available from space, the stellar density in the central Galactic bulge fields observed for the WFIRST exoplanet program is high enough that the photon noise from the blended images of nearby stars can often contribute significantly to the photometry noise budget for many of the WFIRST target stars. Therefore, it is necessary to use simulated images of these crowded bulge fields in order to generate simulated light curves with the appropriate noise properties. The simulated light curves are then searched for planetary signals at the appropriate S/N threshold to generate the values for $N_{\oplus}$, $N_{ff}$, $N_{HZ}$, as described above. For $N_{20\%}$, we also determine the planet and host star masses using the method described by Bennett et al. (2007). Table 2 shows the resulting values for the FoMs for WFIRST. When these values for the FoMs are achieved, as they are for the IDRM, WFIRST will be sensitive to the broad exoplanet discovery space shown in Figure 5.

The projected numbers of planet discoveries listed in Table 2 are also based on these simulations, but additional assumptions regarding the prevalence of exoplanets are also needed. Although there are a number of papers that measure the statistical prevalence of planets based on the radial velocity and transit methods (Cumming et al. 2008; Howard et al. 2010, 2011), it is the ground-based microlensing results that are the best match to the host stars and range of star-planet separations that WFIRST will probe. The combined analyses of Sumi et al. (2010) and Gould et al. (2010) give 0.4 planets per decade of mass and separation centered at a mass of ~80 Earth-masses and a separation of ~3 AU with a mass function slope that scales as $dN/d\log(m) \sim m^{-0.7}$, where $m$ is the planet mass. But, extrapolation to lower masses implies a very large number of low-mass planets. So, we assume that the mass function flattens out to $dN/d\log(m) \sim$ constant for $m < 5$ Earth-masses, which is close to the mass of the lowest ground-based detections. For free-floating mass function, we use the power-law mass function from Sumi et al. (2011), which scales as $dN/d\log(m) \sim m^{-0.3}$. Integrating over these mass functions yields the bound planet results shown in Box 3.

Most of the updates to the Bennett and Rhie (2002) space-based microlensing survey simulations were needed because the Bennett and Rhie simulations assumed CCD detectors instead of the near-IR detectors required for WFIRST. This allowed observations slightly closer to the Galactic center, with a higher microlensing rate but also higher dust extinction, to be used. Another change from Bennett and Rhie (2002) was the model for the input stellar microlensing rate. Bennett and Rhie (2002) used the value measured by observations of main sequence stars (Alcock et al. 2000), as these are the target stars used by WFIRST. But, observations of red clump giant stars consistently found a microlensing rate about 10% smaller, for reasons that weren't understood (Popowski et al. 2005; Sumi et al. 2006; Hamadache et al. 2006). As a result, the current simulations assume a slightly lower microlensing rate than Bennett & Rhie (2002). However, recent work (Kerins et al. 2009) seems to explain why the rate toward main sequence stars should be higher, so the assumed microlensing rate is a conservative one.

### 3.3 NIR Survey FoM

#### 3.3.1 *NIR Survey FoM Description*

The broad range of science potential for the NIR survey aspect of WFIRST presents some challenges in terms of defining an FoM. While the dark energy and exoplanet science programs have relatively specific and well defined questions they are attempting to answer, the NIR survey data will be beneficial for science ranging from studying asteroids in our Solar System to measuring the diffuse NIR background due to the first galaxies to form in the Universe. Examining galaxies and stars in a general NIR survey is also different in method from the very specific questions that an exoplanet or cosmology program is designed to do. Therefore, the goal of the NIR survey FoM is not just to optimize mission design decisions for the handful of specific questions detailed below, but also to ensure a robust and versatile mission that will enable broad science.

We could define an FoM related to the capabilities of the mission, e.g., scaling with the field-of-view, wavelength range, aperture to some power, effective spatial resolution, number of filters. However, by instead using a science-driven FoM, we focus this calculation, and ensure that the FoM is related to unique goals of the mission with large scientific impact. Similar to the discussion in Section 2.3, we focus on two broad categories of WFIRST NIR survey in order to span the range of potential NIR survey FoM's. We first define FoM's related to ancillary science from the HLS survey. We then discuss defining an FoM to quantitatively assess the ability of WFIRST to map the Milky Way Galaxy. This approach, followed for the interim report, ignores the significant ancillary science potential from other core WFIRST observations such as the deep, multi-epoch observations obtained for the supernova and micro-lensing surveys. For now, we also do not attempt to derive an FoM related to guest investigator science. The expectation is that design decisions undertaken based on the NIR survey FOM's defined below will also benefit such science.

The HLS will map many thousands of square degrees of high Galactic latitude sky. Such data will enable a broad range of investigations that are expected to impact a good fraction of the astronomical community. As one example, discussed in detail in Section 2.3, the sensitive, infrared, wide-field observations are ideal and unique for identifying a large census of quasars at the highest redshifts. We define two HLS FoM's related to quasar science:

- $N_{QSO1}$: The large number of quasars detected at z>7. Such sources allow us to probe the environments and clustering of the first black holes to form in the universe, as well as the evolution of their luminosity functions.

- $N_{QSO2}$: The number of quasars brighter than 1 μJy (corresponding to AB=23.9) at z>10. These very rare systems provide the best probes of the early universe.

The HLS data will also be very beneficial for studying galaxy formation and evolution, particularly at the crucial redshift range 1<z<2 when much of the stellar mass in the Universe forms into stars. Studying galaxies in this redshift range requires sensitive NIR data, data that is difficult to obtain over large fields using ground-based facilities. We define two HLS FoM's related to this science:

- $N_{gxy1}$: The number of galaxies at redshift z>1 detected down to the 5-sigma point source depth of the survey.

- $N_{gxy2}$: The fraction of spatially resolved galaxies.

The HLS is also expected to contribute significantly to other science such as cool Galactic sources — e.g., both brown dwarfs and white dwarfs — as well as galaxy clusters and groups out to z>2. Identifying the rare Galactic populations will require robust three-band photometry, similar to the requirements for the high-redshift quasars, while a galaxy cluster FoM would presumably scale directly with $N_{gxy1}$. Therefore, while we recognize and highlight the importance of such science, we do not define FoM's related to such science.

The Decadal Survey also highlighted that WFIRST should spend several months obtaining a sensitive, near-infrared Galactic plane survey. One could imagine a whole range of FoM's that might come from such a survey, such as the accuracy with which WFIRST could detect and study specific source populations such as brown dwarfs, young stellar objects (YSOs), or red giant clump stars across the Galaxy, or the ability of WFIRST to measure structural parameters of the Galaxy such as the pitch angle of spiral arms, the shape and extent of the Galactic warp, or the shape, length and orientation of the Galactic bar. Since confusion will be the limiting factor for WFIRST Galactic surveys towards the bulge, evaluation of these FOM's is a non-trivial task requiring accurate simulated star catalogs as a function of WFIRST pointing, taking into account the WFIRST PSF and filter complement. Such work is on-going, but was beyond the scope of what could reliably be achieved for this interim report. We note, however, that the IDRM wavelength coverage ends at 2.0 μm, which is likely inadequate for disentangling extinction from temperature when studying Galactic populations (e.g., Majewski, Zasowski & Nidever 2011). The longer wavelength infrared data are also important for studying brown dwarfs and YSOs. This issue of extending the long wavelength cutoff of WFIRST is discussed in more detail in Appendix E.

### 3.3.2 *FoM Evaluation of IDRM NIR Survey Science Return*

We evaluate the HLS NIR survey FoM's assuming one year of a weak-lensing DE survey (e.g., "deep" survey) and one year of a BAO DE survey (e.g., "shallow" survey). The deep survey covers 2700 deg$^2$ per year to a relatively uniform limiting depth of 25.9 mag (AB; 5$\sigma$ point source) in all three NIR filters (F1, F2 and F3) and will be very useful for a broad range of NIR survey science. The current shallow survey strategy has been optimized to provide relatively uniform spectroscopic coverage. This will be useful for NIR spectroscopic survey science (e.g., studying very cold Galactic brown dwarfs, to name but one example), though we note that the NIR survey science FoMs listed above focus on the imaging capabilities of WFIRST and this spectroscopic survey science is therefore not captured. The uniform shallow spectroscopic coverage leads to very uneven imaging coverage that is also only obtained in two of the three NIR filters (F1 and F3). This diminution of NIR color information will have some scientific drawbacks in terms of studying source populations. Of greatest concern, only a small fraction (<10%) of the shallow survey has three or more observations in both filters, leading to significant systematic issues with lost pixels due to cosmic rays and array defects, particularly for studies emphasizing extremely rare source populations with unusual colors. We evaluate the HLS FoM's assuming the current IDRM shallow survey strategy which covers 11,000 deg$^2$ per year, of which 5456 deg$^2$ (49.6%) has only one observation in both filters, 3751 deg$^2$ (34.1%) has two observations in both filters, and 979 deg$^2$ (8.7%) has at least three observations in both filters. In order to assure robust photometric catalogs, we consider three exposures as a strict requirement for NIR survey science, though, obviously, significant studies would be possible with the less robust, shallower imaging data. We therefore evaluate the shallow survey FoM's assuming 979 deg$^2$ is imaged to a depth of F1=25.7 and F3=25.5 (AB; 5$\sigma$ point source).

We first evaluate the expected number of high-redshift quasars that will be detected by the WFIRST HLS. We assume the quasar luminosity function (QLF) of Willott et al. (2010), which is based on the Canada – France – Hawaii Telescope Legacy Survey (CFHTLS). The CFHTLS is deeper than the Sloan Digital Sky Survey (SDSS), and therefore is preferred to the Fan et al. (2001, 2004) high-redshift QLF for this calculation. In a single year, the deep survey will detect 904 quasars at z>7, of which 4 will be at z>10 and brighter than 1 $\mu$Jy in F3. In a single year, the shallow survey will detect 261 quasars at z>7, of which 1 will be at z>10 and brighter than 1 $\mu$Jy in F3. This leads to combined quasar NIR survey FoM's of $N_{QSO1}$ = 1165 and $N_{QSO2}$ = 5.

We evaluate the FoM for the galaxy evolution science in the HLS by using data from existing deep near infrared surveys from the Hubble Space Telescope. Our calculations for the number of galaxies are done through the use of counts at 1.6 microns, which is the middle of the WFIRST filter range. The first FoM we consider is the number of galaxies at z > 1 which can be detected in the deep survey, which covers 2700 square degrees, and reaches a depth of 25.9 AB. Here we will detect ~10$^9$ galaxies within this magnitude limit with roughly half at z > 1. Furthermore 33% of these systems will be a z > 2 with most of these at z < 4, although WFIRST will also find a few very high redshift galaxies up to z ~6. For the wider area 11,000 square degree BAO survey, which is up to a magnitude less deep than the weak lensing survey, we find that the total number of galaxies detected will rise to 4x10$^9$, an increase solely due to the larger area surveyed. Within this number, the fraction at high redshifts, z > 1, is slightly lower than the deeper survey, but still just over 50%. The number of galaxies at z > 2 declines to just less than 30%. The faint galaxies between magnitude 25-26 AB are roughly evenly distributed between redshifts and the wider area survey will not detect the fainter systems at high redshift. The FoM for the fraction of spatially resolved galaxies is calculated using Hubble imaging data from previous deep NIR surveys as a basis. At ~0.2 arcsec resolution, the FoM is 0.85, for a slightly less resolved configuration of 0.25 arcsec, the FoM drops to 0.77. Note that although the majority of the galaxies will be resolved, the ones we miss will be typically the faintest and highest redshift systems.

## 3.4 Dark Energy FoM

### 3.4.1 *Scientific Context*

Einstein's theory of General Relativity states that, in general, the effect of gravity depends on both density ($\rho$) and pressure (P), proportional to $\rho c^2$+3P, something that is normally positive for matter. If the observed accelerated expansion of the universe is interpreted using General Relativity it implies this sum of density and pressure is negative on cosmic scales. This is dramatically at odds with our understanding of the na-ture of matter and energy. The missing constituent needed to resolve this discord is known as "Dark Energy".

Einstein described the interaction of gravity (*G*) with matter and energy (*T*) by means of a deceptively simple looking equation $G_{\mu\nu} = 8\pi T_{\mu\nu}$. There are two broad classes of explanations for the acceleration of cosmic expansion: (1) the first includes a negative pressure component in the energy-momentum tensor on the right hand side of Einstein's equation; or (2) the second involves a modification of the metric used on the left hand side of Einstein's equation or otherwise modifying the form of the left hand side of Einstein's equation to give a new fundamental law of gravity. Discovering either of these would have profound implications for our conception of the universe. In the first case, the new component may have a constant energy density, $\varepsilon=\rho c^2$, in which case it is equivalent to Einstein's "cosmological constant". Or, the new component may have a time-varying (dynamical) energy density, in which case it is called "quintessence".

One class of dark energy models posits that there is a negative pressure component to the energy density of the universe. In general, the pressure is related to the energy density by the linear equation of state relationship $P=w\varepsilon$, where $w$ is a number called the equation of state parameter. For radiation, $w=1/3$.; for normal matter or cold dark matter $w\approx 0$; and for a cosmological constant $w=-1$ (i.e., with negative pressure). Any component with $w<-1/3$ leads to a universe with an accelerated expansion.

We do not measure the pressure and energy density of matter directly, but rather infer them indirectly from their effects on the speed and acceleration of the universe's expansion, and how these affect the distances to and between cosmic objects such as supernovae and galaxies. As such, seeing a deviation from $w=-1$ does not imply a resolution to whether dark energy is a new, strange form of matter, or a modification to General Relativity.

Current measurements are completely consistent with a cosmological constant, but a key target of dark energy measurements is to test this model by determining the value of $w$ more accurately and by seeking evidence for the potential time-dependence of $w$. Either a deviation from $w=-1$ and/or and deviation from constancy would invalidate the cosmological constant model of dark energy.

Every technique for studying dark energy is subject to systematic errors – shortcomings of either an instrumental or astronomical nature that limit the accuracy that can be achieved. When systematic errors are present, there is a point beyond which merely acquiring more data of the same kind does not improve the accuracy of the result. It is only a slight exaggeration to say that the most important task facing the WFIRST project is to limit systematic errors. We therefore describe these at length in the sections that follow.

### 3.4.2 *Weak Lensing Considerations*

Weak lensing fundamentally measures the distribution of matter (both baryonic and nonbaryonic) in the universe. Different cosmological models produce different mass distributions, so the expected data can be used to test models. The principal technical challenge for WL is the measurement of galaxy shear with small systematic errors (<0.0003) in the presence of the instrument point spread function (PSF). The latter must be known very accurately to remove its effects, yet it varies as a function of field position, time, and wavelength. The data processing must correct for the instrument PSF and reject bad data while not introducing any new systematic errors of its own. Finally, WL also requires photometric redshifts across the entire range of source galaxy redshifts.

The dependence of the PSF on field position arises from aberrations, jitter, detector effects, and (on the ground) atmospheric turbulence. All of these effects can and often do vary from one exposure to the next, and many parameters are required to describe them. There are three major advantages to putting the telescope in space: (i) the atmospheric turbulence contribution is entirely eliminated; (ii) superior temporal stability of the PSF can be achieved; and (iii) by making the PSF smaller the multiplying factor from PSF ellipticity error to galaxy ellipticity error is reduced. As stated previously, shape measurement errors arising from PSF deconvolution rise as the square of the PSF size. With the guaranteed absence of adverse atmospheric effects, the time dependence of the PSF is dominated by motions of the mirrors (primarily, but not only, the secondary mirror). This source of variation will be controlled, but not completely eliminated on WFIRST. The residual misalignments can be described by a small number of displacement and tilt parameters that correspond with specific patterns in the wavefront error with known field position dependence. This is in contrast to atmospheric turbulence, where the number of parameters required to achieve part in 10,000 accuracy is not known. (The full field/time dependence of the wavefront error or PSF within an exposure cannot be measured; rather, the stellar images provide statistical averages over the much more complex instantaneous PSFs). While the WFIRST imaging optics and focal plane have 36 rigid-body relative-motion degrees of freedom, most linear combinations of these are benign and others are required to be controlled so that their variations on an exposure-by-exposure basis are small. A minimum of 7 telescope degrees of freedom must be taken into account (the nondegenerate secondary mirror displacements and tilts and the tilts of at least one other element) in each exposure, in addition to the jitter pattern (measured on all three axes by the fine guidance system, which occupies the same focal plane and looks through the same optical path as the science array).

Wavelength dependence of the PSF can in principle arise from diffraction, aberrations, atmospheric turbulence, the complex amplitude transmitted by the filter, chromatic aberration induced by refractive elements, and detector effects. Removal of these effects requires both knowledge of how the PSF depends on wavelength and multiband imaging of the source galaxies so that the intrinsic spatial color profile is known. A space telescope again provides a unique opportunity to address these effects. The size and ellipticity of an aberrated diffraction spot depend on wavelength, but for a reflective telescope in space this effect is uniquely determined by the same low-order aberrations that determine the PSF morphology. The WFIRST imaging camera has no refractive elements except for a flat zero-deviation filter substrate and there is no atmospheric dispersion. The filter transmission curve will vary across the pupil due to e.g. angle of incidence effects, which implies a spectral-energy-distribution dependent PSF even in the geometric optics approximation. This will be a challenge for all Stage IV weak lensing experiments but will be most difficult for ground projects with fast beams (versus the f/16 WFIRST beam) and where the filter effects beat against time-dependent contributions to the optical path that may vary across the aperture (e.g. turbulence). Detectors used at wavelengths where light has a long and strongly wavelength-dependent mean free path can exhibit defocus or (in non-telecentric systems) lateral shift as the beam passes through a thick detector; this is a source of color dependent PSF for silicon CCDs used in the far red (especially z+y bands) that is eliminated on WFIRST. Finally, WFIRST provides fully sampled imaging in two shape measurement filters, enabling the color and color variation to be measured for each galaxy. This provides an important built-in cross check.

Weak lensing represents a substantial data processing and algorithmic challenge. One such challenge (one of the few that is more difficult in space) is the need to combine multiple undersampled images to achieve fully sampled output. This was previously identified (both on the JDEM SCG and during the *New Worlds, New Horizons* 2010 Decadal Survey process) as a risk for space weak lensing experiments. The WFIRST project has supported development of algorithms that combine undersampled images without introducing artifacts associated with the pixel grid (Rowe et al., 2011). Another challenge is the automated stacking of several images (ranging from 5-6 for WFIRST through hundreds with some of the proposed ground projects) while rejecting defects, transients, or poor quality data, and avoiding the use of median filters or other highly nonlinear algorithms that produce artifacts in the PSF of the final data. The least risky approach here is to begin with a homogeneous data set that meets systematic requirements on an exposure-by-exposure basis.

Providing accurate photometric redshifts requires access to the near infrared for galaxies whose Balmer or 4000Å breaks have redshifted out of the optical (the *ugriz* and possibly the *y* ) bands, but which do not have an accurately identified Lyman break. Deep broadband wide-field near infrared imaging is only possible from space due to the extremely bright atmospheric OH bands. This limits ground-based photometric redshifts in a wide-field survey to $z \sim 1.3$ (except for the ~1 galaxy per arcmin$^2$ at $z > 2$ that is resolved from the ground). Since weak lensing shear is already an integral over the line of sight, this severely restricts the redshift baseline over which the evolution of the lensing signal can be measured. WFIRST imaging in three near infrared filters will complement ground-based optical photometry, providing both the break position and at least one rest-frame optical filter to $z \sim 3$.

### 3.4.3 *Baryon Acoustic Oscillation Considerations*

Primordial density fluctuations seeded all of the structure in the universe. In the early universe a tightly coupled fluid of photons, electrons, and baryons expanded from regions of initial over-density (and over-pressure). Eventually the photons decoupled from the fluid and are now observed as the cosmic microwave background (CMB). Small over-abundances of baryons were left in spherical shells around the locations of the initially over-densities. These appear as ripples on the power spectrum of large samples of galaxies, $P(k)$. These correspond to a ~1% excess in galaxy correla-tions. WMAP measured the size of the horizon at the drag epoch shortly following decoupling (the radius of the shells) to be 153 co-moving Mpc from the peaks and troughs of the power spectrum that are the signature of baryon acoustic oscillations. The WMAP data, which will be augmented by the currently operating Planck satellite, calibrates a “standard ruler" that is useful for tracing the expansion history of the universe.

The use of the BAO to investigate dark energy requires a large galaxy redshift survey over large regions of the sky. Observations of the spherical sound wave are made radially and transverse to the line-of-sight. The transverse measurement yields the angular diameter distance, which could be measured to an aggregate precision of 0.1%. The radial measurement yields the Hubble parameter, $H(z)$, a direct measurement of the expansion history of the Universe. Most of the power of the BAO measurement of dark energy comes from the radial measurement of the Hubble parameter. Spectroscopic redshifts are required to extract the radial BAO signature, since photometric redshifts unacceptably smear the peak in the radial direction (e.g. at z=1, even an excellent photo-z quality of $\sigma_z/(1+z)=0.02$ produces a 100 Mpc radial smoothing of the density map; the smoothing of the correlation function is a factor of $\sqrt{2}$ greater.

Obtaining the three-dimensional positions of galaxies (sky positions and redshifts) is not particularly challenging. Achieving maximum sky coverage with uniform instrumental properties is an advantage of a space mission. The major limiting factor is sufficient observing time to measure the full available sky to a sufficient depth to avoid excessive shot noise.

One potential systematic error arises only when the measured structures become nonlinear. Due to the increasingly non-linear growth of structure with time, this is a greater error for the ground-based, lower redshift experiments than for WFIRST observation at higher redshifts. Even at low redshifts, the acoustic scale of 153 Mpc is so much larger than the scale of non-linear structure formation that the scale remains a robust standard ruler. Further, since <1% of the volume is non-linear, perturbative corrections for a small degree of non-linearity are possible.

The baryon oscillations are smeared out at low redshift (“smearing” here means that the peak in the correlation function is widened; the Fourier space description is that the amplitude of the wiggles is suppressed at high *k)*. This is thought to be mostly from the effects of large-scale flows: two points that were initially separated by 153 Mpc co-moving may at the present epoch have a somewhat larger or smaller separation. Simple density-field reconstruction techniques can significantly counteract this effect (see e.g. Padmanabhan & White 2009; Noh et al. 2009 for recent studies, including studies of clustered haloes as opposed to simply the dark matter power spectrum). For the large-scale flows to cause the acoustic scale to be shifted would require a consistent mean infall on 153 Mpc scales. This non-linear shift is only 0.2% at z=1 and smaller at higher redshift. Since these shifts can be computed, they can also be corrected.

The acoustic scale could shift from galaxy bias, the offset between the observed baryonic matter distribution and the more significant underlying dark matter distribution. A linear bias does not change the shape of the correlation function, so it cannot shift the scale. Galaxy bias can affect how the convergent and divergent large-scale flows attain positive or negative weights in the two-point function (qualitatively, the source of shift is that large-scale flows tend to bring highly biased objects that trace large overdensities toward each other). The relation between the galaxy field and the large-scale velocity field is the quantity that matters here. Calculations with simple halo occupation models show suggest that this is a small effect (<0.2% at z=1). WFIRST’s large galaxy redshift survey should be able to constrain the bias model to correct this 0.2% effect at $z$ = 1 to < 0.1%. Other, more exotic galaxy biasing effects that might affect the BAO position have been proposed but are not expected to introduce any uncorrectable systematic errors (e.g. Tseliakhovich & Hirata 2010; Yoo et al. 2011).

Because the BAO observations give a differential signature as a function of scale, most errors cancel out. According to the Dark Energy Task Force, “*This is the method least affected by systematic uncertainties...*," and, “*It is less affected by astrophysical uncertainties than other techniques.*" The BAO is the only approach that the NRC's Beyond Einstein Program Assessment Committee (BEPAC) called “*very robust.*" Well-understood cosmological perturbation theory and numerical simulations provide a solid framework with which to control small biases and nonlinearities, especially for $z$ > 1. In the end, the BAO measurement may well be statistics-limited to the cosmic variance limit.

### 3.4.4 *Redshift Space Distortion Considerations*

Weak gravitational lensing tests modified gravity models by comparing the amplitude of clustering measured at two different redshifts to deduce the rate of growth of structure. The growth rate can be measured directly at a single redshift by recognizing that the reason that structure grows is because matter flows into potential minima (clusters). These flows are directly observable as redshift-space distortions in galaxy redshift surveys. While clustering is statistically isotropic, peculiar velocities affect the apparent positions of galaxies only in the line of sight direction. The degree of the observed anisotropy on large scales probes the growth of structure. Redshift space distortion can measure the growth of structure to a level comparable to that from weak lensing and cluster counting, but are complementary in that they measure velocities whereas weak lensing measures the gravitational potential and cluster counting identifies the many-σ peaks of the density field. Moreover the WFIRST redshift space distortion measurements will probe out to z = 2, well beyond the peak of the weak lensing sensitivity.

Large galaxy redshift surveys offer the opportunity to measure growth of structure to high precision. Unlike the more robust BAO, this requires careful modeling of the non-linear velocity field and the effects of galaxy bias. Cosmological simulations are required to help with this modeling. The degree of accurate modeling determines how close to the cluster the matter flow can be used. The minimum scale implies a maximum wavenumber, $k_{max}$. The use of redshift space distortion is a relatively young field, but it is rapidly advancing. It is difficult to predict with precision what $k_{max}$ will be achievable when WFIRST flies, but the proposed hardware design is such that the relevant data are captured during the course of the BAO survey.

### 3.4.5 *Type Ia Supernovae Considerations*

The utilization of Type Ia supernovae as calibrated standard candles led to the original discovery of the acceleration of the expansion of the Universe. In the past decade this technique was used in second generation experiments to further refine our understanding of this phenomenon. These second generation experiments have proven to be very productive, and have performed as predicted before the measurements were carried out. To this day supernovae represent the most mature and predictable method in the study of the acceleration of the universe and the nature of dark energy. The Dark Energy Task Force similarly concluded that: "The SN technique is at present the most powerful and best proven technique for studying dark energy."

The relative peak brightnesses of Type Ia supernovae are used to measure relative luminosity distances across redshift. This method has the advantage of providing a rather direct measurement of the distance. This directness also makes it easier to recognize – and develop controls for – the important sources of systematic uncertainties. These have been extensively studied by the second generation experiments, and systematic error budgets can be found in all recent major results papers (see, e.g., Guy et al. 2010, Suzuki et al. 2011). Almost all of the dominant current systematic uncertainties have reached or are reaching the limits of what can be observed from the ground. To take the supernova technique to its next level of precision, and take advantage of this powerful cosmological tool, will require space.

As we explore a wide range of redshifts, it is important to ensure that the supernovae we are comparing from low to high redshift do not represent different distributions among the subpopulations of SNe Ia, and that they are not being dimmed/reddened by different dust. These two "evolution" systematics are the most important astrophysics considerations in the systematics budget (as opposed to instrumentation/calibration considerations). The concern about supernova population evolution is addressed by measuring identifying characteristics of each supernova (such as the lightcurve timescale/shape and color or specific spectral features) and separately calibrating the subpopulation that is so identified. It is also possible to find specific SN Ia measurements (such as the H band lightcurves) that are empirically more similar across subpopulations. The systematics due to evolution of dust dimming/reddening is also addressed by measuring colors of the supernovae, and by observing in redder bands when possible.

Space observations are required for these measurements for several reasons. The range of restframe wavelengths used for identifying subpopulations and characterizing dust needs to be broad enough to obtain a substantial lever arm. By using the 1- to 2-micron near-infrared capabilities of a space mission it is possible to almost double the rest wavelength lever arm. The low sky background and small PSF available in space also makes possible observations of SN Ia spectral features, providing identification and redshift for every supernova used. In fact, these advantages of space could even allow spectra with excellent signal-to-noise that can further address these systematic uncertainties (see Appendix E for a discussion of how this might be possible with an IFU spectrograph substituting for the slitless prism).

There are also systematics issues related to the instrumentation and the experimental design. Observing the supernovae in different restframe bands at different redshifts has often been necessary in ground-based programs, and the resulting K-correction interpolations and extrapolations have contributed significantly to the systematics budgets. With the wider range of observer-frame wavelengths observable from space, the same restframe wavelengths can be observed at every redshift and K-correction extrapolations can be eliminated; With time-sampled SN spectrophotometry from space the systematics due to K-correction interpolations can be addressed, too.

Perhaps surprisingly, one of the remaining dominant systematic issues is due to the calibration of the photometry. While in principle this systematic can be controlled by careful experimental design on the ground or in space, in practice the thermal and gravitational-load stability of the space platforms, as well as the scheduling predictability (without weather) makes it easier to achieve in space.

The excellent final performance of WFIRST's Type Ia supernova measurements reflects the space-based control of all these systematics. By measuring the light curves of typically a hundred supernovae in each redshift bin of 0.1, distance errors of 0.8 to 1.5 % can be obtained in each bin. (If the optimistic systematics control is obtained there will be a motivation for considering an additional year of observations, since the additional square root of N improvement could then yield strikingly smaller uncertainties.)

### 3.4.6 *The Figure of Merit*

It is useful to be able to estimate the power of different experiments, or combinations of experiments, towards determining or constraining the physical possibilities for dark energy. This is the idea behind a "figure of merit." Over the last few years, the cosmological community, through the JDEM Science Definition Team, the Dark Energy Task Force, and by the JDEM Figure of Merit Science Working Group (FoMSWG) developed the idea of creating a figure of merit to quantify the effectiveness of various dark energy measurements. The figure of merit calculations assume a given set of agreed upon prior information (cosmological values accepted from existing experiments), and test the power of various experiments to distinguish between alternative dark energy models. Ideally, an experiment would distinguish between two models, say A and B, but that is not the case for dark energy, where a very broad range of models, with very different implications for both structure growth and the expansion history of the Universe, are possible. Rather than consider a specific dark energy model or a suite of models, the DETF opted for testing the ability of experiments to test a single representative model.

The figure of merit devised by the Dark Energy Task Force is a simple 2-parameter model, $w(a)=w_0+w_a(1-a)$, where a is the dimensionless scale factor of the universe with $a=1$ today, and a increasingly smaller than unity in the past (it is related to the redshift by a=1/1+z). In this model, the universe then began with $w= w_a$ and the current value is $w=w_0$, with a linear change with scale factor in between. The midpoint of the transition is at $1-a = ½$ or $z=1$. An experiment that places limits on dark energy can then be characterized by an error ellipse in the $w_0$ versus $w_a$ plane. The figure of merit is then inversely proportional to the area enclosed by the 95% confidence contour of this ellipse; the better the constraining power of an experiment, the higher the figure of merit. The DETF figure of merit may be the most standard one in use today. We therefore adopt it for the present purposes, notwithstanding such shortcomings as it might have (see Appendix C).

### 3.4.7 *Comparing Figures of Merit*

While the DETF FoM may be used to compare the outcomes of different dark energy programs, one must do so taking careful account to use identical "priors" and identical assumptions about systematic errors of astronomical (as opposed to instrumental) origin. Priors are previously obtained results that provide additional constraints on the interesting quantities. The DETF proposed a set of "stage III' priors that we adopt here.

As described in the preceding sections, each technique for studying dark energy has systematic errors that result from the method of measurement and others that result from imperfect models for the astronomical systems being measured. In what follows we have adopted two sets of systematic error estimates, a conservative set and an optimistic set. The conservative set represents the collective opinion of the SDT as to what is likely to be achievable.

We can reasonably hope for improved astronomical understanding and better control of instrumental errors between the present writing and the launch of WFIRST. We have therefore included a second set of FoMs for WFIRST that adopt these more optimistic systematic errors. The conservative and optimistic assumptions regarding systematic errors are described below.

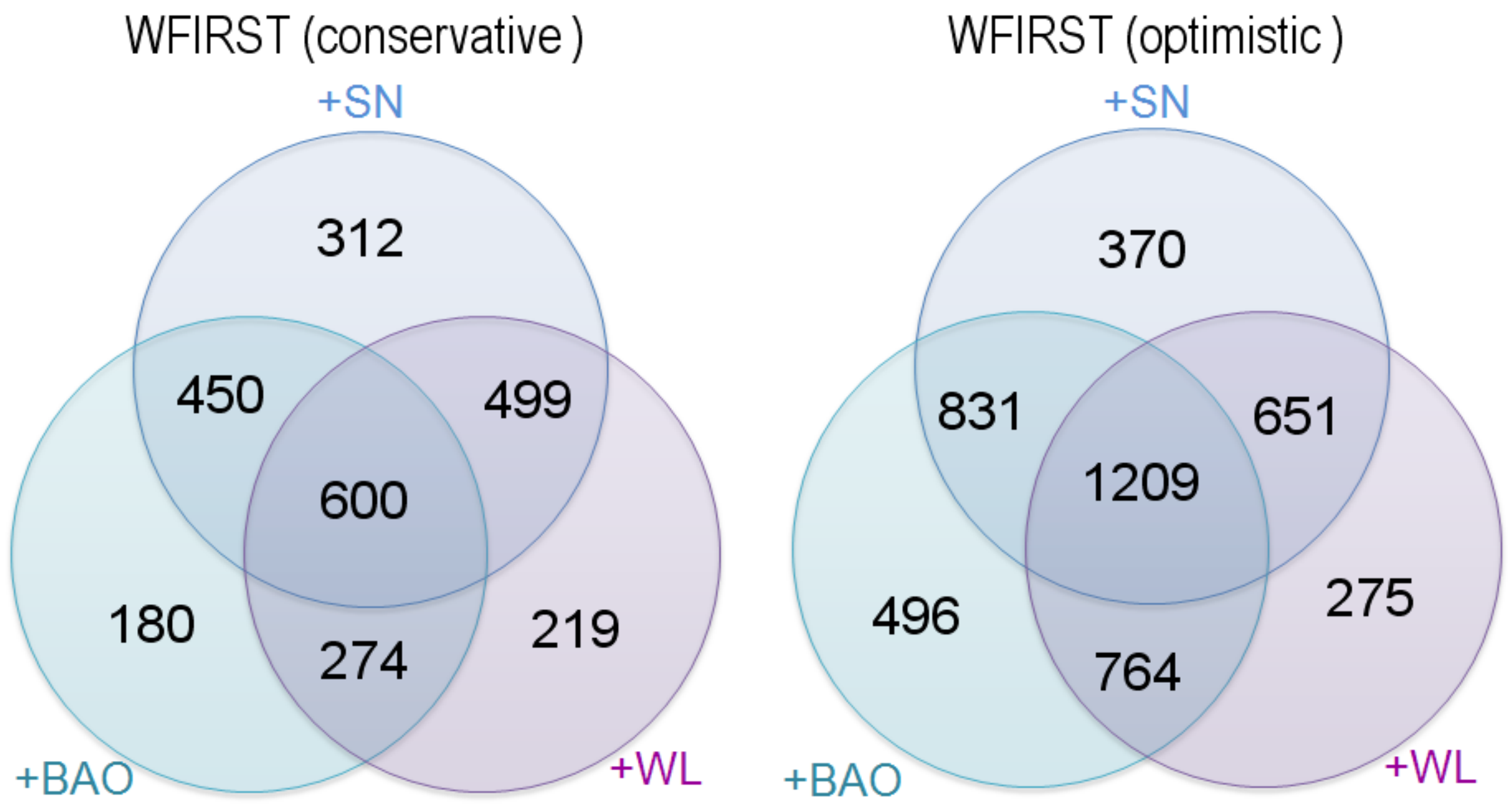

Figure 7: DETF FoM calculations for conservative and optimistic WFIRST assumptions. The stage III baseline is a DETF FoM = 116.

In Figure 7, we show the DETF figures of merit achieved by each of the three methods called out in NWNH, along with the results of combining techniques. They are calculated for a straw man allocation of one year to a deep, weak lensing plus BAO survey, one year to a wide BAO survey, and six months to a supernovae survey. The figure on the left was obtained using conservative systematic error estimates; the one on the right using the more optimistic estimates.

In Appendix C, we compare the results of this straw man program with those for other proposed missions. In so doing we have made every effort to treat systematic errors consistently. It is clear from the figures that such comparisons can only be meaningful if the systematic errors are treated uniformly. Even then, comparisons with different missions is difficult as it doesn't take into account the relative risk of different approaches or the likelihood that design goals will be met.

The expected sky coverage and projected number density of galaxies for the 3 different weak lensing surveys we consider (WFIRST, LSST and Euclid) are given in Table 3.

We assume common survey assumptions for each:

- A galaxy distribution described by $n(z) \propto z^2 \exp(-z/z_0)^{1.5}$ with $z_0=0.64$ for $0<z<2$. We break the galaxies into $N_{ph}=10$ redshift bins, with galaxies evenly distributed between them,
- A photometric redshift uncertainty of $\sigma(z)=0.04(1+z)$.
- A statistical shear measurement uncertainty, $\sigma_\gamma=0.35/\sqrt{2}$.
- 50 logarithmically spaced bins in ell, $10<l<3000$

| | WFIRST | LSST | Euclid |
|---|---|---|---|
| Sky coverage (sq. deg.) | 2640/yr | 15,000 | 15,000 |
| Project galaxy density (per sq. arcmin.) | 30 | 20 | 30 |

Table 3: Summary of survey specifications for WFIRST, LSST and Euclid weak lensing surveys.

We consider two scenarios summarized in Table 4:

- a conservative scenario, in which we consider constraints from lensing shear correlations alone, and with a number of systematic uncertainties included, and
- an optimistic scenario in which galaxy-lensing cross correlation information is also included, and in which we assume better control over both instrumental and astrophysical systematic uncertainties is achievable.

| Assumptions | Conservative | Optimistic |
|---|---|---|
| Data included | Shear-shear only | Shear-shear + Galaxy position - shear corr[lns] |
| Marginalization over: | | |
| Bias: $b_g(k,z)$ and $r_g(k,z)$ | | Y |
| Uncertainties in IA model: $b_I(k,z)$ and $r_I(k,z)$ | Y | |
| Shear calibration offset | Y | |
| Photo-z calibration offset | Y | |

Table 4: Summary of the systematic uncertainties included in the weak lensing scenarios.

Conservative scenario details:

- Intrinsic alignment (IA) contributions to the observed shear field are modeled using a nonlinear alignment model (Hirata and Seljak 2004, Hirata et al. 2007). We marginalize over uncertainties in the amplitude of the IA auto-correlation and cross-correlation with galaxy position using an analogous approach to galaxy bias, marginalizing over 5x5 grids for two parameters $b_I$ and $r_I$ (Joachimi and Bridle 2010). As in the FoMSWG analysis (Albrecht et al. 2009) we include a prior on $\sigma(b_I r_I)$= $0.003 N_{bias}\sqrt{(N_{ph}-1)}$
- A shear calibration $C_{ee}^{ij(obs)}(l)=(1+f_i)(1+ f_j) C_{ee}^{ij}(l)$ with a prior on $\sigma(f_i)=0.001\sqrt{N_{ph}}$ independently in each redshift bin.
- Photometric redshift offsets with a prior $\sigma(\Delta z_{sys}) = 0.002(1+z)$.

Optimistic scenario details:

- We include uncertainties about galaxy bias by introducing a bias amplitude, $b_g$ , $P_{gg}(k,z)^{obs} = b_g(k,z)^2 P_{gg}(k,z)$, and cross correlation coefficient between galaxy position and shear measurements, $r_g$, $P_{ge}(k,z)^{obs} = b_g(k,z)\ r_g(k,z)\ P_{ge}(k,z)$.
- The Fisher analysis includes marginalization over a $N_{bias}\times N_{bias}$ grid with $N_{bias}$=5 logarithmically spaced in z and k of $b_g$ and $r_g$. The values at each scale and redshift the values come from interpolating over the grid.
- Acknowledging that our galaxy bias marginalization may not be sufficient to describe the fully nonlinear regime, we include a cutoff in multipole space for each photometric redshift bin, $l_{max}(z_i)=0.132 z_i$ hMpc$^{-1}$ (Rassat et al. 2008, Joachimi & Bridle 2009). Galaxy position correlations with $l > l_{max}(z_i)$ are excluded from the Fisher analysis.

For the SN survey, the conservative FoM calculations include the following assumptions:

- Imaging/Slitless prism spectroscopy 6 month survey to z=1.2.
- Error assumptions: supernova intrinsic spread in intrinsic luminosity $\sigma_{int}$= 0.11+0.033z, systematic error $\sigma_{sys}$ = 0.02(1+z)/1.8

For the SN survey, the optimistic FoM calculations include the following assumptions:

- Same survey time and depth as the conservative assumptions
- Error assumptions: same as conservative except systematic error reduced to $\sigma_{sys}$ = 0.01(1+z)/1.8

For the SN survey in a five year Dark Energy program, the FoM calculations include the following assumptions:

- Survey time doubled to 12 months and redshift range increased to z=1.5
- Error assumptions: same as optimistic calculations

Supernova on other Dark Energy programs

- Euclid and BigBOSS are not planning on a supernova program.
- LSST will get many supernova but no supernova program has been defined that we are aware of for which we could calculate an FoM. Ground based supernova programs are limited to lower redshifts and shorter wavelengths than possible from space.

For the BAO and RSD survey, the conservative and optimistic scenarios used in the FoM calculations are defined in Table 5. Additionally, the following assumptions were used in all of the BAO and RSD FoM calculations:

- BAO calculations are done in bins of dz=0.1. A 0.36% error is added in quadrature to each of the H and D_A errors in each bin, representing a systematics floor
- The non-linear smearing of linear signal power is modeled following the method of Seo and Eisenstein, 2007.
- The Lagrangian displacement scale that sets the width of the Gaussian smearing kernel is reduced by 50% to represent reconstruction of the linear signal.
- Redshift space distortion is included by measuring $f^2$ P_m(k) marginalizing over bias in the linear theory formula P_{red}(k,$\mu$)=(b+f $\mu^2)^2$ P_m(k), independently each redshift bin.
- Gaussian smearing of information, as in Seo & Eisenstein, again with a 50% reconstruction factor

### 3.5 Growth of Structure Figure of Merit

Dark energy may be the most likely explanation for cosmic acceleration, but it is not the only possibility. One class of alternative explanations involves modifications of Einstein's theory of General Relativity. Modified theories of gravity can have significantly different predictions for how large scale structures, such as galaxies and clusters of galaxies, form and grow, as compared with the predictions of dark energy models. These signatures can be detected by weak lensing measurements and the motions of large scale structure, also known as redshift space distortion, imprinted in galaxy redshift surveys.

A simple extension to the DETF FoM was suggested to include information about large scale structure growth. This assesses an experiments ability to measure the growth rate of fluctuations in density, $\delta=\delta\rho/\rho$, that grow to form galaxies and clusters of galaxies. The growth rate function, f(z)=dln $\delta$/ dlna can be well modeled by $f(z)=\Omega_m(a)^\gamma$ with $\gamma$=0.55 for an acceleration universe with w$\approx$-1 and obeying General Relativity. Measuring a deviation from $\gamma=0.55+\Delta\gamma$ would help determine whether cosmic expansion is due to dark energy or to non-Einsteinian gravity.

A straightforward figure of merit that takes this into account is the inverse of the square of the uncertainty in $\gamma$. As with the DETF FoM, we have computed the contributions of baryon acoustic oscillations, supernovae and weak lensing to this figure of merit, under our conservative assumptions, and show how they combine in Figure 8.

| | WFIRST Conservative | WFIRST Optimistic | WFIRST 5 yr DE Program | Euclid | BigBOSS |
|---|---|---|---|---|---|
| Sky Coverage (sq. deg.) | 13,700 (1 yr deep & 1 yr wide) | Same as conservative | 23,250 (2.5 yr deep & 1.5 yr wide) | 15,000 | 14,000 (Northern survey) |
| Depth Limit (ergs/cm$^2$/sec) @ 1.5 $\mu$m | $2\times10^{-16}$ (deep) $4\times10^{-16}$ (wide) | Same as conservative | Same as conservative | $4\times10^{-16}$ | N/A |
| Redshift Range | 0.7 – 2.0 | Same as conservative | Same as conservative | Same as conservative | ≤ 1.7 |
| Redshift Uncertainty | 0.001(1+z) | Same as conservative | Same as conservative | Same as conservative | Same as conservative |
| Redshift Space Distortion | No | Yes | Yes | Yes | Yes |
| kmax (h/Mpc) | N/A | 0.15 | Same as optimistic | Same as optimistic | Same as optimistic |
| Reconstruction factor | N/A | 50% | Same as optimistic | Same as optimistic | Same as optimistic |
| Bias | N/A | 0.79/D(z) | Same as optimistic | Same as optimistic | 1.7/D(z) for LRGs, 0.84/D(z) for ELGs |

Table 5: Summary of the conservative and optimistic surveys for the BAO/RSD FoM calculations

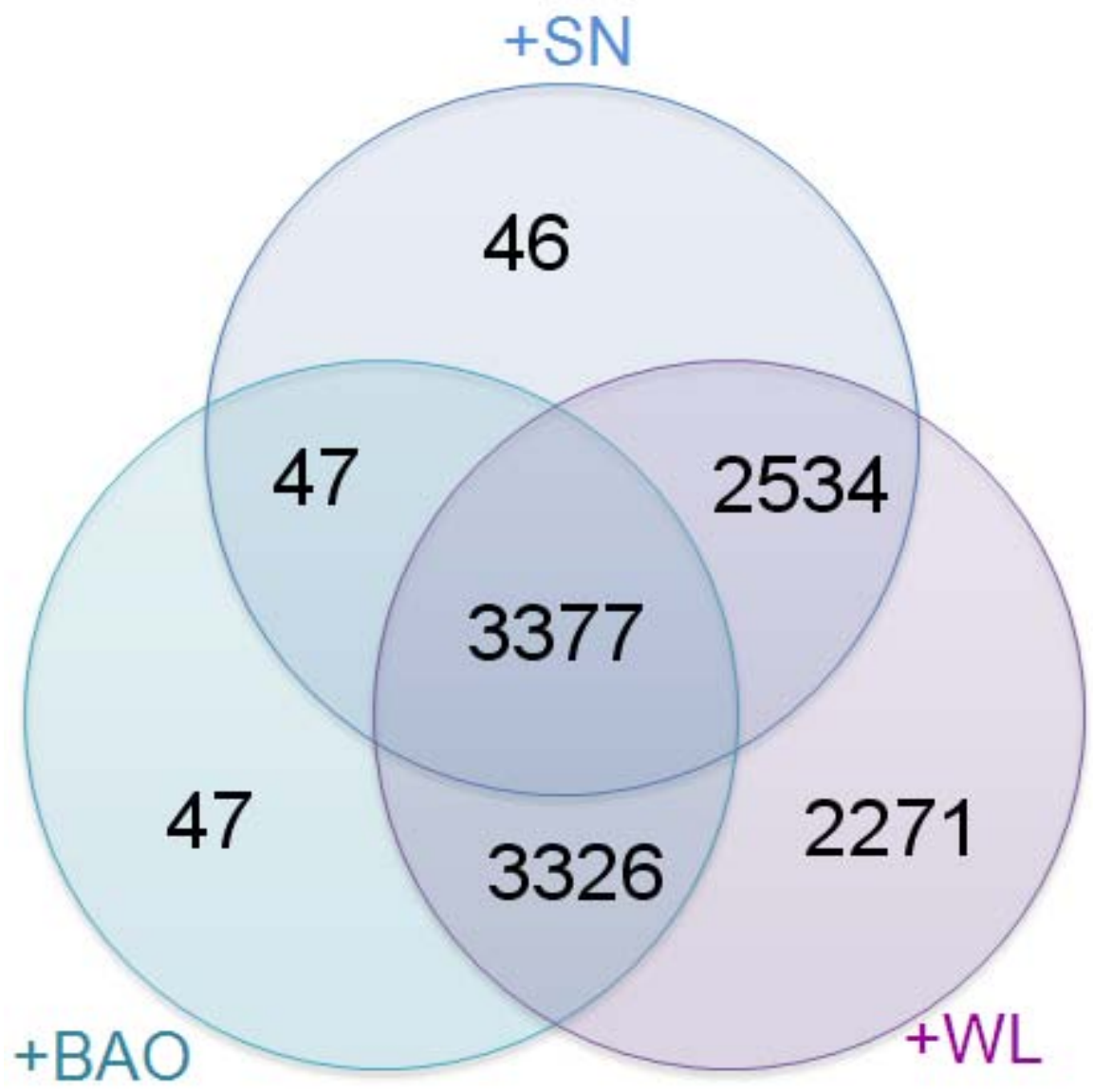

Figure 8: WFIRST conservative $\gamma$ figure of merit = $1/\sigma(\gamma)^2$. The stage III baseline is a Gamma FoM = 46.

Among the three methods, weak lensing is unique in its ability to constrain the growth of structure independent of other measurements. In concert with baryon acoustic oscillation measurements, these constraints are yet stronger.

Opinions vary regarding how much more likely it is that dark energy, as opposed to modified gravity, is the correct explanation for cosmic acceleration. While NWNH discusses the latter alternative, the emphasis is on dark energy. What is clear from the above figure is that a mission that does not involve weak lensing might miss the cause of cosmic acceleration.

# 4 IDRM IMPLEMENTATION

## 4.1 Overview

The WFIRST IDRM payload configuration (see Figure 9 for an optical path block diagram and Figure 10 for a fields of view layout) provides the wide-field imaging and slitless spectroscopy capability required to perform the Dark Energy, Exoplanet and NIR surveys. A 1.3 m unobscured aperture, focal telescope feeds a single instrument comprised of three observing channels: an Imaging Channel (ImC) covering 0.60 – 2.0 μm, and two identical, but oppositely dispersed, Spectrometer Channels (SpCs) covering 1.1 – 2.0 μm. The instrument uses 2.1 μm long-wavelength cutoff HgCdTe detectors already developed as part of a JDEM Engineering Development Unit (EDU) Focal Plane Assembly (FPA). The two SpCs provide the faster survey speeds desirable for a BAO/RSD spectroscopic redshift survey mode and, because they are dispersed in opposing directions, they also provide redshift measurements unbiased by spatial offsets between the line and continuum emission regions, without requiring a later field revisit. The ImC covers the NIR and is optimized to provide good sensitivity down to 0.6 μm in the visible. The ImC provides a high quality point spread function, precision photometry, and stable observations for Exoplanet, SNe, Weak Lensing and NIR surveys, as well as the redshift zero reference for BAO/RSD. Quality pointing accuracy, knowledge, and stability are all required to resolve galaxy shapes and precisely revisit both the Exoplanet and SNe fields. Pointing to between 54° and 126° off the Sun enables the observation of Exoplanet fields for up to 72 continuous days during each of the twice yearly Galactic Bulge viewing seasons. Additionally, the observatory accommodates viewing within 20° of the ecliptic poles to monitor SNe fields in fixed inertial orientations that can be maintained for ~90 days.

The Exoplanet survey requires large light gathering power (effective area times field of view) for precise photometric observations of the Galactic Bulge to detect star/planet microlensing events. Seven fields are observed repeatedly to monitor the (very common) stellar light curves during microlensing events and the (relatively rare) planet lensing signals that may be superimposed on the stellar light curves. Exoplanet monitoring observations are performed in a wide filter spanning 0.97 – 2.0 μm, interspersed ~twice/day with brief observations in a bluer, 0.76 – 0.97 μm, filter for stellar type identification.

To accomplish the wide-field slitless spectroscopic redshift survey, the BAO/RSD measurement requires NIR spectroscopy to centroid Hα emission lines and NIR imaging to locate the position of the galaxy image. High dispersion spectroscopy enables centroiding the Hα emission lines to a precision consistent with meeting the redshift accuracy. To address completeness and confusion issues, prisms are used as the dispersing element and at least 3 roll angles, two of which are approximately opposed, are observed over ~96% of the mapped sky. The bandpass range of 1.1 – 2.0 μm provides the required redshift range for Hα emitters and the 0.45 arcsecond pixels in the SpC provide the area needed to meet the sky coverage requirements while maintaining centroiding accuracy.

The SN measurement also requires large light gathering power to perform the visible and NIR deep imaging and spectroscopy needed to classify and determine the redshift of large numbers of Type Ia SNe. Precise sampling (S/N of 15) of the light curve every five days meets the photometric accuracy requirement and the use of three NIR bands allows measurements of SNe in the range of $0.4 < z < 1.2$, providing better systematics at low z than can be achieved by the ground and extending the measurements beyond the $z > 0.8$ ground limit.

The WL measurement requires an imaging and photometric redshift (photo-z) survey of galaxies to mag AB ~23.7. A pixel scale of 0.18 arcseconds balances the need for a large field of view with the sampling needed to resolve galaxy shapes. Observations in two NIR filters, with ≥5 random dithers each are made to perform the required shape measurements to determine the shear due to lensing, while observations in an additional NIR filter are combined with color data from the shape bands and the ground to provide the required photo-z determinations. Either the ImC or SpC spectrometers with overlapping ground observations are used to perform the photo-z calibration survey (PZCS) needed to meet the WL redshift accuracy requirement.

The three instrument channels along with the auxiliary FGS and the telescope form the payload. The payload and spacecraft together make up the Observatory. The Observatory mass is 2500 kg, including margin. An Earth-Sun L2 libration point orbit has been selected to provide the passive cooling, thermal stability, minimum stray light, and large sky coverage needed to make these precise measurements. An Evolved Expendable Launch Vehicle (EELV) or Falcon 9, with parking orbit insertion and transfer trajectory insertion control to enable eclipse-free launch opportunities vir-tually any day, is used to place the observatory on a direct trajectory transfer orbit to L2. The mission life is 5 years with consumables sized to allow an extension for a total of 10 years. A key requirement is designing a low risk mission using components with flight heritage. Each of the elements of the mission is described below.

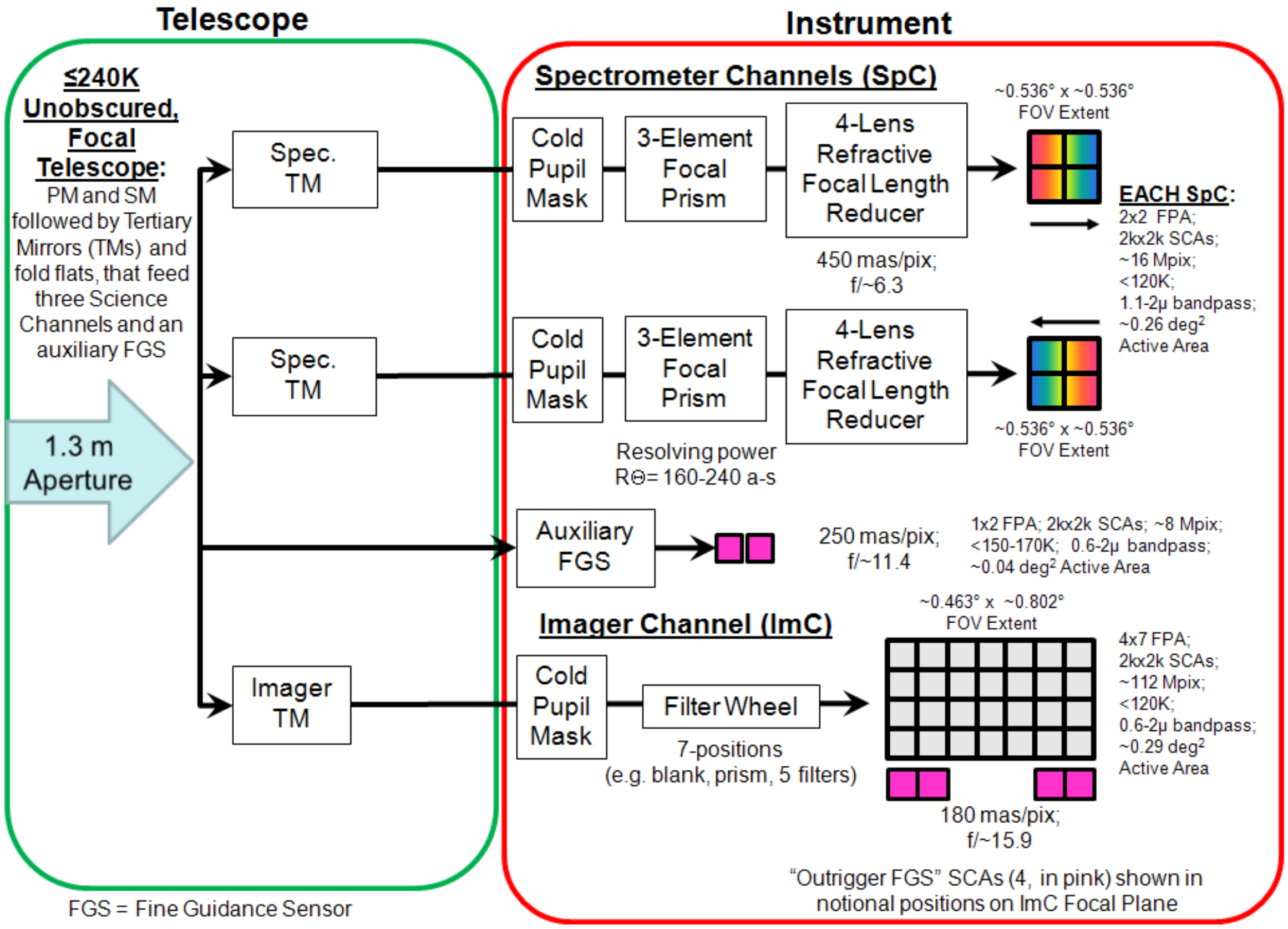

Figure 9: Payload Optical Block Diagram

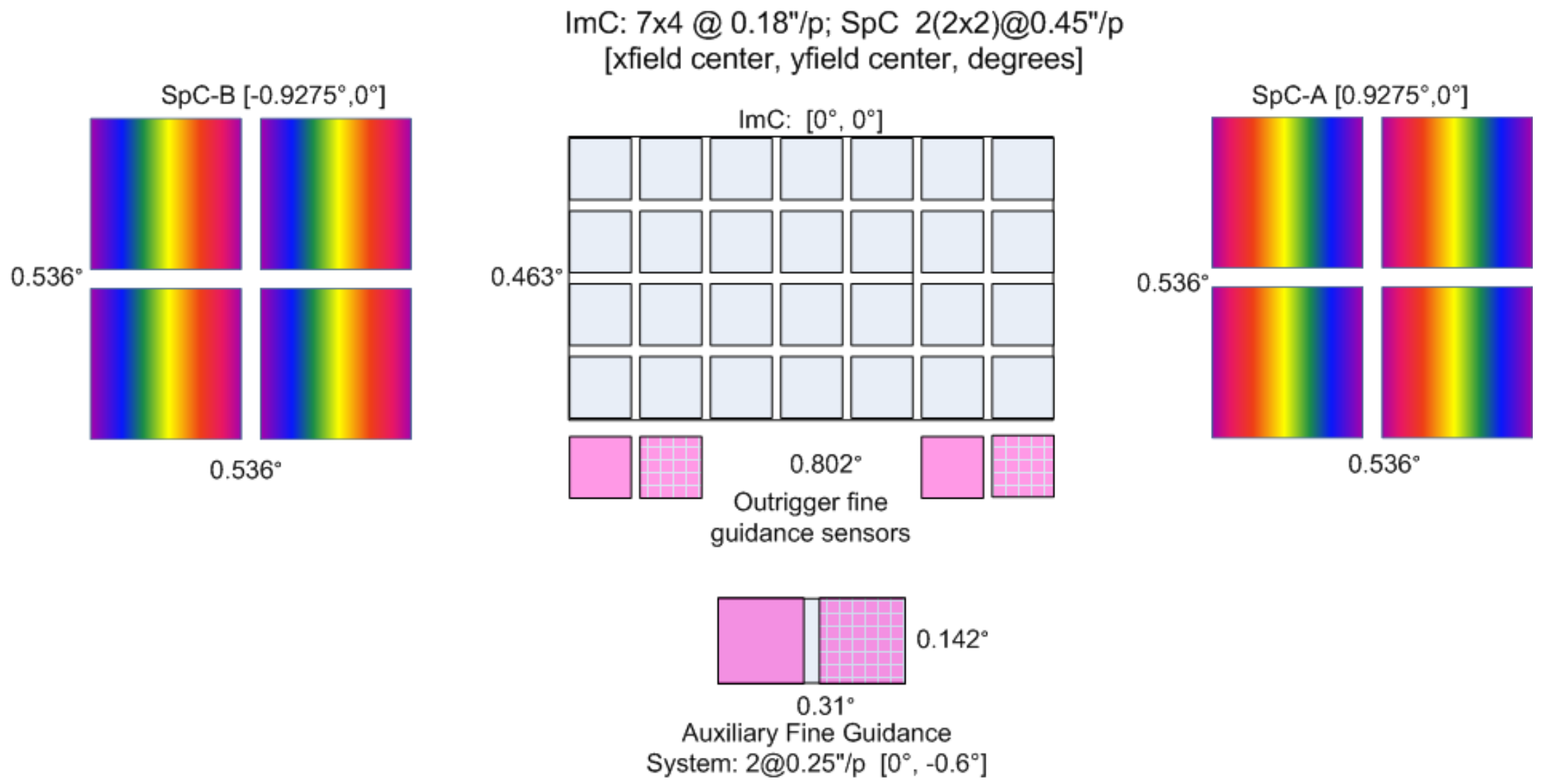

Figure 10: Channel Field Layout

## 4.2 Telescope

The science channels are fed by a Three Mirror Anastigmat (TMA) unobscured telescope, which offers a wide field along with a flat focal surface and good correction of low order aberrations. The design uses a focal TMA working at a pupil demagnification of 11.8 (110 mm pupil diameter with cold mask). A 1.3 meter diameter primary mirror feeds 3 separate tertiary mirrors for the ImC and two SpCs. The unobscured form was selected as the baseline for the IDRM, a change from the JDEM-Omega baseline. Shifting the secondary mirror off-axis eliminates the diffraction pattern created by the secondary mirror supports. It also eliminates the need for a large secondary mirror baffle allowing the primary mirror diameter to be reduced to 1.3m, from 1.5m on JDEM-Omega, while still providing equivalent or better throughput. These improvements provide a lower exposure time to the same limiting flux as compared to JDEM-Omega. The unobscured telescope allows for better alignment of the instrument fields improving the sky tiling schemes and improving the survey speeds. The total field of view extent for all three channels is 0.95 $deg^2$ (0.81 $deg^2$ active area).

The Optical Telescope Assembly (OTA) reflecting surfaces are maintained below 240K to limit the NIR thermal emissions to $\leq$ 10% of the minimum Zodiacal background. The instrument volume is maintained below ~180K to control thermal emissions. The three tertiary mirrors are included in the OTA, so the optical interface for each channel is at a real pupil. The three instrument channels are well separated allowing access for integration of each channel in any order.

The secondary mirror has a 6 degree of freedom mechanism to adjust focus and alignment. The telescope mirrors are made from ULE which allows a highly lightweighted, thermally stable mirror while the closed back design provides the required stiffness. The OTA structure is manufactured from low-moisture composites to minimize mass and thermal distortions while providing adequate stiffness. The telescope structure is insulated to minimize heat transfer from the solar array into the telescope and to minimize heat transfer to the instrument.

## 4.3 Instrument

The WFIRST instrument is divided into three channels, an imager channel and two oppositely-dispersed, but otherwise identical spectrometer channels. The key instrument parameters are shown in Table 6. The ImC is designed to a diffraction limit of 1$\mu$m as required for WL galaxy imaging and SN S/N requirements. The NIR SpCs are designed to a 3$\mu$m diffraction limit to achieve the required dispersion.

The ImC consists of a cold pupil mask, filter wheel, fold mirror, and the HgCdTe FPA. The ImC FPA uses 2k x 2k HgCdTe detectors, with 18$\mu$m pixels and a long-wavelength cutoff of 2.1 $\mu$m. The FPA is arranged in a 7x4 layout with a pixel scale of 0.18 arcseconds/pixel. A 7-position filter wheel provides 5 filters, a blank and an R-75 (2-pixel) dispersing element for executing the SNe program. A description of the filter wheel complement is shown in Table 7.

The two NIR SpCs have identical, refractive lens systems to change the focal length to provide the larger pixel scales required for the BAO/RSD spectroscopic survey. Each SpC consists of a 3 element focal prism of $CaF_2$ and S-TIH1 prisms, four lenses of ZnSe, $CaF_2$ and Infrasil and the HgCdTe FPA. The fixed prisms have a dispersion of R$\Theta$=160-240 (TBR) arcsec and are arranged to disperse in opposing directions on the sky to reduce source confusion without the need to revisit the field later in the mission. The SpCs use detectors identical to those in the ImC. The FPAs for each spectrometer are arranged in a 2x2 layout with a pixel scale of 0.45 arcsecs/pixel. Passive cooling is used to maintain the FPA temperatures for both the ImC and SpCs.

| Channel | Wavelength Range ($\mu$m) | Sky Coverage (active area; $deg^2$) | Pixel Scale (arcsec/pixel) | Resolving Power | FPA Temperature (K) |
|---|---|---|---|---|---|
| Imager | 0.6 – 2.0 | 0.29 | 0.18 | 75 (2-pixel; slit-less prism in filter wheel) | <120 |
| Spectrometer | 1.1 – 2.0 | 0.26 per SpC | 0.45 | 160-240 (R$\Theta$; slitless) | <120 |
| Auxiliary FGS | 0.6 – 2.0 | 0.04 | 0.25 | N/A | <150-170 |

Table 6: Key Instrument Parameters

| Position Name | Bandpass (μm) | Techniques |
|---|---|---|
| Filter, F087 | 0.76 – 0.97 | SN, Exoplanet |
| Filter, F111 | 0.97 – 1.24 | SN, WL photo-z's |
| Filter, F141 | 1.24 – 1.57 | SN, WL shapes |
| Filter, F178 | 1.57 – 2.00 | SN, WL shapes |
| Filter, W149 | 0.97 – 2.00 | Exoplanet |
| Prism, P130 | 0.60 – 2.00 | SN, WL PZCS |
| Blank | N/A | Calibration |

Table 7: WFIRST Filter Positions

The IDRM instrument design incorporates minor modifications from the JDEM-Omega instrument. Two detectors from each of the SpCs have been moved to enlarge the ImC to a 7x4 array, from 6x4. To compensate for the reduced number of SpC detectors, the pixel scale in each SpC is increased to 0.45 arcsecs/pixel, from 0.37 arcsecs/pixel, to provide an equivalent field size for the SpCs. The additional four detectors in the ImC increase the imaging field size by ~17%, benefitting the Exoplanet, NIR, SNe, and WL surveys. Each SpC design is simplified from a collimator, disperser, and camera set to a design with prisms in the converging telescope beam, followed by focal length reducing lenses. This is a more compact and robust design form. Additionally, the bluest filter in the ImC filter wheel is replaced with a broad filter (0.97 μm to 2.0 μm) for the Exoplanet survey. Finally, JDEM-Omega proposed using 2.5μm long-wavelength cutoff detectors but also recognized that a bluer detector cutoff could simplify the thermal and optical system design of the instrument. The scientific requirements do not extend longwards of 2.0 μm, so detectors with sensitivity to 2.5 μm are not required. Reducing the long-wavelength detector cutoff simplifies the design as the detector operating temperatures can be at least 20K warmer for equivalent dark current, reducing the size of the cryogenic radiator. Instead of creating a 2.0 μm cutoff instrument using a 2.5 μm cutoff detector, WFIRST has demonstrated detectors with an intrinsic cutoff of 2.1 μm and these detectors are the baseline for the WFIRST IDRM.

The three instrument channels are kinematically mounted to and thermally isolated from the telescope structure. The instrument optics are maintained below 180K via radiation off of each of the instrument housings. The detectors on the instrument focal planes are cooled to below 120K via dedicated radiators. The Sensor Cold Electronics (SCEs) that power and readout the SCAs are mounted close to, but thermally isolated from them, as they can be held at a warmer temperature. The SCEs are cooled to ~150K via direct radiation off of their mounting plate. Panels on the side of the solar array are used as sunshields to protect the payload structure from solar illumination during the ~90 day inertially fixed SNe observations.

The use of CMOS-multiplexer (readout integrated circuit) based hybrids with non-destructive readouts, supports noise reduction and permits electronic shuttering, eliminating the need for a shutter mechanism. Sample Up The Ramp (SUTR) processing is used during all observations with the raw frames that are generated every ~1.3 sec being combined during the course of the integration period to produce one output image (Offenberg et al., 2005). All detectors in the three instrument channels are identical, simplifying detector production and sparing. The baseline design is 2K x 2K HgCdTe detectors with a 2.1 μm long-wavelength cutoff and 18 μm pixels operating at <120K.

Though not formally a part of the scientific instrumentation, there are additional imaging detectors used for fine guidance. During normal imaging operation, two of the four "outrigger" detectors located on the ImC FPA are used for guiding (the other pair provide redundancy). When the ImC R-75 disperser is inserted by the filter wheel, these outriggers will no longer see undispersed stellar images, so a separate auxiliary Fine Guidance Sensor (FGS) with 2 SCAs provides this function.

### 4.4 Calibration System

Past experience with space imaging and spectroscopic missions leads to the conclusion that WFIRST has stringent calibration requirements in a number of areas. The general WFIRST strategy is to use ground calibration methods to the maximum extent possible, reserving on-orbit calibration to verification of the ground results and extending the calibrations where ground calibration may not be effective. To maintain the calibration requirements over the entire mission, not only are the calibrations important, but so are measurements of calibration stability. The latter will determine the need for and frequency of on-orbit calibrations. The WFIRST calibration program will place strong emphasis not only on the areas requiring calibration, but also on the verification of these calibrations, either on the ground or in orbit, using multiple techniques as cross-checks. The SNe and microlensing fields are observed repeatedly over the lifetime of the mission, providing excellent opportunities to develop and use sky calibration standards.

All optical and detector components will be calibrated at the component, subsystem, and instrument levels. These data will be used to feed an integrated instrument calibration model that will be verified using an end-to-end payload-level thermal vacuum test. This test will involve a full-aperture (1.3 meter) diameter collimated beam that will test for optical wavefront error as well as photometry.

The exoplanet program imposes some constraints on photometric calibration, but also provides a unique opportunity to meet these calibration requirements. Proper measurements of the stellar and planetary light curves require stable relative calibration to 0.1% (relative to nearby stars in the same detector) over the course of the event. These observations use mainly a single filter. Over the course of the mission, the fields are sampled many 10's of thousands of times with a random dither. Most of these observations will be of stars without microlensing events. If the star is not a variable star, then the relative calibration will be monitored during the extensive number of observations to establish stability. Slow variations across the field-of-view and over time can thus be monitored and corrected for. The absolute calibration and occasional color measurements using the bluer color filter have less stringent calibration requirements (1%) that will be met by the more stringent calibration requirements for Dark Energy that are described below.

The three Dark Energy observational methods have different calibration demands on instrument parameters and their accuracy. The SN Survey places the most stringent demands on absolute and inter-band photometric calibration. White Dwarfs and other suitable sky calibration targets will be used to calibrate the absolute flux as well as linearity of the imager over several orders of magnitude. This linearity will be tested on the ground, and verified with an on-orbit relative flux calibration system (if necessary). Observations of astronomical flux standards will be extended across the detector by means of the "self-calibration' techniques described by Fixsen, Mosely, and Arendt (2000) and employed for calibration of Spitzer/IRAC (Arendt et al. 2010). The intra-pixel response function (quantum efficiency variations within a pixel) will be fully characterized by ground testing.

For the WL Survey, the requirement for galaxy ellipticity accuracy places significant demands on both the optical and detector subsystems. The uniformity and stability of the point spread function (PSF) needs to be strictly controlled and monitored to ensure a successful mission. This drives the need to characterize the detector intra-pixel response and the inter-pixel response (capacitive cross-coupling with nearest neighbors) for both magnitude as well as spatial and temporal variations. It is likely that the combined PSF effects, including spacecraft jitter, will have some variability on time scales of a single exposure. These residual effects will be continuously monitored with the observatory attitude control system and field stars and downlinked to provide ancillary information for the scientific data analysis pipeline.

The BAO survey relies primarily on the spectrometer channels, which are not driving the calibration requirements for the mission. Established calibration techniques used for other space missions should be adequate to meet the relatively loose photometric and wavelength calibration requirements. The larger plate scale in the spectrometers may demand some attention to the spatial effects such as intra-pixel response, but not to the degree required by the WL Survey.

### 4.5 Fine Guidance Sensor

The fine guidance sensor (FGS) is used to meet the fine pointing requirements needed for the Exoplanet, WL, and SN surveys. The absolute pointing accuracy requirement of <25 milli-arcseconds is driven by SN requirements for returning precisely to previously observed objects. WL drives the pointing stability of <40 milli-arcseconds over an exposure. The primary Outrigger FGS consists of two pair of HgCdTe detectors, a prime and redundant, located on the ImC FPA and fed through the ImC optical train, including the filter wheel. This guider is used in all observations that include imaging. An additional pair of FGS detectors, the Auxiliary FGS, is fed from a separate field at the telescope intermediate focus via an additional compact reflective relay. The Auxiliary FGS, along with the star trackers, provides pointing control for science observations when the ImC is performing spectroscopy.

### 4.6 Spacecraft

The WFIRST spacecraft has been designed to provide all the resources necessary to support a payload at L2 using mature and proven technology design. The design is based on the Solar Dynamics Observatory (SDO) spacecraft, which was designed, manufactured, tested, and qualified at GSFC. The spacecraft bus design provides cross strapping and/or redundancy for a single-fault tolerant design. *Structures:* The spacecraft bus is an aluminum hexagonal structure, consisting of two modules (bus module and propulsion module) that house the spacecraft and payload electronics boxes and the propulsion tank. The spacecraft bus provides the interfaces to the payload and the launch vehicle. It supports a multi-panel fixed solar array and a sunshield to prevent the sun from illuminating payload hardware during science observations. The structural resonances in the spacecraft and payload are tuned to frequencies that do not overlap with reaction wheel resonances to minimize jitter in the instrument. *Attitude Control:* The spacecraft is three-axis stabilized and uses data from the payload fine guidance sensor, inertial reference unit, and star trackers to meet the coarse pointing control of 3 arcsec, and the fine relative pointing control of 25 mas pitch/yaw and 1 arcsec roll (all values RMS per axis). There are 2 sets of fine guidance sensors. The first is a pair of redundant sensors (4 total) on the imager focal plane and the second is a pair of auxiliary guiders (2 total) for guiding during SNe spectroscopy mode. The star trackers are used for coarsely pointing to within 3 arcsec RMS per axis of a target. After that, the FGS takes over to meet the fine pointing requirements for revisits and relative offsets. A set of 4 reaction wheels is used for slewing as well as momentum storage. The wheels are passively isolated to allow stable pointing at frequencies higher than the FGS control band. The GNC subsystem provides a safe hold capability (using coarse sun sensors), which keeps the observatory thermally-safe, power-positive and protects the instrument from direct sunlight. *Propulsion:* A hydrazine mono-prop subsystem is required for orbit insertion, orbit maintenance, and momentum dumping from the reaction wheels throughout the duration of the mission. The prop subsystem does not have any unique features for WFIRST. This subsystem is well within the requirements of other propulsion systems that have launched. *Electrical Power:* Three fixed, body-mounted solar array panels provide the observatory power. Gallium Arsenide solar array cells operate with 28% efficiency and provide 2500 watts of output for an average orbit usage of ~1400 W. The remainder of the power subsystem is comprised of an 80 A-hr battery and power supply electronics that control the distribution of power and provide unregulated 28 Vdc power to the payload. The solar array is currently sized to provide full observatory power at EOL with 2 strings failed at the worst case observing angles. *Communications:* The communications subsystem uses S-band transponders to receive ground commands and to send real-time housekeeping telemetry to the ground via 2 omni-directional antennas as well as for ranging. A Ka-band transmitter with a gimbaled antenna will downlink stored science and housekeeping data at a rate of 150 Mbps without interrupting science operations. *Command & Data Handling:* The command and data handling subsystem includes a 1.1 Tb solid state recorder (SSR) sized to prevent data loss from a missed contact. The daily data volume is estimated at 0.9 Tb per day, assuming 2:1 lossless compression. The data will be downlinked twice daily to the NASA Deep Space Network (DSN). The C&DH/FSW provides fault management for the spacecraft health and safety as well as being able to safe the payload when necessary. *Thermal:* The spacecraft thermal design is a passive system, using surface coatings, heaters and radiators.

### 4.7 Ground System

The WFIRST Mission Operations Ground System is comprised of three main elements: 1) the facilities used for space/ground communications and orbit determination, 2) the Mission Operations Center (MOC) and 3) the facilities for science and instrument operations and ground data processing, archiving, and science observation planning. For each element, existing facilities and infrastructure will be leveraged to provide the maximum possible cost savings and operational efficiencies. The functions to be performed by the ground system and the associated terminology are shown in Figure 11.

The DSN is used for spacecraft tracking, commanding and data receipt. It interfaces with the MOC for all commanding and telemetry. Tracking data is sent to the GSFC Flight Dynamics Facility.

The MOC performs spacecraft, telescope and instrument health & safety monitoring, real-time and stored command load generation, spacecraft subsystem trending & analysis, spacecraft anomaly resolution, safemode recovery, level 0 data processing, and transmission of science and engineering data to the science and instrument facilities. The MOC performs Mission-level Planning and Scheduling.

The science and instrument facilities maintain instrument test beds, perform instrument & telescope calibrations, assist in the resolution of instrument anomalies, perform instrument flight software maintenance, and generate instrument command loads.

These facilities are also responsible for science planning & scheduling, supporting mission planning ac-tivities carried out by the MOC, running the Guest Investigator (GI) Program, providing Science Team and GI support, and performing EPO activities for the public and the astronomical community. Data handling involves ingesting Level 0 science and engineering data from the MOC and performing Level 1-3 data processing for the Science Teams and GI community and transmitting these calibrated data to the data archive. All data will be archived. Data search and access tools are provided to the science community that enable efficient searches and delivery of archival data that ensure interoperability with NASA data archives and the Virtual Astronomical Observatory.

Approximately 5 dedicated Science Teams will be funded over a 5-year period to execute the primary dark energy, exoplanet, and NIR survey science programs. In this period, the GI program provides funding for ancillary science. Operations costs and grants for the GI in the primary mission are fully included in the lifecycle costs.

## 4.8 ENABLING TECHNOLOGY

The HgCdTe near-infrared detectors baselined for WFIRST have extensive heritage from HST, JWST, and on-going production of commercial off-the-shelf devices for ground-based observatories. WFIRST uses a 7x4 mosaic of these detectors while the largest near-infrared focal plane built for space so far is a 2x2 array of detectors. Although scaling of the focal plane from a 2x2 array to a 7x4 array is not technology development, it is considered an engineering challenge so an EDU was started on the JDEM project and work continues under WFIRST. The EDU FPA includes a 6x3 silicon carbide (SiC) mosaic plate incorporating the 2.1 μm detectors, the Sensor Cold Electronics (SCE), including a repackaged version of the SIDECAR ASIC and associated mounting hardware. The full mosaic plate will be populated with flight-like detectors and qualification testing of the assembly will be completed prior to Phase A, mitigating the technical risk of scaling of a large NIR mosaic.

Anticipating the need to demonstrate acceptable detector performance (in order to meet the scientific requirements) and yield (at an acceptable cost), the JDEM/WFIRST project initiated a detector demonstration program over 3 years ago. To date, two lots have been built, a more experimental one under the Detector Technology Advancement Program (DTAP), and a more WFIRST-specific lot under the FPA EDU detector build. A third lot is currently in process to demonstrate enhanced robustness and margin. These lots have provided test data showing improved detector performance over previous designs and these results are being used to guide mission design and optimization.

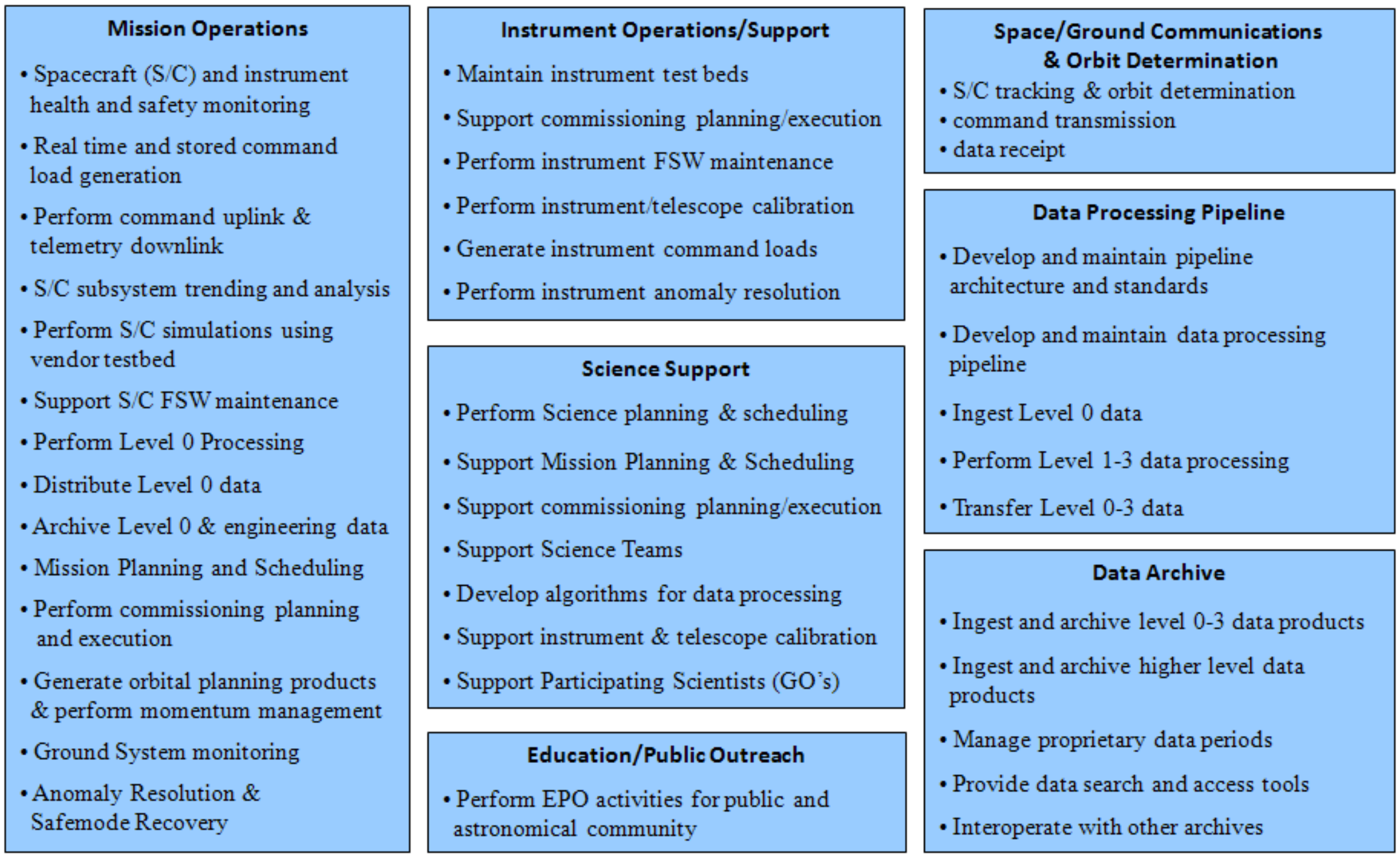

Figure 11: WFIRST Ground System functions and associated terminology

It was recently discovered that some HgCdTe detectors are showing degradation after long term storage. These are manufactured using similar technology to that for WFIRST. This degradation leads to a slow increase in dark current and the number of “hot pixels” during ground processing and testing. The root cause for this degradation has been identified and design and process change features have been identified. The current lot of demonstration parts that the JDEM/WFIRST project is building already includes these design changes. The WFIRST detectors will be tested this summer to validate the design change. There is a minimum of 2.5 years available for WFIRST to complete the performance and qualification testing to validate the design before the procurement process for the WFIRST flight detectors will start, assuming a Phase A start in FY13.

The JDEM Omega RFI identified weak lensing as a goal and the Decadal Survey EOS panel report identified risks associated with weak lensing. Both the SDT and the Project are pursuing the Decadal Survey's recommendation to address challenges related to weak lensing early to mitigate the chances of significant cost growth. Among the efforts underway are ensuring that the telescope design is compatible with controlling systematic effects to the level necessary for weak lensing and end to end simulations of galaxy ellipticity measurements with hardware in the loop. The work leading up to this interim report has affirmed that the pursuit of weak lensing should continue in the upcoming mission development efforts.

# 5 OPERATIONS CONCEPT

The primary focus in designing WFIRST science operations is to optimally acquire the datasets for the four main surveys (the exoplanet microlensing survey, and Dark Energy measurements based on surveys of Supernova SNe-Ia, BAO/RSD galaxy redshifts, and WL galaxy shapes), while considering ways to enhance the general NIR Survey dataset on a non-driving basis. Each survey has its own unique operations concept, but in the case of the WL galaxy shape survey, we are able to simultaneously acquire a deeper BAO/RSD galaxy redshift survey, albeit at a slower than optimal sky coverage rate.

Exoplanet microlensing observations require monitoring of high density star fields in the Galactic Bulge for periods of ≥ 60 days in order to detect these relatively rare but readily observed events. These observations will be carried out in 7 campaigns of ~72 days duration during the twice yearly periods when the Bulge is in the Field of Regard (FOR) of the Observatory, and will provide an exoplanet detection yield rate as specified in Figure 1, Box 3, and Table 2. These campaigns will span the bulk of the planned mission lifetime in order to provide long temporal baseline data for stellar typing through relative proper motion studies of the source and lens stars. Seven fields will be scanned in a continuous pattern at a revisit cadence of 15 minutes. Integrations of 88s with filter W149 will be used to routinely sample microlensing event light curves, but ~every 12 hours a revisit cadence will be executed with filter F087 in place to provide star color information. Revisits to the same field during each campaign will be at the same pointing, to within a random dithering of ~1 pixel rms. All the exposures are downlinked and stored to provide extensive field monitoring data covering pre-, post-, and on-going microlensing signals, as well as to provide relative photometric calibration data at the field star locations.

SNe observations require regular monitoring of two small fields within 20° of an ecliptic pole over an extended period of time (~2 years), and provide a SNe-Ia yield rate as a function of dedicated SNe time as specified in Figure 1. As an example, for a dedicated SNe time allocation of 6 months, the observatory repeatedly (for ~30 hours every 5 days for ~2 years) monitors ~square fields of ~5.8 deg$^2$ for mid z (to z=0.8; Tier 1) SNe-Ia and ~1.4 deg$^2$ for high z (to z=1.2; Tier 2) SNe-Ia. Monitoring consists of observations with filters F111, F141, and F178 (300s @ for Tier 1 and 1100s @ for Tier 2) and with R-75 prism P130 (1300s for Tier 1, and 5300s for Tier 2). Monitoring revisits to each SNe field to within 15 (TBD) milliarcsecs to provide 4 or 9 (TBD) precise sub-pixel dithers are accomplished via the S/C attitude control system (ACS) using the Aux FGS and star trackers when the R75 prism is used, and the ImC FGS when any filter is used. To enable spectral monitoring of the SNe and host galaxy in the same orientation, the observatory roll angle is inertially fixed for ~90 day periods, then rotated ~90 degrees to maintain field monitoring coverage while keeping the sun within 45 degrees of the maximum-power roll angle.

The WL and BAO-RSD surveys require mapping large sky areas as rapidly as exposure times permit, with two mapping modes being planned. A combined WL/BAO-RSD mode, also called the DEEP survey, meets requirements for WL shape and photo-z ImC measurements and simultaneously provides a deep BAO/RSD SpC survey. This mode optimally maps the ImC FOV across the sky in 3 NIR filters (F111, F141, F178) with no gaps in the ImC field at a rate of ~2,700 deg$^2$/yr (~7$^+$ deg$^2$/day). Three of these smooth-filled passes on a target field are required, each with a different filter and with an ~5° roll angle difference (as per BAO-RSD roll requirements). The two passes performed in the two reddest filters are for WL shape/color measurements, and deliver 5 randomly dithered observations of 160 seconds over the full observed field. The third pass is performed with F111 to acquire WL color data only, and only four 160 second exposures over the full observed field are required. A faster WIDE survey mode meets BAO-RSD requirements and also provides photo-z colors for ground based WL programs. This mode optimally rough-fills the SpC FOV across the sky (simultaneously mapping the ImC across the sky in two filters) at a rate of ~11,000 deg$^2$/yr (~30$^+$ deg$^2$/day). Two same-filter passes over a target field are completed with a slight offset in the passes to rough-fill the SCA gaps in the SpC focal planes. A second set of two passes are completed with a second filter at a slight roll angle (~5°) relative to the first pair of passes to limit source confusion. The area covered during the WL and/or BAO/RSD surveys is broken into a series of inertially fixed observations, each comprised of a programmed sequence of small slews/settles. The chosen inertially-fixed fields are ultimately stitched together into a contiguous map, and are selected within the FOR according to a schedule that accounts for availability, Zodiacal brightness, the Galactic plane, and other geometric/thermal constraints, and possibly operational (momentum dumping) considerations. Given the L2 vantage point and the 5 year mission life, the scheduling constraints to achieve the WL and BAO/RSD sky cov-erage are not particularly challenging, but care must be taken to ensure that appropriate observable fields are available at all times.

Combining all of the above sky coverage, monitoring, and cadence requirements with stray light and solar array exposure considerations, the observatory is designed to have a pitch FOR between 54 and 126 degrees off the sun with no azimuthal (yaw) constraint about the sun line. Roll about the observatory line of sight is constrained to ±10° during all observations with the exception of SNe observations, where the requirement is to be able to remain inertially fixed for 90 days while pointed within 20° of either ecliptic pole (max power to occur at the middle of any 90 day period). The WFIRST design provides the flexibility to support different observing strategies, and Figure 1 gives the sky coverage and exoplanet/SNe-Ia detection rates per unit of dedicated observing time. The observing time can be divided in many different ways yet still provide large sky surveys, large numbers of SNe-Ia characterizations, and large numbers of exoplanet detections. Figure 5 illustrates one way in which Exoplanet and SNe ops might be accommodated, then blocks out the large pieces of observing time that remain available for HLS, GI, and Galactic Plane observations (as well as the calibration/training datasets to be defined during the remainder of the SDT study). This flexibility is a key strength of WFIRST, as this ability to accommodate a wide range/order of NIR imaging and spectroscopic surveys is critical to the success of any Facility Class mission.

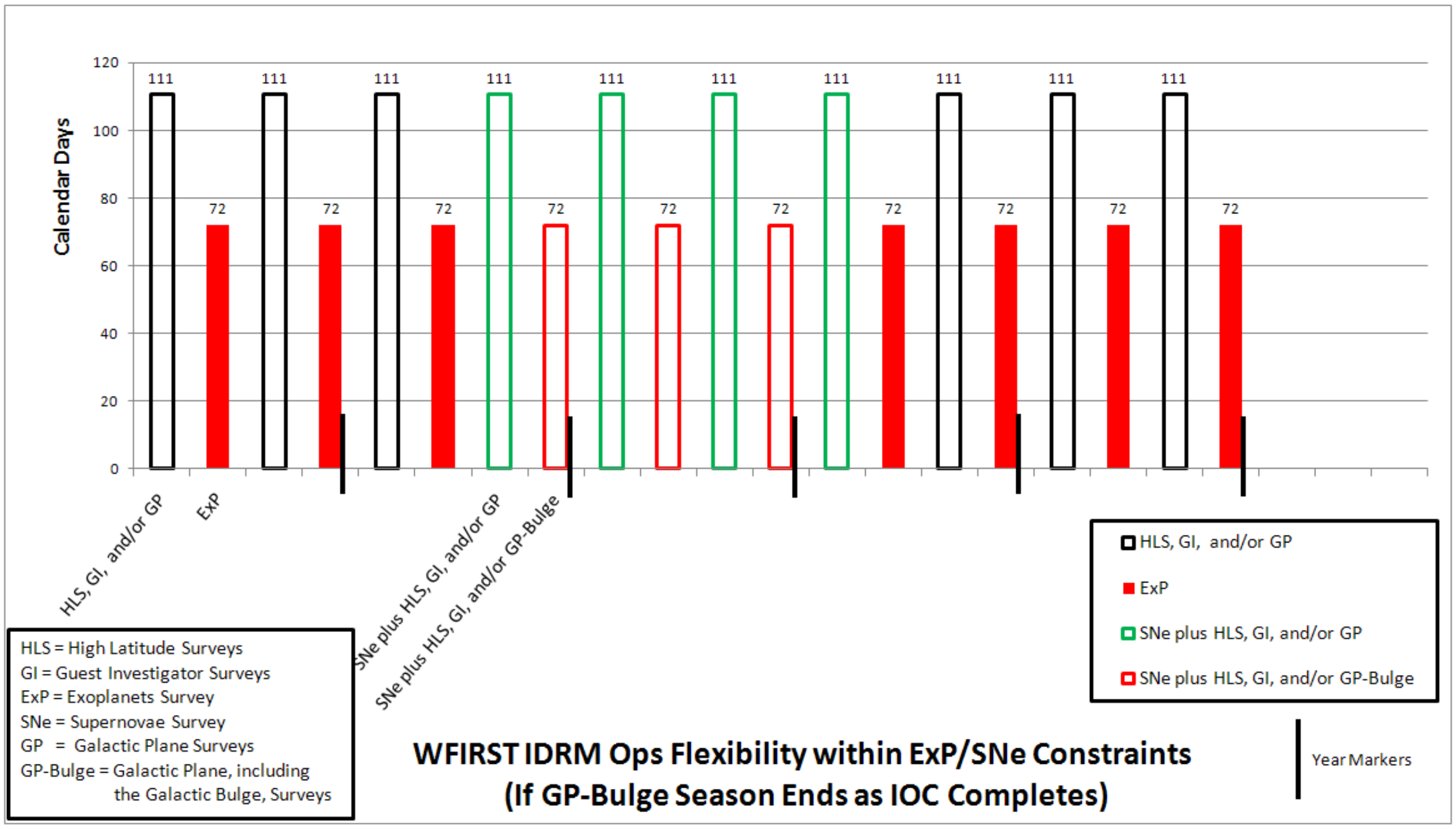

Figure 12: WFIRST exhibits excellent observing mode flexibility in sample ops concept meeting ExP and SNe field monitoring requirements.

## 6 COST & SCHEDULE

The current WFIRST DRM has relatively few changes from the JDEM-Omega reference mission, and thus future independent cost estimates should be very comparable to the NWNH Astro2010 independent cost estimate. The WFIRST mission was estimated at $1.61B by the NWNH independent cost team. These WFIRST optimizations or changes are believed to be cost neutral or provide minor cost savings. Over the summer, the Project will develop a WFIRST LCCE using multiple estimating techniques. Project Management, Systems Engineering, Mission Assurance, Integration & Test, and Public Outreach will be estimated using a grassroots approach and validated against analogous missions. Pre- and Post-launch Science will be projected using guidance from NASA HQ on the expected size of the WFIRST Science Announcement of Opportunity (AO). This estimate will include support for the AO-selected science teams in the mission development phase as well as the 5 years of operations. Additional funds for the Guest Investigator (GI) Program will be included in the Science estimate. The payload and spacecraft will be estimated primarily using parametric estimates. These estimates will be constructed using master equipment lists (MELs) and historical cost databases and are adjusted for mass and complexity factors. Additionally, NASA HQ has requested that an Independent Cost and Schedule Estimate (ICE/ISE) be performed by the Aerospace Corporation. The ICE/ISE is expected to be completed in September 2011.

The development phase of WFIRST spans 82 months, from preliminary design through launch (phase B/C/D). This development phase is preceded by 9 months to fully develop the baseline mission concept (Phase A) and several years of concept studies (Pre-Phase A). The observing phase (Phase E) of the mission is planned for five years. The development schedule is shown in Figure 13. The estimate is at a 70% Joint Confidence Level (JCL) meaning it includes margins for cost and schedule risk. The 70% JCL schedule allocates ten months of funded reserve to the WFIRST development.

The build-up and integration philosophy of the WFIRST observatory is based on the well established practice of building small assemblies of hardware, thoroughly testing them under appropriate flight environments, and then moving on to the next higher level of integration with those assemblies. The WFIRST telescope and instrument will be developed and individually qualified to meet mission environments. The critical path of the mission is through the development of the instrument. The instrument is qualified prior to integration with the telescope. Following ambient check-out, the entire payload is tested at temperature and vacuum at GSFC to verify the end-to-end optical performance. The spacecraft is then integrated to the payload, and checked at ambient. Following a successful ambient checkout, a complete observatory environmental test phase is performed, including a repeat of the optical test with the fully integrated observatory. Upon successful completion of the observatory environmental test program, the observatory is readied for shipment to the KSC, where the launch campaign is conducted.

The parallel build-up of all of the mission elements, allows substantial integration activity to occur simultaneously, increasing the likelihood of schedule success. Because all of the major elements of the observatory (telescope, instrument, and spacecraft) are located at GSFC two years before the planned launch, there is considerable flexibility in optimizing the schedule to compensate for variation in flight element delivery dates. Over the two year observatory I&T period, the Project will have flexibility to reorder the I&T work flow to take advantage of earlier deliveries or to accommodate later ones. Should instrument or telescope schedule challenges arise there are options to mitigate the schedule impacts by reallocating the payload level environmental test period. Should instrument or spacecraft challenges arise, there are options to modify the workflow and pull other tasks forward to minimize risk and maintain schedule. The WFIRST schedule is very achievable. With two years between the delivery of the WFIRST instrument and launch, and given the additional flexibility that is inherent in the I&T flow, the WFIRST observatory I&T program has a high probability of executing within budget and schedule. In summary, the proposed WFIRST observatory I&T flow proposed is very achievable, given the planned schedule reserve and the opportunities available for workaround.

Fifty-one months are allocated to complete the WFIRST instrument, from the start of preliminary design, thru the delivery of the instrument, not including reserve. Additionally, the schedule includes ten months of funded reserve, further increasing the likelihood of executing the plan. An engineering development unit FPA has already been initiated and flight-like detectors have already been delivered. Integration of those detectors into the silicon carbide mosaic plate will occur later this summer and integrated testing will be performed. Thus even before Phase A commences, a large mosaic NIR focal plane will have been demon-strated mitigating the risk of the element that this element would drive the mission critical path.

Early interface testing between the Observatory and ground system is performed to verify performance and mitigate risks to schedule success. Prior to payload integration, interface testing between the spacecraft and the ground system is performed. Immediately following payload integration to the spacecraft, end-to-end tests are performed, including the payload elements. These tests are performed numerous times prior to launch to ensure compatibility of all interfaces and readiness of the complete WFIRST mission team.

Given that WFIRST requires no new technologies, can be built today, has an implementation strategy that is conservative, proven and amenable to workarounds, and has a schedule based on continuously retiring risk at the earliest possible opportunity, the WFIRST mission is executable within the cost and schedule constraints identified in this report, consistent with the New Worlds, New Horizons finding that WFIRST "…presents relatively low technical and cost risk making its completion feasible within the decade…".

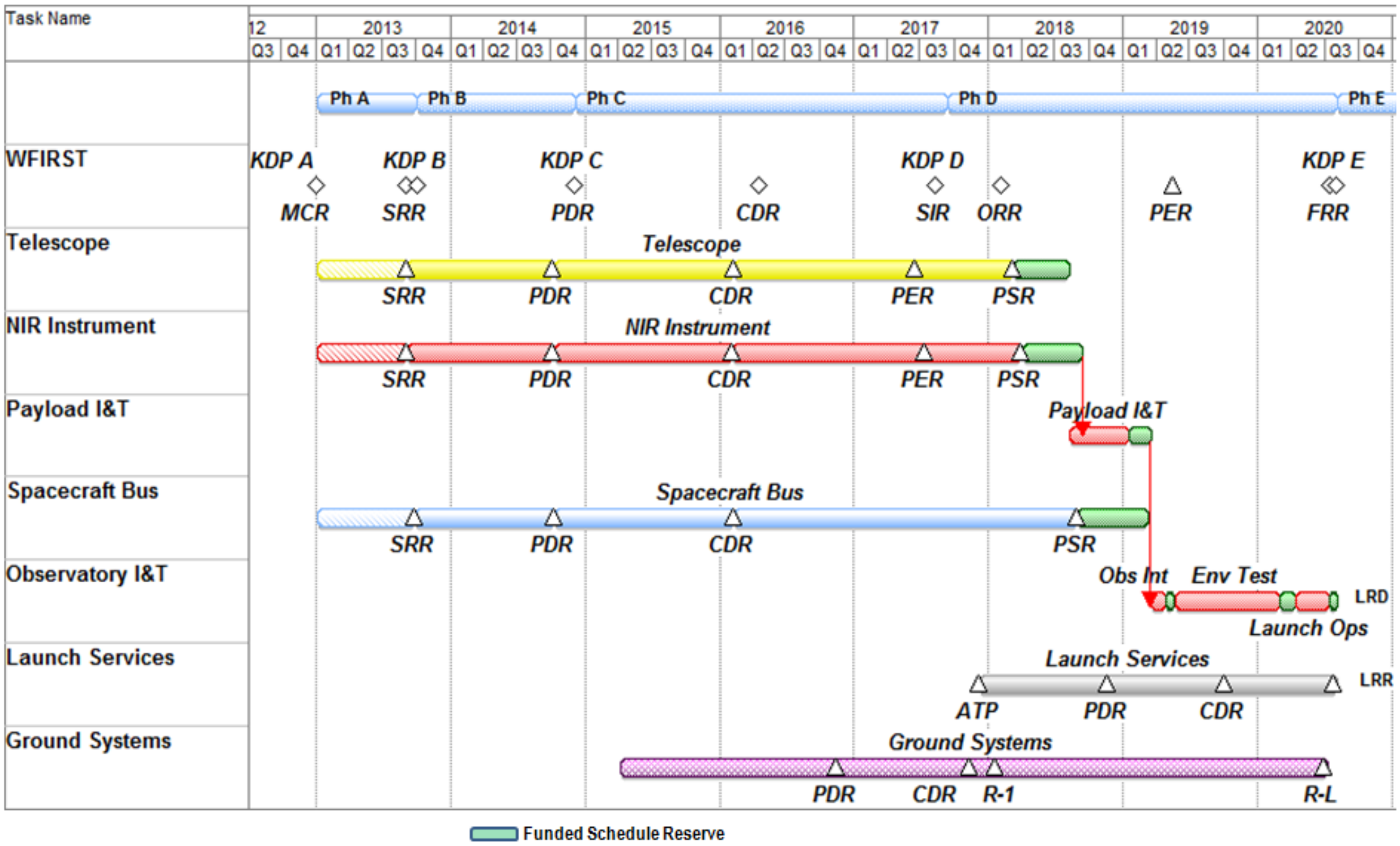

Figure 13: Mission Schedule

# 7 CONCLUSION AND PATH FORWARD

This report describes the work of the WFIRST Science Definition Team and Project Office on defining the WFIRST mission. The SDT has found that the mission objectives given in the NWNH Decadal Survey report are achievable and that the science remains highly compelling. The top-level science requirements have been refined and lower level requirements derived. Figures of Merit have been specified for the dark energy science, based on previous work of the Dark Energy Task Force and the Figure of Merit Science Working Group. New FoMs have been derived for exoplanet microlensing and for sky surveys. The purpose of these FoM's is to provide quantitative means for comparing performance of various mission configurations.

A significant effort by the Project Office and SDT has gone into refining the design concept for the observatory. This report contains the latest Interim Design Reference Mission. The design is similar to that specified in the NWNH report, with a few improvements to boost performance and lower cost. The most important of these is to change from an on-axis "obscured" telescope to an off-axis unobscured concept. The imaging performance is superior for the off-axis design and there is adequate flight heritage to baseline such an approach.

In Europe, the Euclid dark energy mission is in Phase A for the M2 competition. It has overlap with WFIRST in some of its science objectives, but is a significantly different design with different performance characteristics. The WFIRST SDT has compared the missions and found that they are synergistic. Several science objectives of WFIRST, including exoplanet microlensing and supernova dark energy measurement, are not part of the baseline Euclid mission. Also, other areas of dark energy measurement, including weak lensing and baryon acoustic oscillations, are addressed in different and complementary ways in the two missions. WFIRST has more focus on near IR observations and Euclid on visible observations. With this in mind, the SDT has concluded that the WFIRST design should not change, even if Euclid is selected and is launched ahead of WFIRST. However, there would likely be changes made to the observing program of WFIRST to more complement that of Euclid. Should NASA and ESA decide to pursue a joint mission or program, all of the capabilities currently included in WFIRST must be included in the joint effort.

## Planned Studies

The first iteration of an engineering design cycle for the WFIRST mission is planned for the coming year. Some of the more noteworthy aspects of this study are described briefly in this section. These address open trades, liens on the present design, and preparations for the Phase A study. Some of the items listed here are in fact normally addressed in Phase A. However, WFIRST is in a more mature state than is common in pre-Phase A, as a consequence of prior mission concept studies. Hence, we are taking advantage of the opportunity to begin this work now in order to reduce schedule risk downstream.

**Attitude Control** The frequent slews needed for a large-area sky survey, coupled with fine pointing control and the multiple fields of regard, require a capable ACS. An integrated model of the observatory structure and the attitude control system was begun for the JDEM Omega concept, and is now being modified for WFIRST. This includes a finite-element model of the structure and solar arrays, models of the ACS sensors and actuators, and a model of fuel slosh. This serves as a test-bed for development of ACS control laws, performing trade studies of sensor and actuator performance requirements, and providing more accurate input to operations concept development on issues such as settle time following slews.

**Vibration Isolation** The structure must be designed so that vibration modes at frequencies above the ACS control bandwidth do not degrade image quality. At L2, the primary disturbance is vibration of the reaction wheels. The preliminary study of the Omega telescope indicated that the following implementation strategies are likely to be adequate to meet the stability requirements needed for weak lensing galaxy shape measurements, which impose the most stringent stability requirements of all of the required science measurements:

- Passive mechanical isolation of the spacecraft reaction wheels
- Reaction wheel speed limited to <= 45 Hz.
- Possible addition of passive mechanical isolation of the spacecraft from the telescope.
- Resonances in the payload structure detuned from those of the reaction wheels to prevent large jitter motions of the optical metering structure.

This study will be repeated in greater detail for the WFIRST S/C bus and optical metering structure. An important component of managing the science impacts of vibration present during an exposure is the downlinking data from the attitude control system to aid the science data analysis via pointing history reconstructions.

Thermal Design The thermal environment of the metering structure must be controlled so as to maintain the positions of the optical elements to within their budgeted tolerances. The JDEM-Omega studies found that it was sufficient to provide thermal control of the mechanical structure and mirrors to +/-1C. This is less stringent than the stability achieved on Chandra (~0.2 degrees C), indicating that we should have significant design margin given the thermal environment similarities between Chandra's highly elliptical orbit and WFIRST's Earth-Sun L2 orbit. The thermal model will be updated to the WFIRST design, and ultimately integrated into a Structural Thermal Optical analysis.

Optical Design The design meets the image quality requirements with comfortable margin over the entire field of view. A detailed tolerance analysis will be performed to allocate fabrication and alignment tolerances for all of the components. This tolerance budget will later be incorporated into the STOP analysis.

Wavefront Sensing. The addition of wavefront sensors to the imager focal plane is under consideration. Data from wavefront sensors would aid in the analysis of weak lensing observations, and simplify the maintenance of optical alignment by the operations team.

Calibration The optimal combination of astronomical observations and on-board illumination system needed for photometric calibration is still under study. The issues range from wavelength-dependent flat-fielding to detector linearity and filter stability.

Data Simulations High-fidelity simulations are required:

1. Cosmological/Science Simulations to verify the link between our Science Objectives and our Survey Requirements ... if the WFIRST mission delivers the required Surveys, can the Science Objectives actually be met?

2. Image Processing Simulations to verify the link between our Survey Requirements and our Dataset Requirements ... if the mission delivers the required Datasets, can those Datasets be image processed to produce the required Survey information?

3. Observatory transfer function simulations using sky-truth inputs to verify the link between our Required Datasets and our Observatory Design/Ops Concepts ... if we build/operate an Observatory as specified will it in fact deliver the required Datasets?

All of the above levels of simulation are needed to derive/validate requirements at the Survey, Dataset, and Observatory Design/Ops levels, and are critical to giving us the ability to design the right system based on well supported, comprehensive trades. Results from the JDEM mission concept study teams can, in some instances, be applied to WFIRST, but in most cases the existing tools would need to be modified or new tools developed.

Detector and Focal Plane Assembly This is described in Section 4.8 above, and is included in this list for completeness. Additional detector lots would be desirable to mitigate the yield risk, as resources permit.

Independent Cost Estimate (ICE) The ICE is described above in Section 6 and is listed here for completeness.

The SDT and Project Office will continue working on WFIRST after this Interim Report. Several possible modifications to the IDRM have been identified that warrant further study. These include extending the wavelength range redward of 2 microns and adding an Integral Field Unit (IFU) spectrometer for SN spectroscopy. These options are described in more detail in Appendix E. Additional science simulation work is planned to give more detailed assessment of mission performance and aid in further refinement of the mission design. After the ESA M2 decision, there may be other work related to Euclid. A final report of the SDT is due in about 1 year.

# Appendix A Exoplanet Science Goals & Requirements

## A.1. Microlensing Exoplanet Detection Method

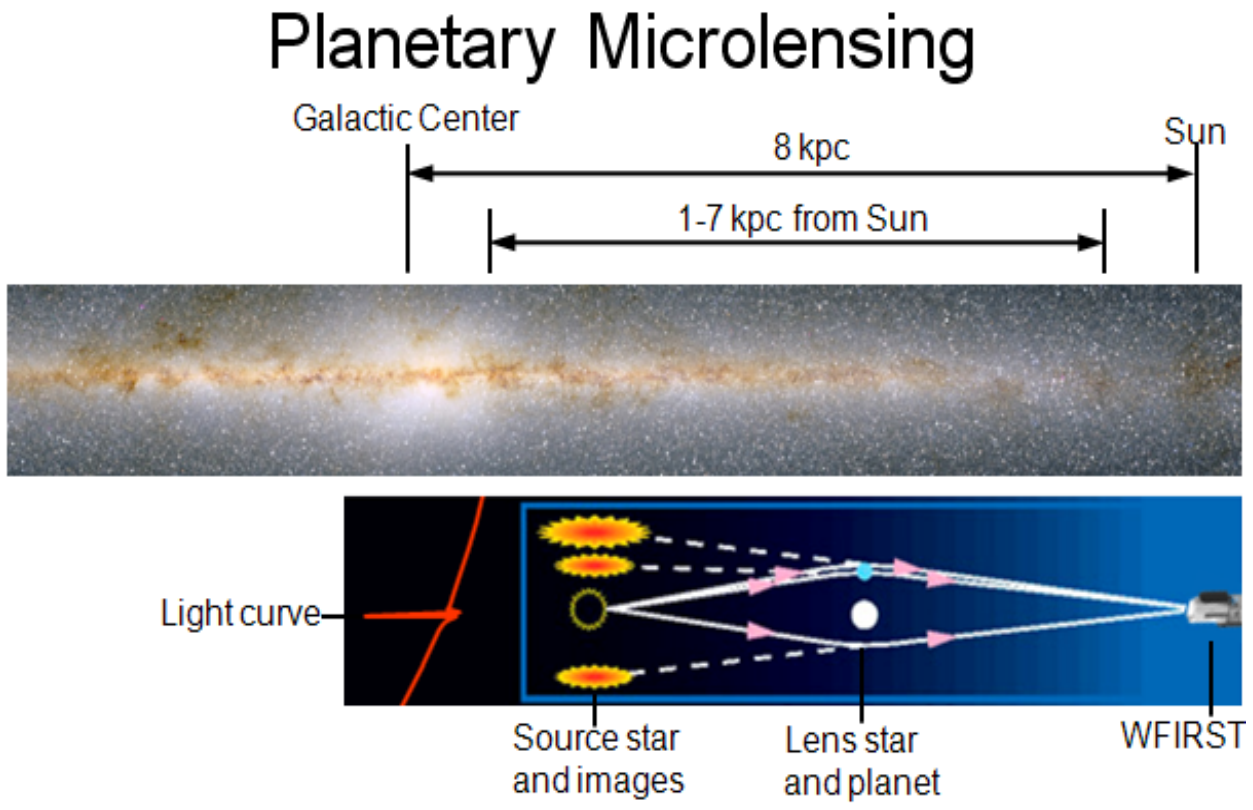

Figure 14: The geometry of a microlensing planet search towards the Galactic bulge. Main sequence stars in the bulge are monitored for magnification due to gravitational lensing by foreground stars and planets in the Galactic disk and bulge.

WFIRST finds exoplanets using the technique of gravitational microlensing. The physical basis of microlensing is the gravitational bending of light rays by a star or planet. As illustrated in Figure 14, if a "lens star" passes close to the line of sight to a more distant source star, the gravitational field of the lens star deflects the light rays from the source star. The gravitational bending effect of the lens star splits and magnifies the images of the source star. The individual images are not resolved, so the observer sees a microlensing event as a transient brightening of the source as the lens star's proper motion moves it across the line of sight. Gravitational microlensing events are characterized by the Einstein ring radius,

$$R_E = 2.0\ \mathrm{AU}\sqrt{\frac{M_\mathrm{L}}{0.5\,M_\odot}\,\frac{D_\mathrm{L}(D_\mathrm{S}-D_\mathrm{L})}{D_\mathrm{S}(1\ \mathrm{kpc})}}\ ,$$

where $M_L$ is the lens mass, and $D_L$ and $D_S$ are the distances to the lens and source stars, respectively. $R_E$ is the radius of the ring image that would be seen with perfect alignment between the lens and source stars. A microlensing event's duration is given by the Einstein ring crossing time, typically 1-3 months for stellar lenses and a few days or less for a planet.

Planets are detected via light curve deviations that differ from the normal stellar lens light curves (Mao & Paczynski 1991). Usually, the planet signal occurs when one of the two images of the source star passes close to the location of the planet, as indicated in Fig. 1 of Gould & Loeb (1992), but planets are also detected at very high magnification where the gravitational field of the planet destroys the symmetry of the Einstein ring (Griest & Safizadeh 1998). The probability of a detectable planetary signal and its duration both scale as $R_E \sim M_p^{1/2}$, but given the optimum alignment, planetary signals from low-mass planets can be quite strong (many tens of percent). The limiting mass for the microlensing method occurs when the planetary Einstein radius becomes smaller than the projected radius of the source star (Bennett & Rhie 1996), resulting in a suppression of the amplitude of the deviation. This suppression is stronger for planets located inside $R_E$. For the F, G or K-dwarf source stars in the bulge that can be monitored with a space-based survey, the sensitivity of the microlensing method extends down to $< 0.1M_\oplus$.

Microlensing is most sensitive to planets at a separation of ~ $R_E$ (usually 2-3 AU) due to the strong stellar lens magnification at this separation, but the sensitivity extends to arbitrarily large separations. Planets well inside $R_E$ have a lower probability of detection. Planets in the habitable zones of their parent star tend to be located at separations that are substantially smaller than $R_E$, and thus are more difficult to detect. This in turn requires the higher photometric precision and higher angular resolution afforded by a space based microlensing mission.

Well-sampled planetary microlensing light curves yield unambiguous planet mass ratios and separations in units of $R_E$, but not absolute planetary masses and separations in physical units (i.e., AU). However absolute values are needed to place the microlensing detections in the context of the discoveries made by other methods. For all but a small fraction of planetary microlensing events, high-resolution (<0.3") imaging is needed resolve out the dense stellar background in the target bulge fields, and so allow for the detection and unambiguous identification of the light from the host stars. This in turn allows the star and planet masses and separation in physical units to be determined (Bennett et al. 2007). A space-based survey with the requisite resolution will obtain this information naturally for the majority of the host stars.

## A.2. Exoplanet Survey Questions and Science Requirements

The WFIRST Exoplanet survey requirements flow down from the Exoplanet Survey Questions discussed in Section 2.2.2:

1. How do planetary system form and evolve?
2. How common are potentially habitable worlds?

WFIRST can address the first of these questions by determining the demographics of planetary systems with its sensitivity to virtually all types of exoplanets that will not be detected by Kepler. Furthermore, due to its sensitivity to planets at substantial distances, WFIRST can determine how the properties of planetary systems depend on their Galactic environment. WFIRST has the unique ability to detect old, free-floating planets of < 1 Earth-mass, which provide important constraints on the dynamical histories of planetary systems. The microlensing is also able to detect planets that orbit stars, which are too dim to be detected. This is an advantage, but if all the planets detected by WFIRST fall into this category, WFIRST's contribution to Exoplanet Survey Question #1 will be significantly compromised. These considerations lead to the following Science Requirements:

A. *Make a definitive measurement of the frequency of bound and free-floating planets with masses extending to less than an Earth-mass and separations greater than 0.5 AU.*
B. *Measure the masses of the planets and host stars for the majority of the detected exoplanetary systems.*

Kepler will address Exoplanet Survey Question 2, which is critical for the planning of future missions, but this is a challenging measurement for Kepler. So, an additional measurement with a very different method by WFIRST is needed. This implies the final WFIRST Science Requirement:

C. *Make a definitive measurement of the frequency of habitable planets.*

## A.3. Exoplanet Survey Requirements

Obtaining a definitive measurement of the frequency of exoplanets of a given type with WFIRST, and so meeting Science Requirements A and C above, translates directly to a requirement on the number of detected planetary deviations of that type. In general terms, inferring the frequency of exoplanets involves first detecting the planetary deviations, then inferring the planetary parameters from those deviations, and finally calibrating the probability of detecting planets of the given type for the ensemble of observed microlensing events. These pieces of information can then be simply combined to estimate the intrinsic planet frequency (e.g., Gould et al. 2010). Calibrating the detection probability is a well-understood procedure incurring negligible uncertainties (Gaudi & Sackett 2000; Rhie et al. 2000), and thus for robustly-measured deviations (as specified below) for which the light curve parameters well-measured, the fractional uncertainty in the inferred planet frequency is due almost exclusively to the Poisson fluctuations in the number of detected planets $N_p$, which we can approximate as $N_p^{-1/2}$. Thus achieving a certain precision in the measurement of the exoplanet frequency requires a minimum number of planet detections.

Of particular interest are planets with mass less than that of the Earth. Microlensing is the only technique available to detect planets as small as the mass of Mars, which are thought to be the largest bodies that can be formed by rapid growth of planetary "embryos". We therefore define:

*Exoplanet Survey Requirement #1:* Planet detection capability to ~0.1 Earth Mass

The primary strength of WFIRST's exoplanet survey is the ability to survey a broad region of parameter space inaccessible to other techniques or ground-based microlensing surveys, and thus there are a large number of specific questions that we seek to address with the mission products. Fortunately, for most regions of planet discovery space, the number of detections scale self-similarly, such that the ratios of the number of detections in different regions remain approximately constant (Bennett & Rhie 2002). Thus we can specify the minimum number of detections at one fiducial set of values of the exoplanet parameters to quantify the primary measurement requirement. With this in mind, we define:

*Exoplanet Survey Requirement #2a: If every star has a planet with a mass of 1 Earth mass and an orbital period of 2 years, detect (at a $\Delta\chi^2$ > 160) at least 120 of them.*

Here $\Delta\chi^2$ is the difference between the $\chi^2$ of the model fit with and without a planet, and the minimum value of 160 is set by the requirement to detect and characterize the planetary perturbation, calibrated from experience with ground-based microlensing planet detections.

Figure 5 shows the WFIRST exoplanet discovery potential when this requirement is saturated. In particular, WFIRST is sensitive to analogs to every planet in our solar system, except for Mercury, clearly demonstrating both its broad discovery space, as well as the ability to address the question of whether or not our solar system is unique. When the survey requirements above are met, WFIRST will measure the mass function of planets with mass $>0.1M_{\oplus}$ and separations in the range of 1.5-5 AU with a resolution of 0.2 dex in mass and a precision of <10% per mass bin. Determining the planetary mass function down to a tenth the mass of the Earth will uniquely allow WFIRST to probe the physics of the assembly of terrestrial, ice, and gas-giant planets.

Owing to its continuous coverage and high photometric precision, the overwhelming majority of the planetary perturbations detected at $\Delta\chi^2 > 160$ by WFIRST will yield measurements of the planetary mass ratios and projected separations in units of $R_E$ at precision of <20%. However, measuring host star masses and distances, and thus planet masses and separations in physical units requires additional measurements, which will only be possible for a subset of the detections. These measurements place quite different requirements on the survey data. In particular, as mentioned above, for most events high-resolution (<0.3") imaging is needed to detect and unambiguously identify the light from the host star, which allows for a measurement of its mass and distance. Therefore, to ensure that Science Requirement B is met we define:

*Exoplanet Survey Requirement #2b: Measure planet masses to 20% accuracy for at least 80 planet events with mass of 1 Earth mass and a period of 2 years, assuming one such planet per star.*

Generally, the number of detections of habitable planets does not scale simply with the number of detections for the fiducial set of values above. This is because habitable planets typically perturb demagnified images, resulting in smaller amplitude deviations whose detectability is more sensitive to the photometric precision and the angular source size. Therefore, to ensure that Science Requirement C is met, we define:

*Exoplanet Survey Requirement #3: If every F, G, and K star has a habitable planet, detect (at a $\Delta\chi^2 > 50$) at least 25 of them.*

When this requirement is met, we can measure the frequency of habitable planets, $\eta_{\oplus}$, to a precision of ~20% $\eta_{\oplus}^{-1/2}$. This is comparable to the expected precision obtained by Kepler assuming an extended mission (Lunine et al. 2008). Habitable planets are detected with a lower $\Delta\chi^2$ threshold than the other planets because they can only be found orbiting relatively bright stars, and this means that the lens-source relative proper motion can be detected directly from the images. This enables an estimation of the angular source size, which would otherwise be degenerate with the mass ratio in such low signal-to-noise ratio planetary deviations, and so could prevent characterization of the planet.

The number of detections of free-floating low-mass planets also may not scale simply with the number of detections for the fiducial set of values above. This is because the robust detection of a free-floating planetary event requires a higher detection threshold than for a planetary deviation associated with a stellar microlensing event, since a much larger number of stars must be searched for these events, therefore increasing the probability of false positives. These high detection thresholds also increase the sensitivity to the photometric noise and the finite size of the source stars. Therefore, to ensure that WFIRST will detect free-floating planets as well as bound planets, we define:

*Exoplanet Survey Requirement #4: If there is one free floating Earth-mass planet per star in the Galaxy, detect (at a $\Delta\chi^2 > 300$) at least 30 of them.*

The detection of free-floating planets requires a higher $\Delta\chi^2$ threshold than is needed for planets detected with a host star because ~$10^8$ light curves must be searched for free-floating planetary events, whereas we expect only a few $\times 10^4$ stellar microlensing events to be searched for the signals of bound planets.

## A.4. Exoplanet Data Requirements

Microlensing events require extremely precise alignments between a foreground lens star and a background source star, and are both rare and unpredictable. Furthermore, the probability that a planet orbiting the lens star in any given microlensing event will give rise to a detectable perturbation is generally much smaller than unity, ranging from a few tens of percent for a Jupiter-mass planet and a typical low-magnification event, to less than a percent for planets with mass less than that of the Earth. These planetary perturbations have amplitudes ranging from a few percent for the lowest-mass planets to many tens of percent for the largest perturbations, but are brief, ranging from a few days for Jupiter-mass planets to a few hours for Earth-mass planets. Also, the time of the perturbations with respect to the peak of the primary event are also unpredictable. Thus detecting a large number of exoplanets with microlensing requires monitoring of a very large number of stars continuously with relatively short cadences and good photometric precisions of ~1%. Practically, a sufficiently high density of source and lens stars, and thus a sufficiently high microlensing event rate, is only achieved in lines of sight towards the Galactic bulge. However, these fields are also crowded, and this high star density means that high resolution is needed to resolve out the individual stars in order to achieve the required photometric precisions and to identify the light from the lens stars.

Because the source star density and event rate are strong functions of Galactic coordinates, and the detection probability of a planet with a given set of properties depends sensitively on the detailed properties of the event (host star mass and distance, event duration, angular source size, photometric precision, cadence), quantitative predictions for the yields of given realization of exoplanet survey dataset require sophisticated simulations that incorporate models for the Galactic distribution of lenses and sources to simulate and evaluate the detectability of events with realistic photometric precision. Such simulations for the WFIRST mission have been performed and are described below, updated from the simulations of Bennett & Rhie (2002). The planet yields from these simulations were ultimately used to determine the set of exoplanet survey data requirements necessary to meet the four measurement requirements listed above. The resulting survey data requirements are as follows:

*Exoplanet Data Requirements:*

- Monitor 2 square degrees in the Galactic bulge for a total of 500 days.
- Signal-to-noise ratio of >100 per exposure for a J = 20.5 star
- Photometric sampling cadence of 15 minutes or better.
- Better than 0.3" angular resolution to resolve the brightest main sequence stars
- Monitor microlensing events continuously with a duty cycle of >95% for at least 60 days to measure basic light curve parameters.
- Measurements in a second filter every 12 hours in order to measure the color of the microlensing source stars.
- Separation of >4 years between first and second observing seasons to measure lens-source relative proper motion.

We can roughly understand the order of magnitude of these requirements. Consider the primary measurement requirement #2a: If every star has a planet with a mass of 1 Earth mass and a period of 2 years, detect (at a $\Delta\chi^2 > 160$) at least 120 of them. The typical detection probability for a planet with $M_p=M_\oplus$ and $P$=2 yrs is ~0.5%, and thus ~100/0.005 = $2\times10^4$ microlensing events must be monitored to detect ~100 such planets. The microlensing event rate in the highest density fields accessible by WFIRST is $\sim2\times10^{-5}$ events/year, and thus $2\times10^4/2\times10^{-5}$~1 billion star-years must be monitored. The typical stellar density down to J=23 is $\sim10^{8.5}$ stars per square degree, and thus at least ~2 square degrees must be monitored. In order to detect, fully sample, and accurately characterize the perturbations due to such planets, which typically last a few hours and have amplitudes of several percent, photometric precisions of <1% and cadences of better than 15 minutes are needed. Finally, given the areal density of $\sim10^8$ stars per square degree, an angular resolution of $10^{-4}$ degrees or 0.3” arcseconds is needed to resolve the faintest stars. The remaining three requirements above ensure the ability to accurately measure the parameters of the primary events, which typically last ~40 days, and allow one to infer the angular size of the source star from its color and magnitude, separate the light from the lens and source, and measure the relative lens-source proper motion, all three of which are required to measure the mass and distance to the primary lens (and thus measure the mass and separation of the detected planets) for the majority of events.

These data requirements then provide the constraints within which hardware and operations concept designs must operate.

# Appendix B Dark Energy Science Goals & Requirements

## B.1. Baryon Acoustic Oscillation/Redshift Space Distortion Requirements

The principal requirement for the BAO/RSD program is the measurement of 3D positions (RA/Dec/redshift) for a large sample of galaxies with a characterizable selection function. Overall one desires to do this over a wide range of cosmic history; we require observations in the $0.7<z<2$ range for WFIRST because the lower redshifts can be obtained efficiently from the ground.

Ideally the survey would cover the entire extragalactic sky (28,000 $deg^2$), and achieve a galaxy density that reaches the sample variance limit out to mildly non-linear scales ($nP \geq 1$ at all redshifts and k~0.2 h/Mpc). This is not feasible within programmatic constraints (which limit the size of the telescope and detector and the available time for BAO/RSD surveys) and so we have attempted to come as close as possible. In particular, this led to the split of the BAO/RSD survey into a DEEP survey (acquired, as a goal only, as a by-product of the WL galaxy shape survey) that samples the entire redshift range well (reaching nP>1 out to z=1.85) and a WIDE mode that covers much more of the sky – 11,000 $deg^2$/yr – but only reaches the sampling variance limit at low k.

The 3D measured positions of galaxies need only be accurate enough not to cause significant degradation of the large-scale structure measurement at mildly nonlinear scales; for Gaussian errors the loss of S/N per mode is $\exp(-k^2\sigma^2/2)$. Thus one needs to know positions to an accuracy of better than 1/k. The transverse requirement is trivially met, but the radial (redshift) measurement requirement is not. We have therefore set a redshift error requirement of 300 km/s rms, which corresponds to a radial error of 3.2-3.5 Mpc/h (depending on z). The degradation of S/N per mode is 20% at $k_{||}$ = 0.2 h/Mpc.

Misidentified lines have two major effects on the power spectrum: there is a suppression of the power spectrum due to introduction of a smoothly distributed component (proportional to the contamination fraction $\varepsilon$), and an additional source of power associated with clustering of the contaminants (proportional to $\varepsilon^2$). In order to keep the degradation of the signal small we require the total misidentification fraction to be $\Sigma\varepsilon<0.1$ (TBR). We have to remove the clustering of contaminants from the true signal and hence we need the total amount of such power (proportional to $\Sigma\varepsilon^2$) to be small enough that there is no significant systematic error after it is subtracted. This flows down to a requirement on the misidentification fraction per contaminant (TBD). The requirement on knowledge of the fraction of misidentified lines is driven by the RSD: the ~$\varepsilon$ suppression of power has no effect on the BAO signal but biases the RSD measurement by a factor of $1-\Sigma\varepsilon$. Therefore the RSD drives the requirement on $\Sigma\varepsilon$ to be known to 0.2% (TBR).

The detailed requirements on the BAO/RSD survey are as follows (DEEP mode is computed at ecliptic latitude 45° and WIDE is at 35°; all fluxes are extinction-corrected at E(B-V)=0.1 and $R_V$=3.1):

- ≥ 11,000 $deg^2$ sky coverage per dedicated year ("WIDE" Survey mode)
- Goal of ≥ 2,700 $deg^2$/yr "DEEP" Survey acquired during the WL Survey
- A comoving density of galaxy redshifts at z=2 of $4.9\times10^{-5}$ $Mpc^{-3}$ (WIDE) or $2.1\times10^{-4}$ $Mpc^{-3}$ (DEEP). [The source density is higher at lower redshifts, peaking at z=1 at $2.2\times10^{-4}$ $Mpc^{-3}$ (WIDE) or $5.9\times10^{-4}$ $Mpc^{-3}$ (DEEP)]
- Redshift range $0.7 \leq z \leq 2$
- Redshift errors $\sigma_z \leq 0.001(1+z)$, equivalent to 300 km/s rms
- Misidentified lines ≤ TBD% per source type, ≤ 10% overall; contamination fractions known to 0.2% (TBR)

The implementation choice for the BAO/RSD was driven by several considerations. A slitless prism approach was selected for the spectroscopy as it eliminates $0^{th}$ order features that otherwise resemble emission lines (thus simplifying data processing), and maximizes S/N by putting all transmitted light in $1^{st}$ order. (Since the SpC is sky-limited, the required exposure time degrades by 2X% if X% of the light is in an undesired order.) The bandpass covers the bright H$\alpha$ line across the redshift 0.7-2.0 range. Since the galaxy redshift survey requires only a detection and not the morphology of the emitting region, there is no requirement to recover full sampling. At fixed detector count it is advantageous to reduce the SpC f/ratio relative to the ImC so that a galaxy occupies only a few pixels per exposure.

At a spectral dispersion $R_\Theta \approx 200$ arcsec, the redshift accuracy of 300 km/s rms corresponds to a requirement to centroid the emission line to 0.4 pixels rms (including the raw centroiding of the emission line, addi-tional issues such as astrometric uncertainties and [NII] or continuum blending, and the mapping from line position to wavelength). This dispersion also supports the splitting of the [OIII] doublet (2900 km/s or 4.3 pixels), which both reduces the raw contamination fraction (by reducing the S/N for [OIII] detections) and makes [OIII] directly identifiable if the S/N is high enough to support marginal detection of the weaker doublet member. This dispersion does not split [NII] from H$\alpha$: the splitting of H$\alpha$ from the stronger [NII] feature is 900 km/s (1.3 pixels). Such splitting would not be desirable since it would reduce the S/N for the combined line detection. (Note, however, that our galaxy yields do not assume any [NII] contribution – we have treated it as margin.)

Oppositely dispersed spectra are advantageous in order to disentangle the astrometric center of H$\alpha$ emission from the redshift (otherwise the former becomes an irreducible source of noise for the latter). WFIRST has considered both options that provide a single dispersion direction in hardware (with the other direction acquired 6 months later by rolling the satellite) and options that provide both directions near-simultaneously (as did JDEM-Omega). Both appear to be viable but the dual dispersion provides a greater range of tiling options.

The ImC is used to provide positions of candidate galaxies, and to provide the location of the bandpass window relative to each emission line (for redshift zero reference).

Variations in the selection function are the principal systematic challenge for the galaxy clustering program. For WFIRST, these include zodiacal light brightness variations, Galactic dust, photometric calibration drifts, stellar density, and variations of the SpC PSF across the fields of view. Repeated visits to calibration fields (e.g. the supernova fields) will track secular changes in the photometric calibration as a function of time. The sky brightness and stellar density variations are critical for WFIRST as they vary by factors of a few over the survey area. However, these are effects that can be simulated by operations on the actual data (e.g. by adding sky photons or stellar traces to an image and then re-processing it); and so we expect that the associated selection function effects will be very well known. Moreover, even if one did not know the amplitudes of these effects, most of their power is in unique patterns on the sky (low-multipole modes and features associated with the tiling) that are very different from the modes used for BAO/RSD, and so removing them via template projection provides a backup option.

Deeper spectroscopic measurements of a small subset of the galaxies will be required to measure contamination fractions. This requirement will be met using some combination of tile overlaps, the SN fields, and the ground assets used for WL-PZCS observations.

The BAO/RSD data set requirements are:

- Spectrometer
  - Slitless prism
  - Dispersion $R_\Theta$ = 195 (TBR) - 240 arcsec
  - S/N ≥ 7 for $r_{eff}$ = 300 mas for H$\alpha$ emission line flux at 2.0 $\mu$m ≥ $1.5\times10^{-16}$ erg/cm$^2$-s (DEEP) or $3.1\times10^{-16}$ erg/cm$^2$-s (WIDE)
  - Bandpass 1.116 ≤ $\lambda$ ≤ 2.0 $\mu$m
  - Pixel scale ≤ 450 mas
  - System PSF EE50% radius 400 mas at 2 $\mu$m
  - ≥ 3 dispersion directions required, two nearly opposed
- Imager (for redshift zero reference)
  - S/N ≥ 10 for $H_{AB}$ ≤ 23.5
  - Approximately equal time in filters F141 and F178

## B.2. Supernova Requirements

The supernova technique relies on Type Ia supernovae as distance indicators to measure the expansion history of the universe and thus yield information about the acceleration of the expansion of the universe and the nature of dark energy. Type Ia's have an intrinsic luminosity spread of around 50%, but can be calibrated using the shapes of their lightcurves to between 12 and 18%, thus providing a measurement of their distances to between 6 and 9%. The low end of this range is achieved by observations in the infrared for the lower redshift supernovae, providing a distinct advantage to observations from space, and by other methods using spectral features at the higher redshifts. It is expected that by the time WFIRST data becomes available, the distance precision will be further improved by methods using other information, such as spectral features. A plot of the calibrated observed luminosity of each supernova versus its redshift, which we call the Hubble diagram, yields the cosmological information.

The requirements of the supernova technique can be divided into four categories, all carried out by the WFIRST supernova survey:

1. Discovery of the supernovae. This is done by the WFIRST imager by observing the same area of sky with a 5 day cadence and carrying

out a pixel by pixel subtraction of previous reference images from each new observation.

2. Selection of the Type Ia 's. Typically one third to one half of the candidates discovered by the imager are good Type Ia 's discovered before peak luminosity. The selection of the Ia's is done by taking a spectrum of each candidate using the prism P130 in the ImC filter wheel.
3. Redshift measurement is obtained from the spectra. A final reference spectrum will be taken after the supernova has faded away, using the narrow lines of the host galaxy for an accurate redshift measurement. The final reference spectrum will also be used to improve the galaxy background subtraction from the supernova spectra.
4. The peak luminosity of each supernova is obtained from the lightcurves provided by the repeated 5 day cadence images from the WFIRST imager.

The precision of the WFIRST supernova data will be further improved by making use of a large, carefully calibrated data set of nearby (z<0.1) supernovae obtained from presently ongoing ground based nearby supernova surveys. These nearby surveys are best done from the ground since WFIRST will not be able to collect a significant number of supernovae below z=0.1 due to the volume effect. In the estimation of the Figures of Merit, a sample of 500 nearby supernovae, as projected by the FoMSWG Panel, was used, although the expectation is that a larger sample will be available by the time WFIRST flies.

The detailed requirements on the survey capabilities and the data set are as follows:

Survey Capability Requirements.

- >100 SNe-Ia per $\Delta z$=0.1 bin for most bins for 0.4 < z < 1.2, per dedicated 6 months
- Redshift error $\sigma \leq 0.005$ per supernova
- Relative instrumental bias≤ 0.005 on photometric calibration across the wavelength range
- Distance modulus error (from lightcurve) $\sigma_\mu \leq$ 0.02 per $\Delta z$=0.1 bin

Data Set Reqirements

- Minimum monitoring time-span for an individual field: ~2 years with a sampling cadence≤ 5 days
- Cross filter color calibration ≤ 0.005
- Four filters: F087, F111, F141, F178
- Slitless prism spec (P130) 0.6-2 μm, $\lambda/\Delta\lambda$ ~75 (S/N ≥ 2 per pixel bin) for redshift/typing
- Peak lightcurve S/N ≥ 15 at each redshift
- Dither with 15 (TBR) mas accuracy
- Low Galactic dust, E(B-V) ≤ 0.02

## B.3. Weak Lensing Requirements

The weak lensing technique uses the shear of distant galaxies by gravitational lensing, which is determined by an integral of the tidal field along the line of sight to the source galaxy. The strength of the lensing signal as a function of the redshift of source galaxies is sensitive to both the growth of cosmic structure at intervening redshifts $0<z<z_{source}$, and to the distance-redshift relation (since the same amount of mass produces more shear on more distant sources). The correlation of lensing shear with foreground galaxies of known redshift $z_{fg}$ probes the relation of these galaxies to the matter field and in combination with RSD allows direct consistency tests of general relativity (e.g. Reyes et al. 2010). Finally the dependence of the cross-correlation signal with the redshift of the source galaxies isolates the purely geometrical information from weak lensing from the information about perturbations (Jain & Taylor 2003).

A weak lensing survey must provide wide angle sky coverage, a high density of usable sources across a wide redshift range (local to z~2), provide photometric redshifts so that the galaxies can be sliced into bins to study evolution of the signal and remove "intrinsic" (not lensing induced) shape correlations, and achieve exquisite control over coherent systematic errors in the shear signal. The latter includes both "additive" shear errors, i.e. apparent shear due to imperfect correction for instrument effects (aberrations, anisotropic jitter, geometric distortions, etc.) that is not present in the sky; and "multiplicative" shear errors where the calibration of the observed shear is incorrect. Some sources of error (e.g. failure to recover full sampling) can lead to both additive and multiplicative errors.

The detailed requirements on the weak lensing survey are as follows:

- ≥ 2,700 $deg^2$ sky coverage per dedicated year (in "DEEP" Survey mode)
- Effective galaxy density ≥ 30 per $arcmin^2$, shapes resolved plus photo-z's
- Additive shear error ≤ $3x10^{-4}$
- Multiplicative shear error ≤ $1x10^{-3}$

- Photo-z error distribution ≤ 0.04(1+z), error rate <2%
- Goal is for the WL Galaxy Shape Survey to be taken in a manner such that concurrent spectroscopy also meets the BAO survey requirements on source density, redshift errors and fraction of misidentified lines.

Control over systematic errors in the shear measurement implies several requirements on the data set, in the areas of (i) knowledge of the PSF; (ii) multicolor imaging; and (iii) sampling.

The PSF ellipticity must be known in order to ensure to meet the additive shear error requirement, and the PSF second moment ($I_{xx} + I_{yy}$) must be known in order to meet the multiplicative shear error requirement. The data set requirements on PSF knowledge were calculated to use 50% (in a root-sum-square sense) of the shape measurement error budget (which also includes other terms, such as residuals from the data processing algorithms and detector effects). In setting this requirement, we conservatively assume no decorrelation (or "averaging down") of systematic shear errors in the 10 exposures used to measure shapes.

The PSF of any telescope will be color-dependent; for WFIRST diffraction will dominate this effect. In order to solve it, it is required to know the intraband color variations of the galaxies and hence fully-sampled imaging is required in two filters.

We also require a sufficient number of dithered exposures to recover full sampling of the sky image out to the band limit set by the optics (D/λ). This imposes a joint requirement in the space of number of dither positions, the dither pattern (ideal vs random), and the f/ratio. Given other pressures on WFIRST, we chose to implement the shape measurement in the two reddest filters (F141 and F178) where the sampling requirements are easiest to meet. Simulations show that 4-5 random dithers are sufficient (Rowe et al. 2011). F111 is strongly undersampled at 0.18"/pix and implementing full sampling would have required more dithers or a higher f/ratio, either of which would negatively impact the survey speed.

A further advantage of this choice is that due to the red spectral shapes of the source galaxies the S/N is greater for a redder filter. The principal disadvantage is the larger PSF; however the quadrupole moments (e.g. $I_{xx}$-$I_{yy}$) of the PSF induced by a given wavefront error depend only very weakly on wavelength.

The photometric redshifts require multi-filter imaging with coverage across the optical and NIR bands. This implies the need for ground-based imaging in griz or a similar filter set. It also implies the need for imaging in the WFIRST F111 filter to bridge the gap between the ground optical filters and WFIRST F141 in order to provide photo-z's for objects whose Balmer or 4000Å features are in or near this gap (roughly 1.3<z<2.0).

Finally, the photo-z error distribution needs to be calibrated against spectroscopic redshifts for a representative sample of the WL source galaxies. The sample should contain ≥ $10^5$ galaxies (Bernstein & Huterer 2010) spread over several fields observable from WFIRST, the ground-based telescopes that provide optical photometry, and any assets needed to complete the spectroscopic survey. Bright emission line searches (using NIR slitless spectroscopy from WFIRST and an optical multi-fiber spectrograph such as proposed for low-z BAO on the ground) would likely measure a robust redshift for ~50% of the source galaxies, with the remainder to be done via deeper spectroscopy. The detailed trade of how much of the program to do from WFIRST before switching to sensitive ground spectrographs with lower multiplex factors must await decisions about which ground spectrographs will exist by the time WFIRST flies. We have *provisionally* allocated time for WFIRST to cover 5 fields using one arm of the SpC to a depth of $3\times10^{-17}$ erg/cm$^2$/s (7σ). These fields would be at ±20° ecliptic latitude, accessible from ground assets either hemisphere.

In summary, the WL data set requirements are:

- From Space: 2 shape/color filters, F141 and F178, and one color filter, F111
- S/N ≥ 18 per shape/color filter for galaxy $r_{eff}$ = 250 mas and mag AB = 23.7
- PSF second moment ($I_{xx} + I_{yy}$) known to a relative error of ≤ $9.3\times10^{-4}$ rms (shape/color filters only)
- PSF ellipticity ($I_{xx}$-$I_{yy}$, $2I_{xy}$)/($I_{xx} + I_{yy}$) known to ≤ $4.7\times10^{-4}$ rms (shape/color filters only)
- System PSF EE50 radius ≤ 170 mas for filter F141, and ≤ 193 mas for filter F178
- 5 random dithers required for shape/color bands and 4 for F111 at same dither exposure time
- From Ground: 4 color filter bands ~0.4 ≤ λ ≤ ~0.97μm
- Complete an unbiased spectroscopic PZCS training data set containing ≥ 100,000 galaxies with ≤ mag AB = 23.7 for F141 and F178 and covering at least 4 uncorrelated fields; redshift accuracy required is $\sigma_z$<0.01(1+*z*)

## Appendix C Shortcomings of and Alternatives to the Dark Energy Task Force Figure of Merit

The idea that the performance of a dark energy experiment can be boiled down to a couple of numbers is overly simplistic. There was considerable discussion within the SDT of the shortcomings of the DETF FoM, and whether they were so severe as to negate any benefit there might be in adopting it. Not everyone agreed to its adoption. The shortcomings are spelled out here in the interest of representing the range of opinions held by the members of the SDT.

- Potential models to be tested are so varied and unconstrained that the simple specific DETF model form (with its two parameters and linear transition) is not generally representative of the full suite of potential models. Quoting from the JDEM Figure of Merit Science Working Group (FoMSWG*), "The FoMSWG finds that there is no single number that can describe the scientific reach of a JDEM.*"

- Figures of merit estimates depend on an accurate assessment of the quality of the experimental data and associated systematic errors, in many cases years before the experiment is even fully defined. Figures of merit only reflect the information provided so the result is dependent on the accuracy of the projected data quality. In particular, systematic errors and the performance of instruments play central roles.

- Astrophysical systematic errors must be estimated. There can be (and often are) reasonable differences of opinion between scientists on how to quantify these uncertainties. and it is likely that these estimates will change with time, understanding, and new measurements. These changes could be for the better, but also could be for the worse as new complexity is discovered.

- The wide-ranging auxiliary science that could come from the survey is given no credit.

- Prior astrophysical knowledge must be appropriately assumed, but projections of that knowledge are both uncertain and changeable as the field continues to evolve. As the FoMSWG wrote, "*Dark energy remains a compelling astrophysical question (perhaps the most compelling) and the creativity and imagination of astronomers and physicists will continue to be directed toward investigations into the nature of dark energy. Predictions of what will be known about dark energy, or what will be known about systematic uncertainties associated with dark energy measurements, eight years in the future are inherently unreliable."*

- The FoM formalism does not properly reward the cross checks on consistency enabled by performing multiple measurements of different observables with very different systematics that probe the same aspects of dark energy (e.g. measuring the expansion history with both BAO and SN).

- Redundancy in the data set (e.g. measuring more degrees of freedom in a detector than exist on the sky) and internal checks within a given technique are often the keys to reducing systematic errors. This level of detail is generally not present in parameteric FoM optimization studies, and consequently such tests should be considered a factor outside the FoM.

The DETF figure of merit is strongly model-dependent, so it provides a test of experiments against an *assumed* form of dark energy evolution, whereas the very core of the measurement objective is to test for non-standard forms. An improved figure of merit would effectively test the ability of experiments to constrain the *function* $w(z)$. This was the motivation for the JDEM FoMSWG to define a figure of merit with 36 uncorrelated redshift bins of piecewise continuous parts to represent the *function* $w(z)$.

A central flaw of the DETF figure of merit for our purposes is easily seen by noting that excellent measurements at only two redshifts purport to determine the entire evolution of dark energy over all of cosmic time. Additional measurements are rendered irrelevant in this framework. The great bulk of dark energy measurements to date have been made at redshifts $z<1$ so the standard figure of merit devalues high redshift measurements.

Experiments do not directly measure $w(z)$. Supernova measurements provide a luminosity distance, $D_L(z)$ (either assuming a value for space curvature, or marginalizing over the uncertainty). Baryon Acoustic Oscillations provide the angular diameter distance, $D_A(z)$ as well as the Hubble expansion rate $H(z)$. The luminosity distance and angular diameter distance are related by $D_L = D_A(z)\,(1+z)^2$ and both the luminosity and angular diameter distances are integrals over the inverse of *H(z)*. The weak gravitational lensing technique effectively measures the three-dimensional distribution of matter in the universe, from which one can derive the evolutionary properties of the universe.

The measurement techniques all have potential systematic effects. These can be entered into the Fisher matrix machinery of the figure of merit calculations, but this must be done by hand. The machinery does not add value, but only reflects the information supplied to it. For this reason, and due to the sensitivity of the results to the priors and systematic error models, *it is common for different scientists to arrive at very different figure of merit values when attempting to calculate the power of identical experiments.*

Due to the reliance on priors, the figures of merit are also time-sensitive. As the JDEM FoMSWG wrote, "*However, we would like to make the important point that although considerable effort went into constructing the Fisher matrices, they should be used with caution, since any data model purporting to represent the state of knowledge eight years in the future is highly suspect.*" (Joint Dark Energy Mission Figure of Merit Science Working Group, Dec 7, 2008).

In the end, the figure of merit has provided important general lessons on the dependencies and interdependencies of different methods. Beyond that, it is not at all helpful when designing an instrument for minimum systematic errors or in estimating the values of those systematic errors. It does not replace the need to calculate, simulate, and innovate. It does not even allow for the cross-comparison with different experiments using the same "technique" (e.g. SN, WL, or BAO) unless the exact same priors, exact same systematic error estimates (or basis for such estimates), the exact same astrophysical systematics, and the exact same statistics (e.g. emission line luminosity functions or SN Ia rates) are agreed upon.

Figure of merit calculations for WFIRST alone cover an enormous range, depending on assumptions made about the inclusion or exclusion of various techniques, the level of systematic errors, and observing time allocations. It is important to recognize that the figure of merit *responds* to these inputs and does not create new information, so the substantial uncertainties in the figure of merit based on the choice of input assumptions render it less than useful.

It is a major strength of the WFIRST approach that the mission is designed with the capability to execute a wide range of possible programs, so that progress in dark energy research in the coming decade can influence the balance of the observations to be performed. In this manner, the WFIRST program can adapt to ground-based progress, numerical simulation progress, new modeling capabilities and insights, and any other knowledge that enhances or casts doubt on various aspects of the program. The WFIRST SDT does not pretend to have the insight to maximize the dark energy impact of the mission from today's perspective other than to design and build in capabilities and to minimize instrumental systematic errors. Rather, the SDT opts to maximize WFIRST's dark energy accomplishments at launch and even after initial data are captured and analyzed. This adaptive approach guarantees a powerful dark energy program that is complementary with intervening dark energy measurements and accomplishments, whatever they may be.

# Appendix D WFIRST in the Context of Other Missions and Projects

WFIRST will launch seven years from the beginning of a new start. We must consider the astronomical context in which WFIRST might be operating 7-12 years hence. At least some of the astronomical questions that WFIRST will address might also be addressed by another space mission (in particular ESA's Euclid) and by ground-based programs (in particular DES and LSST for weak lensing and BigBOSS for BAO). In this appendix, we briefly consider the extent to which these missions might or might not pre-empt one or another part of the WFIRST dark energy science program.

It is the great strength of the WFIRST mission that it will carry out many different science programs. It was precisely for this reason that it ranked more highly than more narrowly focused missions in the decadal survey. Were one or another piece of the WFIRST program were to be pre-empted, more time might be devoted to other equally important parts of its program.

## D.1. Weak Lensing

WFIRST would carry out weak lensing observations at the "deep" rate of 2,700 square degrees per year. Our straw man allocation devotes one year to this mode, not taking possible results from DES, LSST and Euclid into account.

LSST is expected to produce 15,000 square degrees of weak lensing data. Despite the overlap between LSST and WFIRST, the decadal survey recommended LSST as its highest priority ground-based program and WFIRST as its highest priority space mission. Weak lensing can constrain both dark energy and alternative theories of gravity, making it especially important. But for weak lensing to constrain dark energy or modified gravity, there must be a factor of ten improvement in accuracy over present day measurements. The thermal stability of a space telescope, the absence of gravitational stresses, and most importantly, the absence of an atmosphere with constantly changing "optics" makes space the natural place from which to carry out weak lensing.

The LSST PSF will have contributions from telescope jitter, telescope optics and detector effects, as will WFIRST, but in addition will suffer from much larger changes in thermal and gravitational stresses, and largest yet, contributions from atmospheric seeing. The hope for LSST is that these much larger variations can be averaged out by observing each field a few hundred times. As is shown below, if LSST does as well as is hoped, it might do as well as the WFIRST weak lensing program. But if not, WFIRST will be positioned to carry out a strong program. WFIRST's shape measurements would in any event be subject to different instrumental systematic errors and, since they are to be carried out at longer wavelengths, somewhat different astronomical systematics.

The Euclid mission, a candidate for selection as an ESA M-class mission, would carry out weak lensing observations of 15,000 square degrees at optical wavelengths (like LSST and unlike WFIRST) over the course of a 5 year mission. There are reasons to prefer weak lensing observations in the infrared over those in the optical. Much more of the galaxy light is emitted in the infrared. Galaxies are also less misshapen in the infrared. On the other hand, the diffraction limited PSF is bigger in the infrared than in the optical.

One can therefore measure more galaxies per square arcminute in the optical than in the infrared if one has enough photons to do so. But if Euclid is to take full advantage of its smaller diffraction limit, it must satisfy more stringent jitter and optical aberration constraints than the already demanding requirements for WFIRST. As is shown below, if Euclid does as well as is hoped, it might do a better job of weak lensing than WFIRST. But if not, WFIRST will again be positioned to carry out a strong program. And WFIRST's shape measurements would, again, be subject to different instrumental systematic errors and, since they are to be carried out at longer wavelengths, somewhat different astronomical systematics.

## D.2. Baryon Acoustic Oscillations

WFIRST will carry out baryon acoustic oscillation observations at the "wide" rate of 11,000 square degrees per year, and also at the "deep" rate of 2,700 square degrees per year, covering the redshift range $0.7 < z < 2$.

The proposed Prime Focus Spectrograph on the 8-meter Subaru telescope would carry out a BAO survey at $0.7 < z < 1.5$ over ~80 nights using the [OII] emission feature. The FoMSWG "Stage III" forecasts already assume a survey (WFMOS) very similar to this one.

The proposed BigBOSS survey (Schlegel et al. 2011) would use the Mayall 4-m telescope on Kitt Peak to carry out BAO observations of galaxies at optical wavelengths over ~14,000 square degrees, covering the redshift range $z < 1.7$. It would also obtain observations of quasar Lyman-alpha absorption lines in the range $2 < z < 3$; this will provide a BAO measurement in a complementary redshift range to WFIRST as well as a wealth of data on the intergalactic medium (IGM). (The Lyman alpha forest is also affected by redshift space distortion; however unlike for galaxies, there is no relation between the bias b, RSD amplitude beta, and growth rate f, so determination of the latter relies on accurate modeling of the IGM by hydrodynamic simulations.

The WFIRST 1 year Deep + 1 year Wide survey covers the same solid angle of sky as BigBOSS. The WFIRST Wide survey component achieves the same number density of galaxies (and nP) at z=1. BigBOSS performs better at lower redshift (in part because the BigBOSS luminous red galaxies are available, and these contribute more per object to nP than star forming galaxies), whereas WFIRST performs better at high redshift: by z=1.3 WFIRST-Wide is providing more than 3 times the source density of BigBOSS. WFIRST-Deep provides a source density that is 3-4 times higher, albeit over only 2700 deg^2.

The Euclid mission under consideration by ESA would carry out BAO observations over 15,000 square degrees, covering the redshift range $0.7 < z < 1.6$. By virtue of its smaller aperture and its less efficient spectrometer (including the grism: transmitted $0^{th}$ order light adds background but not signal), Euclid fares less well per unit time. But by investing more time into BAO, we show in Figure 15 that Euclid performs almost as well as WFIRST.

### D.3. Supernova

By virtue of carrying out its supernova program in the infrared, WFIRST is likely to be superior to optical supernova programs. While LSST will doubtless identify many supernovae, in the absence of a coherent program of spectroscopic followup we cannot assign any figure of merit to an LSST supernova effort. Additionally, neither Euclid nor BigBOSS are planning on a supernova program.

### D.4. Caveats Regarding Comparisons

As we have emphasized often and at length, the accurate characterization of the accelerating expansion of the universe hinges critically on limiting sources of systematic error. For astronomical sources of systematic error, it is relatively straightforward to adopt consistent estimates for different projects. It is more difficult to be consistent in estimating instrumental sources of systematic error.

In some cases we can independently estimate these from the specifications for the instrument. In other cases we must rely upon the numbers reported by the different project teams.

This also applies to the sensitivities adopted. To the extent that different projects use different approaches to estimating sensitivities, results may not be commensurable. Short of carrying out the detailed analyses ourselves, we must again accept the numbers presented.

### D.5. Systematic Errors

The systematic errors adopted for Euclid, LSST and BigBOSS are given in Table 3, Table 4, and Table 5 in section 3. We have adopted the optimistic assumptions in all cases. The BAO figures of merit therefore include both classical BAO measurements and redshift space distortion, but not measurement of the transition from radiation to matter domination. For LSST we have assumed an effective useful galaxy density of 20 per square arcminute. A more conservative estimate, adopting the same RMS ellipticity requirement for LSST as for WFIRST and allowing for the larger LSST PSF would give 15 galaxies per square arcminute.

### D.6. Comparison

Figure 16 shows 3 different Venn diagrams: one for the five year Euclid program, one for the combined programs of LSST+BigBOSS and one for our straw man 2.5 year program for WFIRST. In each case Stage III priors have been assumed.

### D.7. Interpretation

Restricting attention to the combination of weak lensing and baryon acoustic oscillations, Euclid and LSST+BigBOSS, were they to perform as hoped, would produce better figures of merit than WFIRST. But with the addition of supernovae, WFIRST outperforms them both. Supernovae measure the equation of state parameter, w, at low redshift, complementing the measurement of w at higher redshift by BAO and weak lensing. This leads to a dramatic improvement in the DETF FoM.

### D.8. Five Years of Dark Energy with WFIRST

It might be fairer to compare Euclid's five year dark energy program with a five year WFIRST dark energy program. We have done that, by increasing the time allotted to each of the three methods of measuring dark energy (a 2.5 year deep survey, 1.5 year wide survey, and a 1 year supernovae survey), again using the optimistic assumptions. The DETF FoM improves for WFIRST improves from 1335 to 1770. As we have not tried to optimize such a program one might do yet better by redistributing the observing time. Our purpose in carrying out this exercise is not to recommend that more time be allocated to dark energy at the expense of exoplanet microlensing and IR surveys, but rather to emphasize that WFIRST is very well suited to dark energy studies.

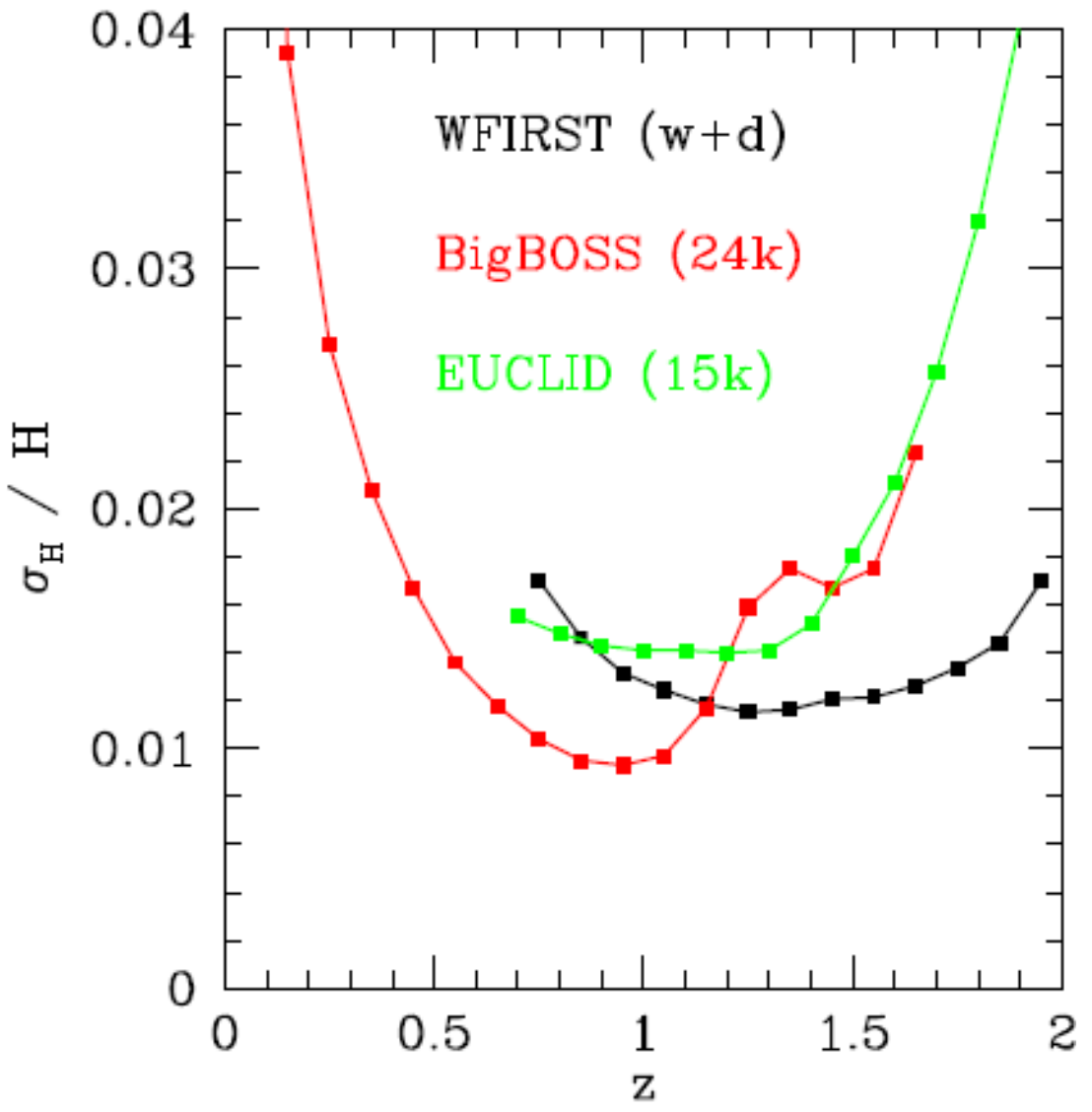

Figure 15: The fractional error in the Hubble expansion rate, H(z), as a function of redshift, z, driven by the precise measurement of the Baryon Acoustic Oscillation (BAO) scale in the spectroscopic galaxy redshift survey. The constraints for the proposed WFIRST 1 year deep + 1 year wide survey, are shown alongside those for prospective Euclid and BigBOSS surveys. Each survey has been evaluated per the "optimistic" scenario outlined in 3.4.7, using the same assumptions about systematic uncertainties, and using the stated galaxy yields or limiting fluxes estimated by that survey. WFIRST's capabilities in the IR allow it to achieve superior measurements at z>1, providing strong complementarity with ground based observations at z<1.

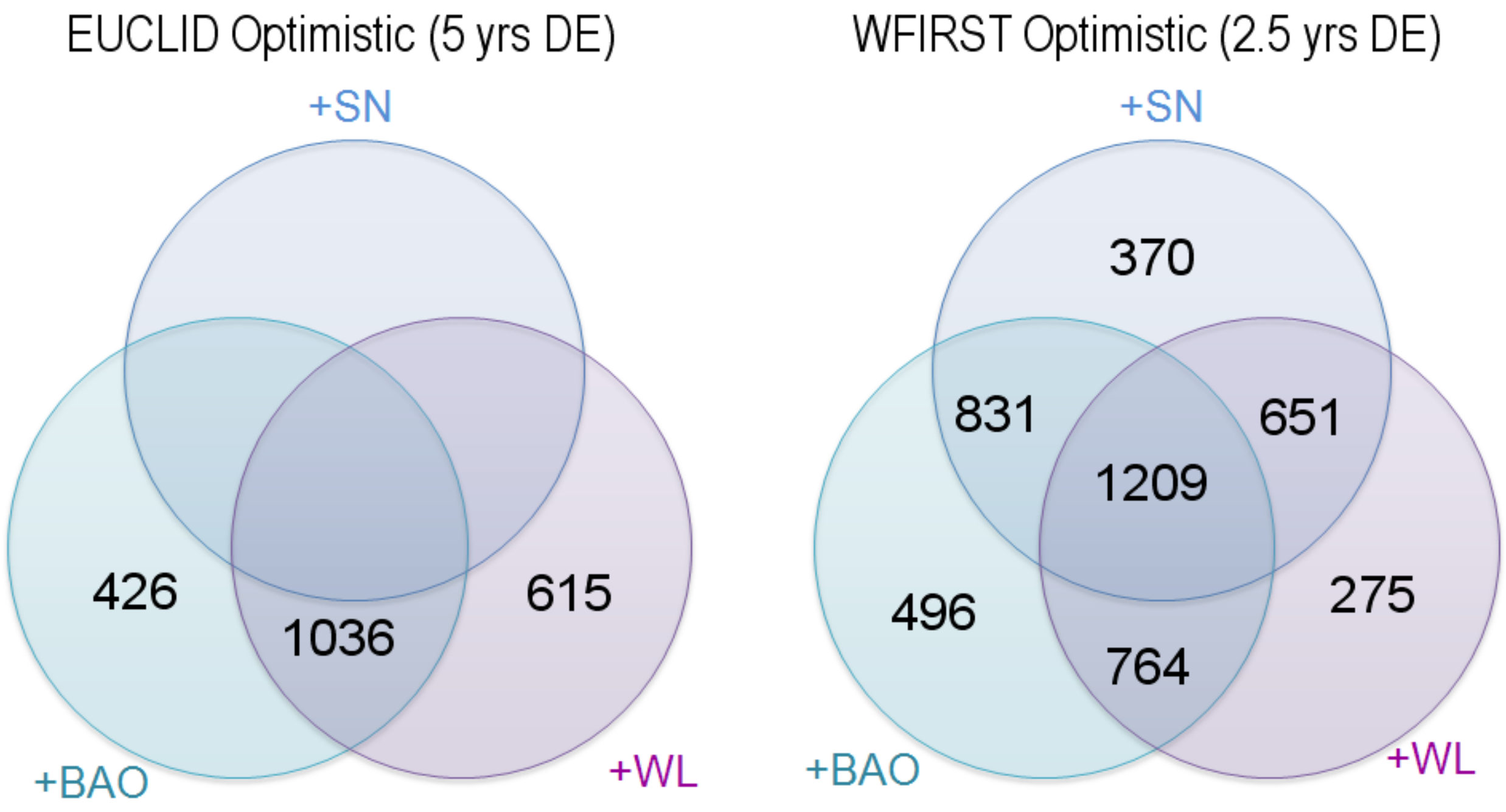

(a)

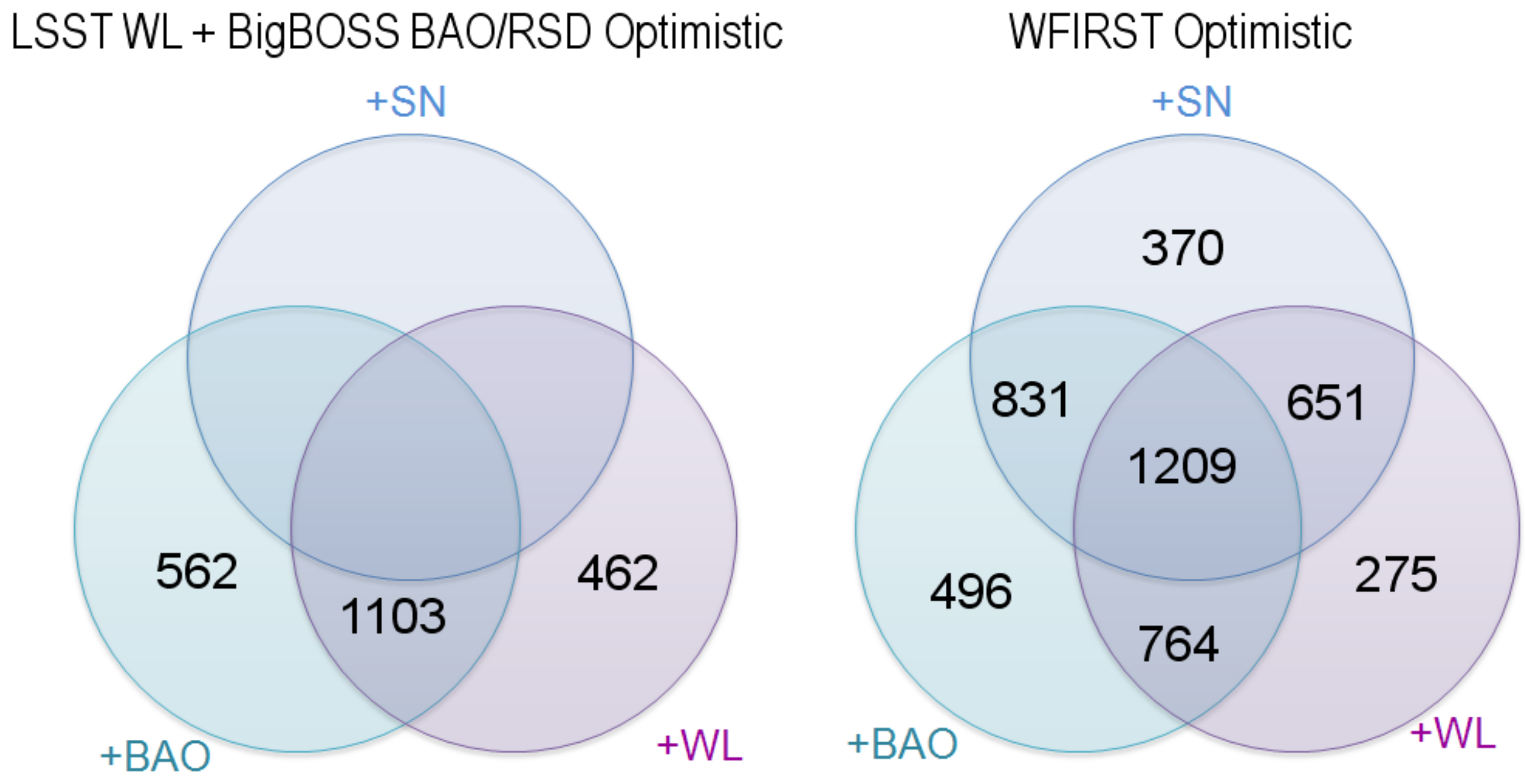

(b)

Figure 16: FoM comparisons of WFIRST with (a) Euclid and (b) LSST + BigBOSS

## D.9. Exoplanet Microlensing

The WFIRST Exoplanet survey will complete the statistical census of exoplanets begun by the Kepler mission with sensitivity extending outward from the habitable zone to infinity (unbound planets). In addition to Kepler, ground-based radial velocity surveys will extend to longer periods with somewhat improved velocity sensitivity, which will enable the detection of planets below a Saturn mass in ~10 year orbits, but the radial velocity method is unlikely to be sensitive to lower mass planets due to intrinsic stellar velocity noise. A space-based astrometry mission could be sensitive to planets down to about an Earth mass, but such a mission would not be sensitive over nearly as wide a range of orbital separations as WFIRST, and now that the Space Interferometry Mission (SIM) has been canceled, it is highly unlikely that such a mission could fly prior to WIFRST.

This leaves only other microlensing planet search programs that could detect some of the low-mass planets that are the primary goal of the WFIRST Exoplanet program.  As shown in Table 2, the number of detections expected from the ground is quite modest. The only observing program that could compete with the WFIRST exoplanet program is Euclid, as it is a telescope with a similar size and field-of-view. However, the exoplanet program on Euclid is considered secondary science, and neither the thermal design nor the filters are optimized for an exoplanet search. As a result, Euclid's exoplanet survey is limited to only a few months of observations and less than 10% of the exoplanet yield of WFIRST. Thus, if Euclid is selected, it will not contribute significantly to the exoplanet science planned for WFIRST.

# Appendix E Possible Modifications to the WFIRST IDRM

The NWNH took JDEM-Omega to be the template for WFIRST, saying that JDEM-Omega seemed nearly ideally suited for the WFIRST science goals. The SDT took JDEM-Omega as the starting point for its design reference mission. It heard proposals for changes, some of which have been adopted. Other changes are still under consideration, in particular the substitution of an integral field unit (IFU) for the slitless prism used in supernova spectroscopy and the extension of the sensitivity of the detectors redward of 2 microns.

## Supernova Spectroscopy Option: Substituting an Integral Field Unit Spectrograph (IFU) for the Slitless Prism and its Supporting Instrumentation

The knowledge that can be extracted from any survey is limited by the uncertainties in the data. These uncertainties can be statistical in nature, in which case additional data points (e.g. more SNe ) will reduce the final error, or through systematic errors, which, by definition, cannot be reduced through the accumulation of additional, similar data. WFIRST is a powerful survey telescope, and will collect such large and deep surveys that we expect the final data products to be systematics limited. The only way to increase the science return in such a case is to reduce the systematic errors.

Systematic errors can be created by the instrument, and these are addressed by design and calibration. They can also be astrophysical in nature. For example, it is known that not all Type Ia SN are identical, and that sub-classes exist with different light curves. Ignoring this subtlety will eventually limit the strength of the scientific conclusions that can be drawn from a large data set that cannot remove this systematic uncertainty.

To address this issue, the SDT is considering the potential inclusion of an Integral Filed Unit Spectrograph (hereafter, IFU) to more accurately classify the observed SNe and reduce the systematic errors in the analysis.

The baseline supernova program uses a prism in a filter wheel to obtain slitless spectroscopy of each supernova. An IFU uses a compact splayed arrangement of mirrors to slice a small image (including, e.g., a supernova, its host galaxy, and some background galaxies) into separate elements that each get dispersed. The resulting data cube of flux at each position and wavelength has many times higher signal-to-noise than a slitless spectrum with the same exposure time – or, equivalently, significantly reduced exposure times for the same signal-to-noise. In contrast, a slitless spectrum includes contributions from the full sky in each spectral bin, making it more difficult to isolate the faint SNe signal.

While the inclusion of an additional channel clearly involves some increases in complexity, there are potential practical advantages to using an IFU instead of a slitless prism approach. The IFU is expected to be less demanding in its pointing and stability requirements compared to the baseline slitless approach, and it is expected to eliminate the need for certain calibration instruments. Whether these advantages outweigh the additional complexity will be studied over the coming months.

### *Science Improvement with an IFU*

From a science perspective, the signal-to-noise gain with an IFU is quite important. The supernova program that can be accomplished with the lower signal-to-noise slitless spectroscopy is only sufficient to recognize the supernovae as “Type Ia” and provide its redshift. With the dramatically higher signal-to-noise of an IFU the spectral features of the supernova can be used to:

1. Distinguish intrinsic color variations from the effects of dust (Chotard et al., 2011). Currently, these two sources of reddening and dimming are not distinguished at high redshift, so the mix of these two effects is assumed to stay constant over the redshifts studied. For the precision measurements of $w(z)$ it would be important to separate them, since both are important corrections in the distance modulus calculation. This systematics control remains important even in the redder observer wavelengths (out to 2 microns) that WFIRST can reach.

2. Improve the “standard candle” calibration of the Type Ia supernova. Bailey et al. (2009) showed that with spectral feature ratios of sufficient signal-to-noise the magnitude dispersion of SN Ia distance modulus in the restframe optical can be reduced from 0.16 mag to 0.12 mag dispersion. This is as good as the improvement in dispersion using restframe H band observations. With IFU spectroscopy, however, this distance modulus improvement can be obtained over a large redshift range (beyond $z$ = 1.7), while the restframe H band photometry is only available to redshift $z$ = 0.1 (and J band only to $z$ = 0.4) even with an instrument observing out to 2 microns.

3. Compare the detailed composition and physical state of high-redshift supernovae to that of low-redshift supernovae. Type Ia supernovae are not all identical, but we can find spectroscopic matches of subsets of SNe Ia. If surprising cosmologies are inferred from supernova distance measurements over a range of redshifts it will be important to show that the effect is not simply an artifact due to a population of SNe Ia that is demographically drifting from one distribution of these spectroscopic subsets to another. With IFU spectroscopy it is possible to obtain the signal-to-noise sufficient to distinguish the different matching spectroscopic subsets.

4. Remove the K-correction systematics from the measurement. The slitless grism-based program uses only three filters over the whole redshift range studied, introducing the need for K corrections. The low signal-to-noise slitless grism spectra must then be combined to statistically remove any systematic biases in these K corrections (although this approach has not yet been studied to see what systematics will remain). With IFU spectrophotometry providing the lightcurves there would be no K corrections at all.

While the control of systematic uncertainties is the primary motivation for considering an IFU, its shorter exposure times can also be used to improve the depth of the survey and therefore the Figure of Merit. An example six-month-observing-time program has been developed (similar to the supernova program studied in detail by the ISWG) that yields significantly improved FoM over the slitless spectrograph six-month-observing-time program described as the baseline program above. The systematic uncertainties discussed in the FoM section of this document may be significantly reduced with the IFS spectroscopy. One study has found that the FoM for the supernova measurement combined with WL and BAO will increase by 20% (see Figure 17).

*SDT Future Studies*

Prior to the conclusion of the final report, the SDT will conduct a trade study concerning the utilization of an IFU for SNe redshift determination and sub-classification instead of the current baseline plan. This will involve the development of consensus FoM's for comparison of the potential science gains, as well as the programmatic impacts from a cost, integration, system level requirements, and calibration perspective. The positive and negative impacts on observational implementation will also be considered. These will used to reach a final SDT consensus recommendation on the inclusion of an IFU in the final SDT baseline mission.

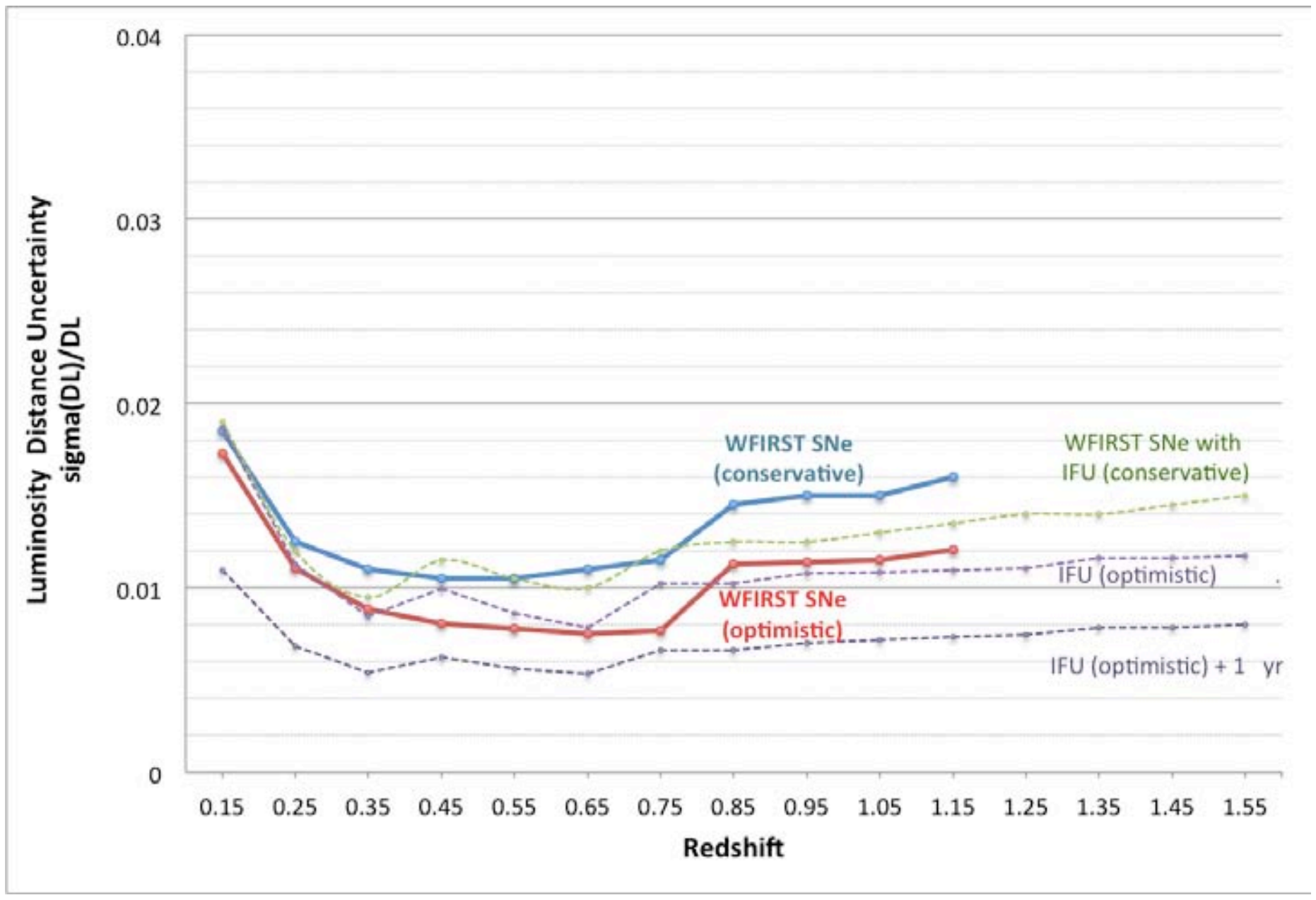

Figure 17: Projected WFIRST measurements of the expansion history of the universe using Type Ia supernovae as calibrated standard candles. Approximately 100 SNe Ia are measured for each redshift bin. The solid lines show the results from the current baseline WFIRST supernova program for conservative and optimistic (red) projections of the systematic uncertainty control achieved. Also shown (dotted lines) are possible alternative programs using an IFU instead of the slitless prism spectroscopy.

## Extended Bandpass Option: Extending the longwavelength cutoff beyond 2.0 μm:

In the course of its deliberations, a number of SDT members repeatedly expressed the opinion that one or another part of the WFIRST science program would be carried out more effectively if the WFIRST detectors had sensitivity redward of 2.1 microns and filters that cut off redward of 2 microns. But the SDT also understood that if the DRM calls for hardware that is more expensive than for JDEM-Omega, it is less likely to be funded. This IDRM adheres to the same wavelength cutoffs as in JDEM-Omega. In the year between this report and the final DRM, the SDT will consider further the tradeoffs associated with incorporating redder detectors and filters. In what follows we present some of the arguments presented by the SDT in favor of redder wavelengths.

*Scientific gains from redder response*

The cosmological science return of WFIRST would be yet greater were it sensitive beyond 2 microns. The zodiacal light foreground is a minimum at 3.5 microns, and distant galaxies and QSOs are red sources. For a partially resolved galaxy with an exponential scale length of 0.05 arc-seconds, and a spiral SED redshifted to z=2, the SNR with a 1.3 meter unobscured telescope is 2.05 times better at 3.54 microns compared to 1.77 microns. For a QSO with a constant $\nu F_\nu$, which is not quite as red as the z=2 galaxy, the SNR is 1.39 times better at 3.54 microns.

Even though shear measurements require resolution of sources, the SNR for measuring shear is 1.1 times better at 2.5 microns than at 1.77 microns for the galaxy above. For a larger galaxy with 0.1" exponential scale length the optimum SNR for shear is at 3.54 microns. Note that for an exponential disk, the half-encircled energy diameter is 3.32 times the scale length, while for an unobscured circular aperture the half-encircled energy diameter is $\approx\lambda/D$ = 0.33" when $\lambda$ = 2.1 microns for a 1.3 meter diameter. WFIRST will have a better PSF than DES or LSST out to 4 microns, both of which are specifically designed to measure weak lensing.

The improved SNR for detecting galaxies will facilitate the detection of high redshift clusters. A wider range of wavelengths will ease the separation of QSOs from stars.

A 3.5 micron capability for WFIRST will never suffer in comparison to ground-based observations. While the foreground at 2 microns is 1600 times smaller in space than on the ground, the foreground is 2.7 million times smaller from space in the thermal infrared. Hogg et al. 2000 spent 15 hours of integration time on the 10 meter Keck telescope [many nights with overheads] to find 11 sources at 3.2 microns, while the 40 cm WISE telescope catalogued 28 sources per second of wall clock time in its Preliminary Data Release.

*The advantages of redder response for a survey of the plane of the Milky Way*

NWNH and the EOS panel recommended that six months of the WFIRST mission should be devoted to a galactic plane survey, but did not elaborate on the nature or goals of that survey. The most important galactic plane science topics WFIRST could address generally fall into two broad categories:

1) galactic structure (spiral arm geometry and extent; disk scale length and scale height; disk warp) and

2) star formation (extent of the star-forming disk; star forming efficiency, IMF, and triggering mechanism as function of galactocentric distance and metallicity).

The best tracers for these goals are red-clump giants and young-stellar objects (YSOs) with IR excesses due to circumstellar dust disks. In the inner galaxy, source crowding near b = 0 is so large that even with WFIRST's small pixel size confusion will limit the gains over ground based surveys. However, over much of the outer galactic plane, source crowding is a small enough effect that WFIRST's great intrinsic sensitivity can be mostly achieved, allowing a wide area survey much deeper than any that currently exist or likely could be accomplished from the ground.

Even in the outer galaxy, however, extinction is still a strong impediment to one's ability to do the desired science, because color cannot be directly equated to stellar effective temperature. In order to identify the red clump giants and YSOs, it is necessary to have a filter set that allows one to break the degeneracy between effective temperature and reddening. The standard JHK filter set accomplishes this goal remarkably well for late type stars because J-H remains essentially constant for Teff < 5000 K, whereas H-K becomes redder for cooler stars (yielding the well-known hook-shaped or step-shaped locus of stars in the J-H vs. H-K color-color diagram). In this color-color diagram, therefore, stellar effective temperature (for spectral type > K0) and reddening vectors are well-separated, allowing one to accurately determine the extinction to each star. Similarly, YSO IR excesses typically begin at K band because the inner edge of the disk is set at the dust sublimation temperature of about 1300 K. This leads to a distinct YSO locus in the J-H, H-K diagram, allowing YSOs to be identified in wide-area NIR surveys even in the face of large and varying extinction.

The ability of a JHK filter set to do the desired galactic plane science has been well-documented in the published literature (see particularly Lucas et al. 2008; Meyer et al. 1997). Can one accomplish the same goals with a filter set that does not extend to 2.2 microns? There are some shorter wavelength color-color diagrams that provide some separation between reddening and Teff. However, any such survey will be depth limited by extinction to that of the shortest wavelength band - limiting the radial depth of the survey and diminishing or eliminating the advantage of WFIRST over a dedicated ground-based survey. Because the current WFIRST longest wavelength filter does not extend far enough to the red, it neither samples far enough outside the H- opacity minimum to yield a good temperature sensitive color, nor is sensitive to the IR emission from dust in the inner disk of most YSOs. Without a longer wavelength filter, WFIRST's ability to do the unique galactic plane science envisioned by the EOS report will be correspondingly limited.

## Appendix F References

## Appendix G Acronym List

| | |
|---|---|
| 2MASS | 2 Micron All Sky Survey |
| A&A | Astronomy and Astrophysics |
| ACS | Advanced Camera for Surveys |
| ACS | Attitude Control System |
| AJ | Astronomical Journal |
| ANGST | ACS Nearby Galaxies Survey Treasury |
| AO | Announcement of Opportunity |
| ApJ | Astrophysical Journal |
| ApJL | Astrophysical Journal Letters |
| ApJS | Astrophysical Journal Supplement Series |
| ASIC | Application Specific Integrated Circuit |
| AU | Astronomical Unit |
| BAO | Baryon Acoustic Oscillations |
| BEPAC | Beyond Einstein Program Assessment Committee |
| BigBOSS | Big Baryon Oscillation Spectroscopic Survey |
| C&DH | Command and Data Handling |
| CaF2 | Calcium Fluoride |
| CCD | Charged Coupled Device |
| CDR | Critical Design Review |
| CFHTLS | Canada – France – Hawaii Telescope Legacy Survey |
| CMB | Cosmic Microwave Background |
| CMOS | Complementary Metal Oxide Simi-conductor |
| COBE | Cosmic Background Explorer |
| DE | Dark Energy |
| DES | Dark Energy Survey |
| DETF | Dark Energy Task Force |
| DOF | Degree of Freedom |
| DRM | Design Reference Mission |
| DSN | Deep Space Network |
| DTAP | Detector Technology Advancement Program |
| E(B-V) | Extinction (B-V) |
| EDU | Engineering Development Unit |
| EE | Encircled Energy |
| EELV | Evolved Expendable Launch Vehicle |
| ELG | Emission Line Galaxy |
| EOL | End of Life |
| EOS | Electromagnetic Observations from Space |
| EPO | Education and Public Outreach |
| ESA | European Space Agency |
| ExP | Exoplanet Microlensing Survey with WFIRST |
| FGS | Fine Guidance Sensor |
| FoM | Figure of Merit |
| FoMSWG | Figure of Merit Science Working Group |
| FOR | Field-of-Regard |
| FOV | Field-of-View |
| FPA | Focal Plane Array |
| FSW | Flight Software |
| FY | Fiscal Year |
| GHz | Gigahertz |
| GI | Guest Investigator |
| GLIMPSE | Galactic Legacy Infrared Mid-Plane Survey Extraordinaire |
| GNC | Guidance Navigation and Control |
| GP | Galactic Plane |
| GSFC | Goddard Space Flight Center |
| HgCdTe | Mercury Cadmium Telluride |
| HLS | High Latitude Survey |
| HST | Hubble Space Telescope |
| I&T | Integration and Test |
| IA | Intrinsic Alignment |
| ICE | Independent Cost Estimate |
| IDRM | Interim Design Reference Mission |
| IFU | Integral Field Unit |
| ImC | Imager Channel |
| IMF | Initial Mass Function |
| IR | Infrared |
| IRAS | Infrared Astronomical Satellite |
| JCL | Joint Confidence Level |
| JDEM | Joint Dark Energy Mission |
| JWST | James Webb Space Telescope |
| KDP | Key Decision Point |
| KSC | Kennedy Space Center |
| L2 | Sun-Earth $2^{nd}$ Lagrangian Point |
| LCCE | Lifecycle Cost Estimate |
| LRD | Launch Readiness Date |
| LRG | Luminous Red Galaxy |
| LSST | Large Synoptic Survey Telescope |
| mas | Milli-Arc-Seconds |
| Mbps | Megabits per Second |
| MCR | Mission Concept Review |
| MEL | Master Equipment List |
| MNRAS | Monthly Notices of the Royal Astronomical Society |
| Mpc | Megaparsec |
| MOC | Mission Operations Center |
| MPF | Microlensing Planet Finder |
| NASA | National Aeronautics and Space Administration |
| NIR | Near Infrared |
| NIRSS | Near-Infrared Sky Surveyor |
| NRC | National Research Council |
| NWNH | New Worlds, New Horizons in Astronomy and Astrophysics |
| OTA | Optical Telescope Assembly |
| PASP | Publication of the Astronomical Society of the Pacific |
| PDR | Preliminary Design Review |
| Photo-z | Photometric Redshift |
| PSF | Point Spread Function |
| PSR | Pre-Ship Review |
| PZCS | Photo-Z Calibration Survey |
| QE | Quantum Efficiency |
| QLF | Quasar Luminosity Function |
| QSO | Quasi-Stellar Object (Quasar) |
| RFI | Request for Information |
| RMS | Root Mean Square |
| RSD | Redshift Space Distortion |
| S/C | Spacecraft |
| S/N | Signal/Noise |
| SCA | Sensor Chip Assembly |
| SCE | Sensor Cold Electronics |
| SCG | Science Coordination Group |
| SDO | Solar Dynamics Observatory |
| SDT | Science Definition Team |
| SDSS | Sloan Digital Sky Survey |
| SED | Spectral Energy Distribution |
| SiC | Silicon Carbide |
| SIM | Space Interferometry Mission |
| SIR | Systems Integration Review |
| SINGS | Spitzer Infrared Nearby Galaxies Survey |
| SN | Supernova |
| SNe | Supernovae |
| SNR | Signal to Noise Ratio |
| SpC | Spectrometer channel |
| SRR | System Requirements Review |
| SSR | Solid State Recorder |
| SUTR | Sample Up The Ramp |
| TBD | To Be Determined |
| Tbits | Terabits |
| TBR | To Be Resolved |
| TMA | Three Mirror Anastigmat |
| TRMD | Transition from Radiation to Matter Domination |
| UKIDSS | UKIRT Infrared Deep Sky Survey |
| US | United States |
| vDC | Volts Direct Current |
| VISTA | Visible and Infrared Survey Telescope for Astronomy |
| WD | White Dwarf |
| WFIRST | Wide-Field Infrared Survey Telescope |
| WISE | Wide-field Infrared Survey Telescope |
| WL | Weak Lensing |
| WMAP | Wilkinson Microwave Anisotropy Probe |
| YSO | Young Stellar Objects |
| ZnSe | Zinc Selenide |